\documentclass[useAMS,fleqn,usenatbib]{mnras}
\usepackage{setspace}  
\usepackage{newtxmath,newtxtext}

\usepackage[T1]{fontenc}
\usepackage{siunitx} 
\DeclareRobustCommand{\VAN}[3]{#2}
\let\VANthebibliography\thebibliography
\def\thebibliography{\DeclareRobustCommand{\VAN}[3]{##3}\VANthebibliography}

\usepackage{graphicx}	

\newcommand{\MKGPS}{\mbox{SMGPS}}
\newcommand{\SMGPS}{\mbox{SMGPS}}
\newcommand{\HI}{H\textsc{i}}
\newcommand{\HII}{H\textsc{ii}}
\newcommand{\Jb}{\mbox{Jy~beam$^{-1}$}}
\newcommand{\muJb}{\mbox{$\mu$Jy~beam$^{-1}$}}
\newcommand{\kms}{\mbox{km\,s$^{-1}$}}
\newcommand{\Msun}{\mbox{M$_\odot$}}

\title[SARAO MeerKAT GPS]{The SARAO MeerKAT 1.3 GHz Galactic Plane Survey}

\author[S. Goedhart et al.]{S.~Goedhart,$^{1,2}$\thanks{sharmila@sarao.ac.za} W.~D.~Cotton,$^{3,1}$ F.~Camilo,$^{1}$ M.~A.~Thompson,$^{4}$ G.~Umana,$^{5}$ M.~Bietenholz,$^{6,7}$ P.~A.~Woudt,$^{8}$
\newauthor
L.~D.~Anderson,$^{9,10,11}$
C.~Bordiu,$^{5}$
% ...
\\
\\
The authors' affiliations are shown in Appendix \ref{sec:affiliations}.
}

\date{Accepted XXX. Received YYY; in original form ZZZ}

\pubyear{2023}

\begin{document}
\label{firstpage}
\pagerange{\pageref{firstpage}--\pageref{lastpage}}
\maketitle

\begin{abstract}

We present the SARAO MeerKAT Galactic Plane Survey (SMGPS), a 1.3 GHz continuum survey of almost half of the Galactic Plane (251\degr\ $\le l \le$ 358\degr\ and 2\degr\ $\le l \le$ 61\degr\ at $|b|\le 1\fdg5$). SMGPS is the largest, most sensitive and highest angular resolution 1 GHz survey of the Plane yet carried out, with an angular resolution of 8\arcsec\ and a broadband RMS~sensitivity of $\sim$10--20 \muJb. Here we describe the first publicly available data release from SMGPS which comprises data cubes of frequency-resolved images over 908--1656 MHz, power law fits to the images, and broadband zeroth moment integrated intensity images. A thorough assessment of the data quality and guidance for future usage of the data products are given. Finally, we discuss the tremendous potential of SMGPS by showcasing highlights of the Galactic and extragalactic science that it permits. These highlights include the discovery of a new population of non-thermal radio filaments; identification of new candidate supernova remnants, pulsar wind nebulae and planetary nebulae; improved radio/mid-IR classification of rare Luminous Blue Variables and discovery of associated extended radio nebulae; new radio stars identified by Bayesian cross-matching techniques; the realisation that many of the largest radio-quiet WISE \HII\ region candidates are not true \HII\ regions; and a large sample of previously undiscovered background \HI\ galaxies in the Zone of Avoidance.

\end{abstract}

\begin{keywords}
catalogues --
Galaxy: general --
radio continuum: ISM --
radio continuum: stars --
radio lines: galaxies -- 
surveys
\end{keywords}





\section{Introduction}
\label{sec:intro}

Our understanding of the physical processes within the Milky Way Galaxy has seen steady progress through a succession of multi-wavelength surveys of the Galactic Plane. These surveys have been increasingly sensitive and at higher angular resolution, taking full advantage of new observatories such as VISTA and \emph{Herschel} and/or upgrades to existing facilities such as the Jansky Very Large Array.  The end result is a rich archive of both imaging and spectroscopic data from radio to X-ray wavelengths that covers a large fraction of the Milky Way Galaxy \citep[e.g.][]{Grindlay+2005,Carey+2009,Molinari+2010,Schuller+2009,Hoare2012,Irabor+2023}.

Radio-wavelength surveys are particularly powerful as the Galaxy is largely optically thin in the radio which means that we can study objects across (or even beyond) the Galaxy regardless of Galactic latitude. Moreover, radio photons are emitted via a range of physical processes (thermal brehmsstrahlung, synchrotron or gyro-synchrotron, recombination lines, ro-vibrational lines and hyperfine transitions) which enable us to study many different astrophysical environments in main sequence and evolved stars, young stellar objects, \HII\ regions, supernova remnants and all phases of the interstellar medium from ionised to molecular \citep[e.g.][]{Umana+2015b,Thompson2015}. 

Recent radio surveys such as CORNISH \citep{Hoare2012}, CORNISH-South \citep{Irabor+2023}, GLOSTAR \citep{Brunthaler+2021}, THOR \citep{Beuther+2016}, the Methanol Multi-Beam survey \citep[MMB;][]{Green+2009}, GLEAM \citep{Hurley-Walker+2019b} and SCORPIO \citep{Umana+2015} have given insights into the Galactic star formation rate \citep{Wells+2022}, identified new planetary nebulae and supernova remnants \citep{Fragkou+2018,Ingallinera+2019,Hurley-Walker+2019,Dokara+2021}, enabled the assembly of complete samples of ultracompact and compact \HII\ regions \citep{Urquhart+2018, Kalcheva+2018,Djordjevic+2019}, underpinned the discovery of new variable 6.7 GHz methanol masers \citep{Maswanganye2017} and revealed new optically thick hypercompact \HII\ regions \citep{Yang+2019}. The key features of these surveys are their high angular resolution ($\sim$ 2\arcsec--20\arcsec) and sensitivity ($\sim$ 0.1--2 mJy in continuum 1 $\sigma$). This range of
angular resolution is comparable to that
in visible to far-infrared (IR) surveys (e.g.\ IPHAS/VPHAS+, GLIMPSE, Hi-GAL), which enables straightforward multiwavelength analyses.
Milli-Jansky sensitivity is crucial to trace the bulk of the population of Galactic massive star formation regions via compact and ultracompact \HII\ regions that were the primary targets of CORNISH, CORNISH-South and GLOSTAR. 

One limitation that applies to many of the interferometric surveys mentioned above is that to cover a wide area in a reasonable length of observing time places constraints on the Fourier-transform plane,
or $uv$, coverage of the observations with resulting effects on image fidelity and dynamic range. This can be particularly apparent at low frequencies ($\sim$1 GHz) where the increased brightness of non-thermal Galactic and extragalactic sources places dynamic range limitations on surveys. 
For this and other reasons, there is a significant gap in the coverage of the Milky Way at GHz frequencies, particularly in terms of angular resolution and sensitivity.  In Quadrant I the deepest survey at 1--2 GHz is the THOR survey \citep{Beuther+2016}, which has an angular resolution of $\sim$20\arcsec\ and a 1-$\sigma$ point-source sensitivity of $\sim$0.4 m\Jb\ in a 128 MHz spectral window \citep{Bihr+2016}. Similarly, in Quadrant IV the Molonglo Galactic Plane Survey at 843 MHz has an angular resolution of 45\arcsec\ and a 1-$\sigma$ point-source sensitivity of $\sim$ 1 m\Jb. More recently, the Rapid ASKAP Continuum Survey \citep[RACS;][]{McConnell+2020} has observed the Galactic Plane as part of its all-sky survey programme with an angular resolution of 10\arcsec--47\arcsec\ (dependent on declination) and median sensitivity of 0.2 mJy PSF$^{-1}$ at 1357 MHz \citep{Duchesne+2023a}. 887 MHz RACS data are also available, although the Galactic Plane has not been catalogued due to source complexity \citep{Hale+2021}.

Going beyond these limitations is crucial for understanding the non-thermal radio populations within the Milky Way, especially stellar radio sources \citep{Umana+2015b} and compact, potentially young supernova remnants \citep{Gerbrandt+2014,Ranasinghe+2021}. Moreover, since constrained $uv$ coverage also results in limited image fidelity as well as dynamic range, our knowledge of extended low surface brightness populations (e.g.~\HII\ regions and old supernova remnants) is also affected. Indeed, the power of new facilities with dense instantaneous $uv$ coverage is exemplified in the MeerKAT Galactic Centre image \citep{Heywood+2022} which revealed both the striking complexity of this region and new populations of previously undiscovered radio sources.

It is with these goals in mind that we present the SARAO (South African Radio Astronomy Observatory) MeerKAT Galactic Plane Survey (SMGPS), an 8\arcsec\ angular resolution, $\sim 10$--$20 \, \mu$Jy\,beam$^{-1}$ root-mean-square (RMS) sensitivity, 1.3 GHz survey of almost half  of the Galactic Plane. The SMGPS was designed to cover the bulk of Galactic radio emission and to exploit the tremendous capabilities of the MeerKAT array to explore the 1 GHz radio population
with hitherto unavailable
sensitivity, angular resolution and image fidelity. SMGPS is primarily a continuum imaging survey using the 4096-channel continuum correlator mode of MeerKAT. However this spectral resolution also makes extragalactic \HI\ studies feasible, as shown in Section~\ref{sec:gps-hi}.

SMGPS is distinguished from the MPIfR-MeerKAT Galactic Plane Surveys \citep[MMGPS;][]{Padmanabh+2023}, which uses the subsequently developed commensal observing mode of MeerKAT to feed simultaneous pulsar search and imaging pipelines. The MMGPS L-band survey (MMGPS-L) is broadly similar to SMGPS in imaging terms, but shallower, with poorer $uv$ coverage, and covering a wider range of Galactic latitudes ($|b| <5\fdg2$ vs $|b| <1\fdg5$) and a narrower range of Galactic longitudes (260\degr\ $\le l \le$ 345\degr\ vs 251\degr\ $\le l \le$ 358\degr\ and 2\degr\ $\le l \le$ 61\degr).  Together the two surveys are highly complementary, not least for variability studies with a multi-year baseline between the two surveys, but also as  pathfinders to design and optimise future Square Kilometre Array observations.

In this paper we present the survey, including the initial SMGPS Data Release and some science highlights, describing the observations in Section~\ref{sec:observations}, the calibration, imaging and mosaicing procedures in Section~\ref{sec:dataredux}, and the principal data products and a through assessment of their quality in Section~\ref{sec:products}. In Section~\ref{sec:highlights} we present a selection of the Galactic science highlights of the \SMGPS, paying particular attention to areas where the SMGPS data makes a unique contribution to the state of the art. This includes the discovery of a new population of radio filaments (Section~\ref{sec:filaments}), new candidate supernova remnants (Section~\ref{sec:SNR}), new potential pulsar-wind nebulae
of youthful pulsars (Section~\ref{sec:PWNe}), planetary nebulae (Section~\ref{sec:pne}) and radio stars (Section~\ref{sec:stars}). In Section~\ref{sec:gps-hi} we describe the re-analysis of the SMGPS data to identify and study \HI\ galaxies in the Zone of Avoidance. Finally in Section~\ref{sec:conclusions} we present a summary and conclusions.

\section{Observations}
\label{sec:observations}

\begin{figure*}
    \centering
   \includegraphics[width=\linewidth, trim=1cm 0cm 1cm 0cm,clip]{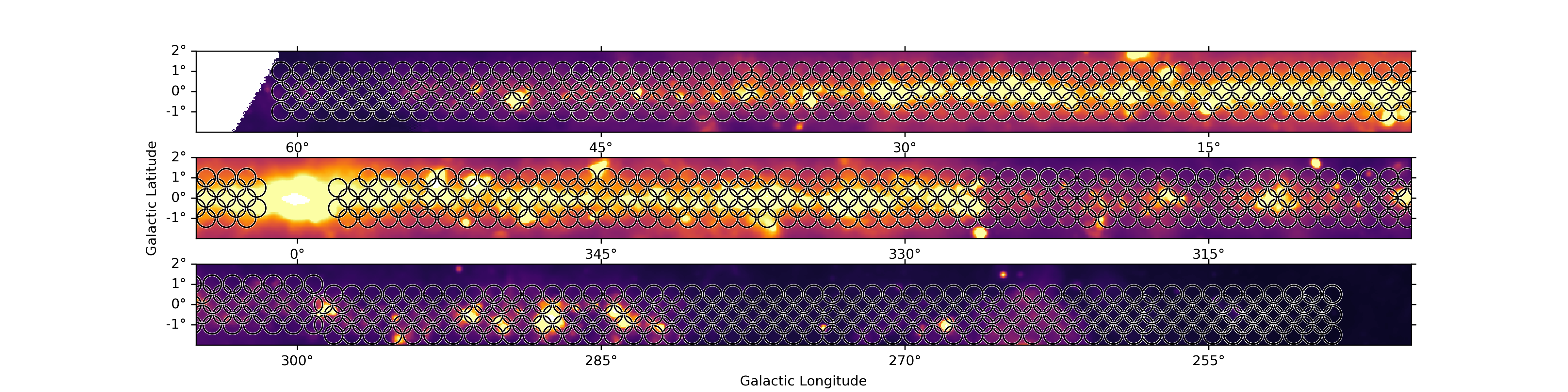}
    \caption{The area of the Galactic Plane covered by the SARAO MeerKAT Galactic Plane Survey. Individual pointing centres are shown by black circles, overlaid on a single-dish 1.4 GHz colourscale image from the CHIPASS survey \citep{CalabrettaSB2014}.} 
    \label{fig:svy_area}
\end{figure*}

The observations were carried out with the 64-antenna MeerKAT array in the Northern Cape Province of South Africa, which is described
in \citet{Jonas2016}, \citet{Camilo+2018}, and \citet{Mauch+2020}. The survey area was chosen to cover two contiguous blocks in Galactic longitude of 251\degr\ $\le l \le$ 358\degr\ and 2\degr\ $\le l \le$ 61\degr. Each block covers a Galactic latitude range of approximately $|b| \le 1\fdg5$, with the first block chosen to follow the Galactic warp in a similar manner to the Hi-GAL survey \citep{Molinari+2010}. The \SMGPS\ survey area is shown in Fig.\ \ref{fig:svy_area} with individual pointing centres represented as circles. The Galactic Centre was not observed as part of this survey, being instead observed separately and described in \citet{Heywood+2022}.

The \SMGPS\ observations were made between 2018 Jul 21 and 2020 Mar 14 using the L-band receiver system, covering a frequency range 856--1712 MHz with 4096 channels, and an 8 second correlator integration period. 
The correlated data consists of all four combinations of the two orthogonal linearly polarized feeds. The observations were executed over a series of $\sim$10 hour sessions, cycling among 9 pointings of a hexagonal grid spaced by $0\fdg494$ to provide uniform sensitivity. Each pointing was visited multiple times over the session, providing good $uv$ coverage and
an on-source time of $\approx$1 hour per pointing. 
The individual pointings were then formed into mosaics,
extending to $|b|=1\fdg5$, having full sensitivity 
to $|b|\approx 1\fdg2$, and slightly reduced beyond (for the warp-offset fields these latitude ranges are shifted southwards by 0\fdg5).

Observations in the fourth quadrant were initially chosen to have scan durations of 3 minutes per pointing, with a complex gain calibrator observed every 30 minutes for 65 s. Later observed scans for each pointing were extended to 10 minutes. The  band-pass and flux calibrators PKS~B1934$-$638 or PKS~J0408$-$6545 were observed for five minutes every 3 hours.
The polarization calibrator 3C\,286 was observed for some sessions, when visible to the MeerKAT array. 
Typically at least 60 antennas were online for each session. A few observations were split into multiple sessions due to scheduling constraints. 

Since these data were taken during an active development and commissioning stage of the MeerKAT array, a number of instrumental and calibration issues were found to affect the quality of different subsets of the observations. The main effects combine to result in systematic errors in the astrometry of the data at around the arcsecond level. We have corrected the astrometry errors as much as possible in our data reduction and post processing (see Section \ref{sec:astrometry} for a full discussion of the astrometric accuracy of the processed data), but we describe the underlying instrumental issues here so that users fully understand the limitations of the survey data. 

The earliest data 
taken had a labelling error of 2 seconds in time and a half channel in frequency, resulting in incorrectly calculated $u,v,w$ baseline coordinates. These errors result in rotated and mis-scaled images with apparent position errors of up to to 2\arcsec\ at the edge of each pointing. After the discovery of the labelling errors, subsequent data were corrected but earlier data were not. The affected data lie in the fourth quadrant between $l = 320$\degr\ and 358\degr\ (mosaics G321.5 to G357.5 --- in this paper we refer to mosaics by their center longitude). The labelling errors are mitigated to some extent by the mosaicing process carried out in the data reduction. The mosaics place low weight on the outer parts of the pointing images and so the labelling errors are only a significant contribution at the extremes of Galactic latitude covered by the survey where the mosaics are dominated by single pointings.

A second source of astrometric error was discovered in the initial MeerKAT calibrator list, which included calibrators with position errors of up to several arcseconds. This resulted in a constant positional offset in the pointings to which these calibrators were applied. The offset was corrected by modifying the reference pixel of the World Coordinate System of the affected pointings prior to the mosaicing process.

Finally, a more subtle problem resulted from the low accuracy of the correlator model used in the delay tracking of the observations. An insufficient number of mostly Earth orientation terms were included in the model. In order to get calibrators that dominate the field of the MeerKAT antennas, many were 10\degr\ or more from the target pointing, which can result in constant position offsets of up to several arc-seconds in the target-pointing images. Since nearby regions of the Galactic Plane used the same calibrator these errors will be spatially correlated. We have not applied a correction for this potential offset, but as will be seen in Section \ref{sec:astrometry} we do not see a substantial error in the astrometric accuracy of the survey data.

\section{Data reduction}
\label{sec:dataredux}

The observational data were calibrated and imaged with a simple and straightforward procedure as described in \citet{Mauch+2020} and \citet{Knowles+2022}. All calibration and imaging used the Obit package\footnote{\url{http://www.cv.nrao.edu/~bcotton/Obit.html}} \citep{Cotton2008}.  We describe the calibration, imaging and mosaicing process in the following subsections.

\subsection{Calibration and editing}

Data affected by interference and/or equipment malfunctions were identified using the procedures outlined in  \citet{Mauch+2020} and were removed from further analysis. The remaining data were calibrated for group delay, band-pass and amplitude and phase as described in \citet{Knowles+2022}. The reference antenna was picked on the basis of the best signal-to-noise ratio, S/N, in the band-pass solutions. The flux-density scale is based on the \citet{Reynolds1994} spectrum of
PKS~B1934$-$638: 
\begin{equation}
\begin{split}
  \log(S) = -30.7667 + 26.4908 \bigl(\log \nu\bigr)
  - 7.0977 \bigl(\log \nu\bigr)^2 \\
  + 0.605334 \bigl(\log \nu\bigr)^3,
\end{split}
\end{equation}
where $S$ is the flux density in Jy and $\nu$ is the frequency in MHz. 
After small time and frequency offsets were discovered (see Section \ref{sec:observations}), subsequent data-sets were corrected before imaging. From long-term observatory monitoring of calibrators, the flux density calibration uncertainty is believed to be $\sim$5\%.

\subsection{Imaging}
\label{sec:imaging}

Individual pointings were imaged with the  wide-band, wide-field Obit imager MFImage \citep{Cotton2019}. Each individual pointing was gridded in the J2000 FKS equatorial coordinate system. MFImage corrects for the curvature of the sky using facets. Multiple frequency bins were imaged independently and CLEANed jointly to accommodate the frequency dependencies of the sky brightness distribution and the antenna gains. A resolution of $7.5\arcsec$--$8\arcsec$, nearly independent of frequency, was obtained using a frequency dependent taper. 

For each pointing, the sky within 0\fdg8--1\degr\ radius was fully imaged with outlying facets to cover bright sources from the SUMMS 843 MHz catalog \citep{Mauch+2003} within 1\fdg5 of the pointing center. Two iterations of 30 second solution-interval phase-only self calibration were used.  Amplitude and phase self calibration were added as needed. The final Stokes $I$ CLEAN  used 250,000 CLEAN components with a loop gain of 0.1 and CLEANed typically to a depth of 100--200 \muJb. No direction dependent corrections were applied. Robust weighting ($-1.5$ in AIPS/Obit usage) was used to down weight the
very densely sampled inner portion of the $uv$ coverage. The resulting resolution, as mentioned, was in the range 7\farcs5--8\farcs0. Imaging was done using 14 channel images each with 5\% fractional bandwidth.
 
A small number of mosaics were also processed with polarisation calibration as described in \citet{Knowles+2022} and \citet{plavin+2020}. This calibration used 3C\,286 as a polarized calibrator, and PKS~B1934$-$638 as an unpolarized one, and involved jointly solving for the polarization of the
session complex gain calibrator and the instrumental polarization parameters
using all calibrators. This calibration was applied to the data for
which individual pointings were imaged in Stokes $I, Q, U$ and $V$ and
formed into mosaics as the rest of the survey was.  
Although the initial survey observations were not specifically designed with polarisation calibration in mind, the instrumental stability of the MeerKAT system means that it is possible to recover the polarisation calibration from initial noise injections performed at the start of each observation. 
For those tiles that were processed, Stokes $I, Q, U$ and $V$ mosaics are included in the data release presented in this paper to demonstrate the potential of a future data release with full polarisation calibration.

All four polarization products are available in the raw data for all
pointings, it is only due to computational limitations that we only
calibrated a small subset of the data for polarization.  As an example
of what is possible, we show in Appendix \ref{sec:W44} images from a single pointing centred approximately on the supernova remnant W44, which was
recalibrated for polarization.

The dynamic range can be limited by very strong sources ($>$ few hundred m\Jb), especially if several are present in a given pointing. The self calibration applied cannot correct for direction dependent gain effects (DDEs),
which may be the cause of some of the
remaining artifacts seen in the individual images. The DDEs are thought to be dominated by pointing errors, asymmetries
in the antenna pattern, and ionospheric refraction. For images that are not dynamic range limited, the off-source RMS brightness is $\sim$10--15 \muJb.

MeerKAT has extensive short baseline coverage allowing the imaging of
extended emission. However, there is a minimum baseline length, which in wavelengths is frequency dependent; more extended emission is recovered at lower
frequencies than higher ones. Regions of bright extended emission which are not well sampled by the $uv$-coverage will have negative bowls surrounding them. These bowls will be deeper at higher frequencies introducing an artificial steepening of the apparent spectrum. Angular scales up to 10\arcmin\ are generally well recovered although the estimate of the spectral index may be considerably in error and care must be taken to derive accurate spectral indices from the data. We discuss the inherent limitations of spectral indices derived from the data in Section \ref{sec:spi}.

\subsection{Mosaics}
\label{sec:mosaic}

The individual pointing images were collected into linear $3\degr \times 3\degr$ mosaics. To allow convenient cross-comparisons with other Galactic Plane surveys the mosaics were gridded on the IAU 1958 Galactic coordinate system. The mosaic formation process for each image plane is given by the summation over overlapping pointing images: 
\begin{equation}
M(x,y,\nu)\ =\ {\frac{\sum_{i=1}^{n}A_{i,\nu}(x,y)\
    I_{i,\nu}(x,y)}{\sum_{i=1}^{n}A^2_{i,\nu}(x,y)}},
\end{equation}
where $A_{i,\nu}(x,y)$ is the array gain of pointing $i$ in direction $(x,y)$ and at frequency $\nu$, $I_{i,\nu}(x,y)$ is the pointing $i$ pixel value interpolated to direction $(x,y)$ in frequency plane $\nu$, and $M$ is the mosaic cube\footnote{Note that the antennas have already applied one power of the antenna gain to the sky during the observations.}. Quality checks were applied to each frequency plane of each pointing before being added to the mosaic; frequency planes with more than 50$\times$ the mean RMS were excluded from the mosaic. The images of the individual pointing were convolved to 8\arcsec\ FWHM resolution before combining into mosaics.

\begin{figure*}
    \centering
    \includegraphics[width=\linewidth]{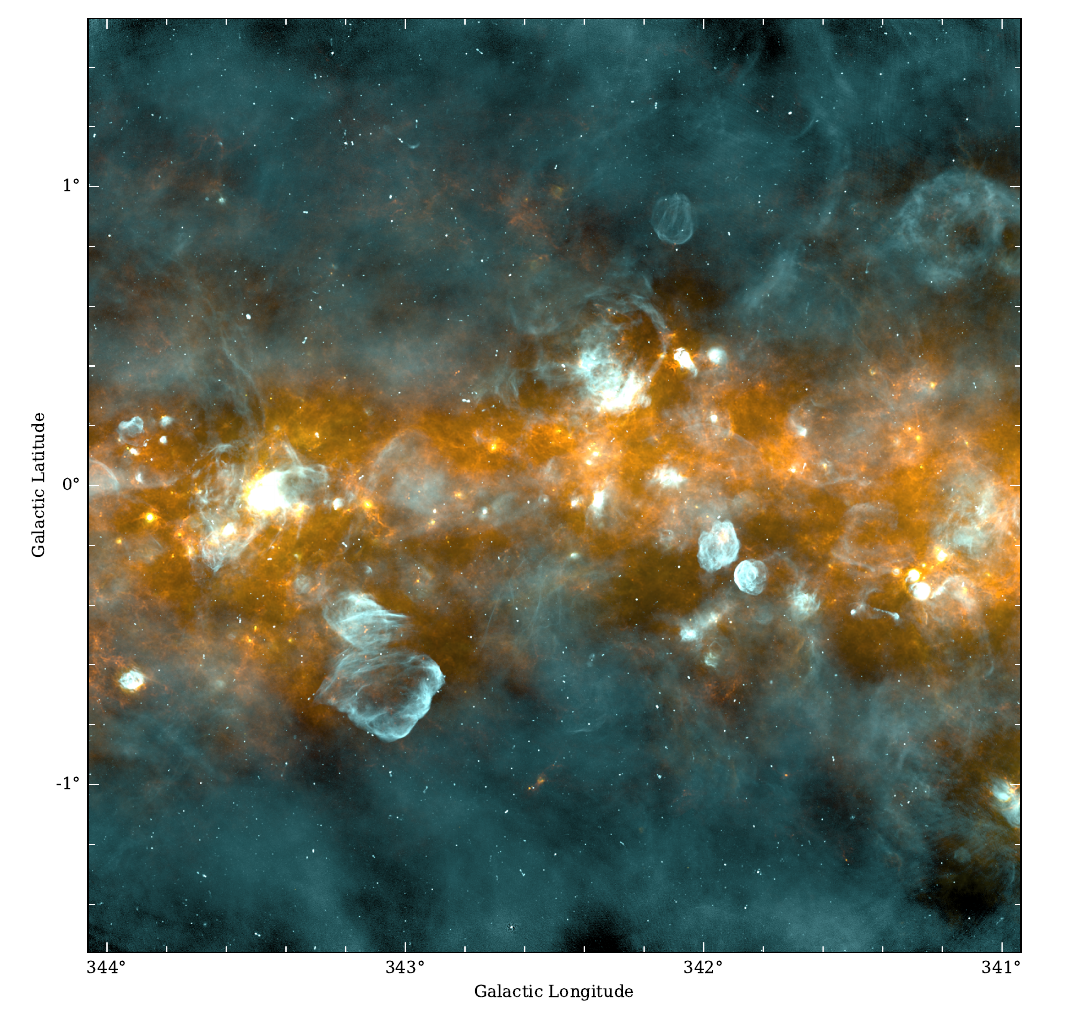}
    \caption{An example Moment 0 mosaic from SMGPS Data Release 1. The image is a combination of SMGPS 1.3 GHz (white, blue and cyan, which encode bright, mid-range and low brightnesses respectively) and \emph{Herschel} Hi-GAL 70 $\mu$m (yellow) and 250 $\mu$m (orange) emission. This colour combination was chosen to differentiate between SNRs and \HII\ regions; the former appear mostly cyan whereas the latter appear with white cores surrounded by yellow/orange emission.}
    \label{fig:DR1example}
\end{figure*}

As part of the mosaic formation process a correction was applied for the primary antenna beam shape as part of the weighting of overlapping pointing images. The shape of the individual antenna power pattern is reasonably well understood although the array's effective power pattern is less so. Although array effective beam studies in \citet{Mauch+2020} show that the inner beam is very close to that of an individual antenna, application of this pattern in \citet{Knowles+2022} further out in the beam results in physically implausible spectra. The effect is similar to that expected from the known pointing errors of the antennas in the array. The primary beam corrections are not thought to be accurate past a radius of $\sim36$\arcmin. This region of the beam is given a low weight in the mosaic formation so the mosaics should not be adversely impacted by errors in the assumed array beam pattern except possibly for the extreme values in latitude which are dominated by data far from the nearest pointing center.

\section{\SMGPS\ Data Release}
\label{sec:products}

With the publication of this paper we make available the first data release (DR1) of the \SMGPS\@. DR1 consists of mosaiced data cubes assembled from the individual pointings, and `zeroth-moment' integrated intensity images derived from the mosaics. The data cubes and images are presented in a common $3\degr \times 3\degr$ format with $7500 \times 7500$ pixels
of $1\farcs5$ each.  Fig.~\ref{fig:DR1example} presents an example of one of the mosaics (G342.5), combining a MeerKAT 1.3 GHz Moment 0 image with \emph{Herschel} Hi-GAL 70 $\mu$m and 250 $\mu$m images to illustrate the thermal and non-thermal emission present in the image.
Fig.~\ref{fig:DR1example} reveals a multitude of complex and striking emission on multiple angular scales, from large supernova remnants and \HII\ regions to compact radio galaxies. 

Advanced data products will be published separately and include catalogues of compact point-like sources, 
extended sources, and filamentary sources. All
DR1 data products are available through a DOI\footnote{\url{https://doi.org/10.48479/3wfd-e270}.
When using DR1 products, this paper should be cited, and the MeerKAT telescope acknowledgement included.}
and the raw visibilities are also hosted on the SARAO Data Archive\footnote{\url{https://archive.sarao.ac.za/}} under project code SSV-20180721-FC-01.

Here we describe the individual data products in DR1 (Sections \ref{sec:cubes} and \ref{sec:mom0}), discuss the accuracy of the flux density calibration (Section~\ref{sec:fluxcal}) and astrometry (Section~\ref{sec:astrometry}), and the limitations of in-band spectral indices (Section~\ref{sec:spi}).

\subsection{Data Cubes}
\label{sec:cubes}

As mentioned, the standard data product is a set of 3\degr\ $\times$ 3\degr\ overlapping
FITS mosaics, assembled from the individual pointing images as described in Section \ref{sec:mosaic}. Two versions of the Stokes $I$ mosaics are included in DR1: (1) cubes containing observed flux densities for each of the 14 frequency planes described in Section~\ref{sec:imaging} and Table \ref{tbl:channelfreqs}; and (2) 
fitted parameter cubes, containing planes of broadband flux density, spectral index and their associated errors,  obtained using a power-law fit to the frequency planes (described below).
For the Stokes $Q$, $U$ and $V$ mosaics, only the first of these are provided.

\begin{table}
\begin{center}
\caption{Central frequencies and bandwidth of the 14 fractional bandwidth flux density images contained in the SMGPS data cubes. Note that channels 7 and 8 are completely flagged throughout all data cubes due to persistent radio frequency interference from GPS satellites.}
\label{tbl:channelfreqs}
\begin{tabular}{ccc}\hline\hline
Channel No. & Central Frequency & Bandwidth \\
 & (MHz) & (MHz) \\ \hline
  1 &  908.142 & 43.469 \\                 
  2 &  952.446 & 43.469 \\
  3 &  996.751 & 43.469 \\
  4 & 1043.563 & 48.484 \\
  5 & 1092.884 & 48.484 \\
  6 & 1144.712 & 53.500 \\
  7 & 1199.048 & 53.500 \\
  8 & 1255.892 & 58.516 \\
  9 & 1317.751 & 63.531 \\
 10 & 1381.700 & 62.695 \\
 11 & 1448.157 & 68.547 \\
 12 & 1520.049 & 73.562 \\
 13 & 1594.446 & 73.563 \\
 14 & 1656.724 & 49.320 \\ \hline
\end{tabular}
\end{center}
\end{table}

The data format of the frequency plane cubes is that outputted by MFImage and is described in detail in
Obit Memo 63\footnote{\url{https://www.cv.nrao.edu/~bcotton/ObitDoc/MFImage.pdf}}. In brief, the frequency plane cubes contain 16 data planes comprising a weighted average flux density plane at an effective frequency of 1359.7 MHz, a spectral index plane, and each of the individual frequency planes in  order of increasing frequency.  
The central frequencies and bandwidths of each of the frequency planes are given in Table \ref{tbl:channelfreqs}. Since the steep spectrum Galactic emission contributes a
significant portion of the system temperature, the weighting used in the broadband flux density plane was by
1/RMS of the frequency plane off-source background RMS brightness rather than the more usual
1/$\mbox{RMS}^2$.  The weights used were the average RMS values over all
mosaics (scaled to the overall RMS in each mosaic). This weighting
results in an effective frequency of 1359.7 MHz. 

The first two planes of the fitted parameter cubes, broadband flux density at 1359.7 MHz and spectral index (Stokes $I$ only), $\alpha$, are identical to the first two planes of the frequency plane cubes described above. We use the $S_\nu \propto \nu^\alpha$ definition of $\alpha$. Values of $\alpha$ were obtained by fitting by nonlinear least squares to each frequency plane of the frequency plane cubes using the weighting used for Stokes $I$\@.
This fitting was done only for pixels with at least 500 \muJb\ of
broadband Stokes $I$, with the remainder being blanked.  The resulting
fitted values of $\alpha$ were accepted if the addition of $\alpha$  in the
fit did not increase the $\chi^2$ per degree of freedom, with any
pixels failing this criterion also blanked. The third plane contains the error estimate for the broadband flux density plane. The fourth and fifth planes are the least squares error estimate and $\chi^2$ of the spectral index fit. These planes  are blanked for pixels with no valid spectral index.

To maintain a fixed effective frequency and consistent spectral index fit, the 1359.7 MHz flux density and spectral index values have been calculated only for those pixels containing information in the highest frequency plane (i.e.~channel 14 at 1656 MHz). As the primary beam FWHM reduces with increasing frequency, outlying parts of the mosaics do not contain data across the full range of frequencies (a graphical explanation of this is shown in Fig.~\ref{fig:momzero}). 

The filename convention for the frequency plane cubes is $<$pos$>$ I\_Mosaic.fits where $<$pos$>$ is the
Galactic coordinate of the center of the mosaic, e.g.\ G339.5+000. 
The mosaics with polarization calibration additionally include separate frequency plane cubes for $Q, U$ and $V$ with ``Q'', ``U'', or ``V'' substituted for ``I'' in the file name.
Frequency plane cubes with fixed timing and labelling errors (Section~\ref{sec:observations}) are denoted as IFx instead of I; all polarized products are corrected. The fitted parameter cubes are named as $<$pos$>$I\_refit.fits, with again Fx denoting cubes with fixed timing and labelling errors.

\subsection{Zeroth moment images}
\label{sec:mom0}

\begin{figure*}
    \centering
    \includegraphics[width=\linewidth]{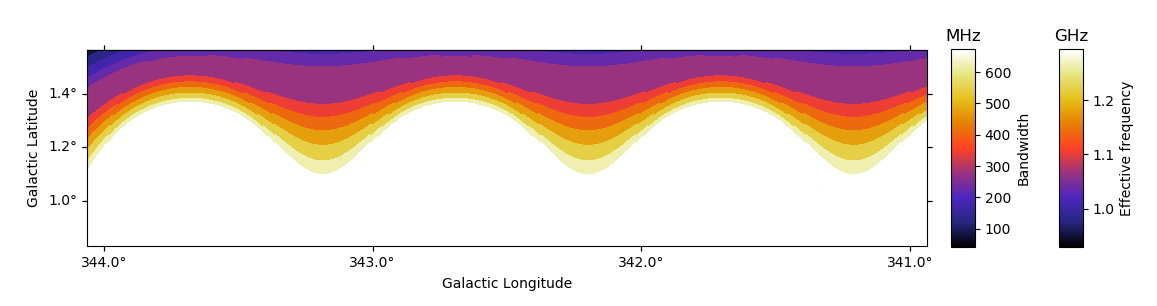}
    \vspace*{-0.5cm}
    \caption{A portion of the G342.5 mosaic to illustrate the variation in effective bandwidth and frequency of the zeroth moment images caused by the variation in the primary beamwidth.}
    \label{fig:momzero}
\end{figure*}

In addition to the data cubes discussed in the previous section we also make available zeroth moment integrated intensity images. The rationale behind these images is to provide an easily accessible standard FITS data product that encompasses the largest possible sky area. As described in Section~\ref{sec:cubes}, the weighted average Stokes $I$ plane contained within the data cubes is computed only for pixels where there is a measurement in the highest frequency plane images. The large difference in the primary beamwidth between the lowest and highest frequencies observed by MeerKAT means that the Stokes $I$ plane presented in the survey data cubes misses the extremes in Galactic latitude that were only observed at the lower frequencies.

The zeroth moment images were calculated on a pixel by pixel basis by summing the product of the flux density and bandwidth in each fractional bandwidth image, which was then weighted by the total bandwidth of all the fractional bandwidth images. The zeroth moment $M_{0}$ is then given by
\begin{equation}
M_{0} = \frac{\displaystyle\sum_{i}\, S_{i} \Delta\nu_{i}}{\displaystyle\sum_{i}\, \Delta\nu_{i}},
\end{equation}
where $S_{i}$ and $\Delta\nu_{i}$ are the flux density and bandwidth of each pixel in frequency plane $i$. This represents a bandwidth-weighted integrated intensity over all available pixels in each frequency plane in the frequency plane cubes. We note in passing that this is the astronomical definition\footnote{The astronomical definition of image moments is `off-by-one' compared to the mathematical definition of moments, e.g.~\url{https://casa.nrao.edu/docs/casaref/image.moments.html}} of moment zero as integrated intensity ($\int S \,{\rm d}\nu$) as opposed to the mathematical definition as the mean ($\left<S\right>$). As the zeroth moment images are derived from the data cubes described in Section~\ref{sec:cubes}, they have the same world coordinate system and size (i.e. $3\degr \times 3\degr$).

Pixels that were flagged (e.g.\ because they lie outside the imaged area or are affected by RFI) do not contribute to the zeroth moment. It is important to note that as each pixel in the zeroth moment can contain contributions from different frequency plane images, different pixels can have differing total bandwidths or effective frequencies. This effect is most apparent at the extremes in latitude of each image where contributions are predominantly from lower frequencies (see Fig.~\ref{fig:momzero}). To enable these effects to be taken into account in future data analysis we provide corresponding images of the effective central frequency and total bandwidth for each zeroth moment image as part of our data release. In general, the zeroth moment images have largely constant effective frequencies and bandwidths of 1293 MHz and 672 MHz for Galactic latitudes $|b| \le 1\fdg1$ (except near some bright sources as described below). These values steadily decrease towards the edges of the zeroth moment images as the higher frequency planes do not contribute to the zeroth moment, reaching values of 908 MHz and 43 MHz respectively at the extremes (i.e.\ where only the first 908 MHz frequency channel is present). Caution must also be taken near bright emission where flagging of particular channels or regions may also reduce the effective frequency and bandwidth of the moment zero images near bright sources.

\subsection{Flux density calibration}
\label{sec:fluxcal}

In this section we assess the accuracy of the flux densities extracted from the \SMGPS\ by comparing to the JVLA THOR survey \citep{Beuther+2016}. THOR covers a similar observed frequency range to the \SMGPS\ and, in particular, the 1.31 GHz channel of THOR is close in frequency to the 1.29 GHz median effective frequency of the MeerKAT GPS zeroth moment images. The \SMGPS\
observations are tied to the primary flux calibrator PKS~B1934$-$638 whereas the JVLA THOR observations are tied to the JVLA primary flux calibrator 3C\,286 \citep{Beuther+2016, Wang+2020} 
with the two calibrators being tied to the same flux density scale
by \citet{Reynolds1994}.
Comparing the two surveys thus enables an independent measurement of the systematic calibration errors between JVLA and MeerKAT.

Samples of bright (S/N $\ge$ 10), isolated (separated by at least 1\arcmin\ from the nearest neighbour) point sources were extracted from the THOR catalogue given in \citet{Wang+2020}, and a catalogue of \SMGPS\ point sources was constructed by running the Aegean algorithm \citep{Hancock+2012} on the zeroth moment images. The isolation constraint was to preferentially select radio sources that were not associated with extended complexes and for which the flux densities could be determined more accurately. We restrict our analysis to point sources in THOR and SMGPS to ensure that we are comparing like-for-like flux densities. The THOR catalogue contains only peak brightness values in Jy\,beam$^{-1}$, but for point sources this is also equal to their integrated flux density in Jy.

The SMGPS and THOR  catalogues were then cross-matched against each other with a matching radius of 2\arcsec. We made no attempt to correct for any variability between THOR and \SMGPS\ epochs (THOR was observed during 2012--2014 and \SMGPS\ during 2018--2020) or spectral index variations between the 1.31 and 1.29 GHz effective observing frequencies (which would amount at most to $\sim$2\% for the most extreme spectral indices).

\begin{figure}
    \centering
    \includegraphics[width=\linewidth]{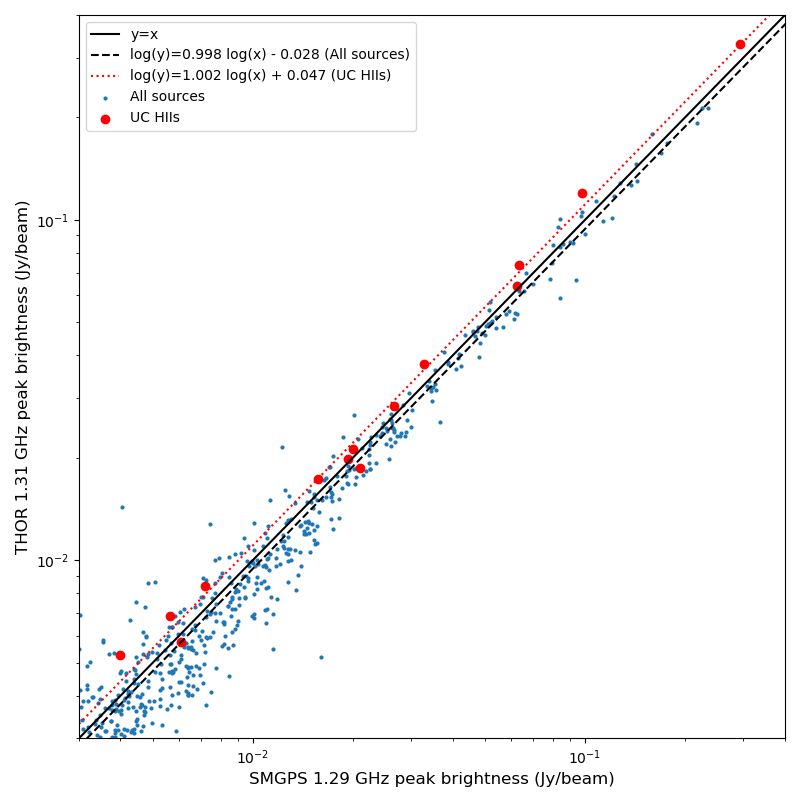}
    \caption{A comparison of the flux densities  
    of isolated point sources from the \MKGPS\ and THOR surveys (at 1.29 GHz and 1.31 GHz respectively). Each survey catalogue was filtered to retain only sources with S/N $>10$. The line of equality is shown by a solid black line, the best orthogonal distance regression fit to all sources by a dashed line, and the corresponding fit to the UC\,\HII\ region subsample by a dotted line. The values for the slopes and intercepts derived from the regressions are shown in the figure label.}
    \label{fig:fluxcomp}
\end{figure}

Fig.\ \ref{fig:fluxcomp} shows a plot of the THOR 1.31 GHz peak flux density
against the corresponding \SMGPS\ 1.29 GHz peak flux density, together with the line of equality 
and a best fit regression. We use an orthogonal distance regression to take into account the errors in both THOR and SMGPS peak flux densities and carry out the fitting in log--log space to avoid bright sources dominating the fit.  As can be seen in Fig.\ \ref{fig:fluxcomp}, the relationship between the THOR and \SMGPS\ peak flux densities is linear and close to $y=x$, deviating only at the level of a few percent. There is increasing scatter towards lower flux densities, which may be due to the presence of uncorrected negative bowls in the SMGPS images. THOR includes zero-spacing information from Effelsberg 100m single-dish mapping and is thus less affected by negative bowls. Alternatively this may indicate source variability between the observation epochs of THOR and SMGPS. The dominant constituent of the sample at $\sim$ mJy flux densities are likely to be radio galaxies \citep{Hoare2012,Anglada+1998,Condon+1984} of which Active Galactic Nuclei (AGN) are well known to be variable on timescales from minutes to years \citep{dennett-thorpe2002}. The increased scatter at low fluxes could be due to intrinsic source variability, particularly as our isolation and flux constraints may have led to a selection bias toward these source types.

In order to check the latter hypothesis we analysed subsets of the THOR and SMGPS samples, selected to be known UC\,\HII\ regions from the CORNISH survey \citep{Hoare2012,Urquhart+2013}. UC\,\HII\ regions are known to be variable, but not on the sub-decade timescale between THOR and SMGPS measurements. The UC\,\HII\ subsample again shows a close to linear relationship, but with a much smaller RMS scatter (see Fig.~\ref{fig:fluxcomp}), confirming the hypothesis that much of the scatter in the larger sample is due to intrinsic source variability. For the UC\,\HII\ subsample the  intrinsic scatter in the relation between THOR and SMGPS flux densities is around 4\% and so we consider this to be the flux calibration uncertainty of the SMGPS.

\subsection{Astrometric accuracy}
\label{sec:astrometry}

In Section \ref{sec:observations} we outlined two issues that could affect the astrometric accuracy of the data cubes and images (timing and frequency labelling errors, incorrect calibrator positions). In this section we investigate the astrometric accuracy of the \SMGPS\ both through an internal cross-comparison and comparisons with CORNISH \citep{Hoare2012}, CORNISH-South \citep{Irabor+2023} and the International Celestial
Reference Frame \citep[ICRF;][]{Charlot+2020} catalogues.

In summary we find that the astrometry of the \SMGPS\ is  accurate to an overall  level of $\sim$ 0\farcs5, with possibly a small systematic offset of around 0\farcs1--0\farcs3. The positions of individual sources and particularly sources at high or low Galactic latitude in those mosaics affected by timing and frequency errors may be affected up to $\sim$1\farcs5 (19\% of the \SMGPS\ FWHM resolution). Users who require much greater astrometric precision are advised to take particular care with their analysis.

\subsubsection{Timing and frequency labelling errors}

One of the main sources of astrometric error in the SMGPS is due to a timing and frequency labelling error of 2 seconds of time and half a channel of frequency. This results in an apparent rotation of each affected pointing image of up to 2\arcsec\ at the edge of the image. The affected data were the earliest data to be taken and lie between longitudes of 320\degr\ and 358\degr. Collecting the affected pointings into mosaics is expected to mitigate the errors due to the low weight placed on the edges of each individual pointing. We investigate these potential errors using the G321.5 mosaic, which was processed twice --- once with the timing and frequency errors uncorrected and once with the errors corrected. Comparing the positions of point sources in both versions of this mosaic allows us to quantify the potential astrometric error resulting from this effect.

\begin{figure}
    \centering
    \includegraphics[width=\columnwidth]{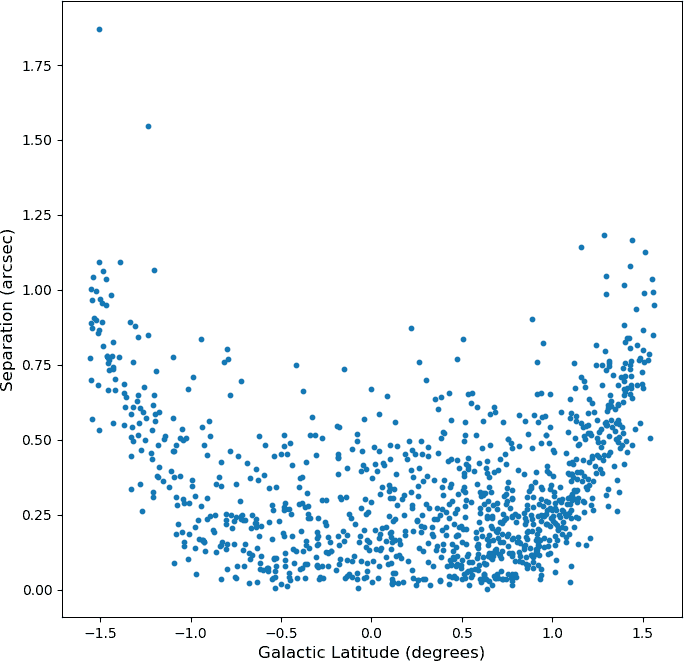}
    \caption{Angular separation of bright point sources on the timing and frequency corrected mosaic G321.5 from their apparent positions on the corresponding
    uncorrected mosaic, plotted against Galactic latitude. A small systematic offset at the high latitude edges of the mosaics can be seen, but is $\le$ 2\arcsec\ even at the extremes.}
    \label{fig:tfcomp}
\end{figure}

 Aegean \citep{Hancock+2012} catalogues of the corrected and uncorrected G321.5 mosaics were cross-matched with a maximum  radius of 8\arcsec. Each catalogue was prefiltered to contain point sources with a signal-to-noise ratio greater than 10 to reduce the intrinsic positional uncertainty. Fig.\ \ref{fig:tfcomp} shows the separation between matched point sources plotted against Galactic latitude. As can be seen the overall effect of the timing and frequency errors is small, with a median source separation of 0\farcs3. As expected, the source separation increases towards the high and low latitude edges of the mosaics where the mosaic mostly depends upon a single pointing. However, even at its most extreme value the positional shift resulting from the timing and frequency error --- which affects only the data in the longitude range 320\degr--358\degr\ --- is less than 1\farcs9.

\subsubsection{Cross-comparisons with CORNISH, CORNISH-South and the ICRF}

We estimate the overall astrometric accuracy across the entire SMGPS survey by comparing to the CORNISH/CORNISH-South surveys \citep{Hoare2012,Irabor+2023} and a sample of sources drawn from the ICRF
\citep[third realization;][]{Charlot+2020}.

Aegean catalogues of isolated (by more than 1\arcmin) point sources from the SMGPS were cross-matched against similar isolated point sources taken from the reliable (S/N $\ge$ 7) catalogues of 5 GHz point sources drawn from the CORNISH \citep{Hoare2012} and CORNISH-South \citep{Irabor+2023} surveys. The point sources were selected to be isolated to avoid confusion with nearby objects and in addition the CORNISH and CORNISH-South catalogues were prefiltered to only include source types expected to have similar compact or point-like morphologies at the different frequencies of SMGPS and CORNISH/CORNISH-South (ultracompact \HII\ regions, planetary nebulae and radio stars). This latter point takes into account the much greater sensitivity of SMGPS to the unresolved lobes of radio galaxies, which can introduce a systematic shift in the measured positions of these sources as compared to CORNISH/CORNISH-South.  

\begin{figure}
    \centering
    \includegraphics[width=\columnwidth,]{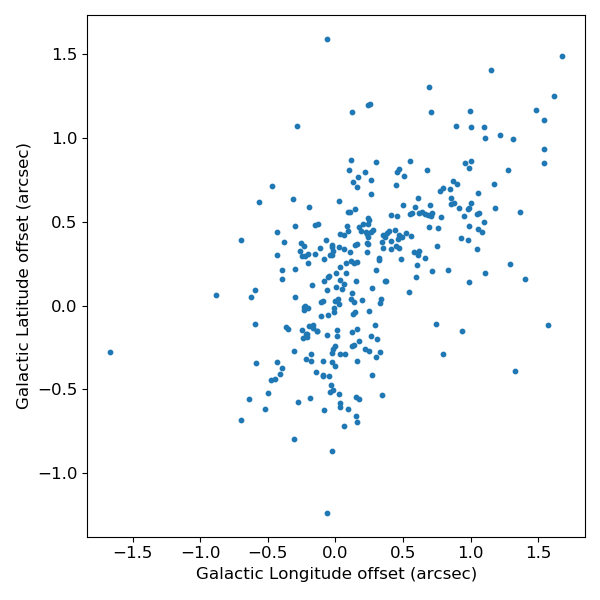}
    \caption{Angular separation between the positions of matched 
    sources in the \SMGPS\ and CORNISH/CORNISH-South surveys. To avoid confusion from lobes of radio galaxies that are preferentially detected by MeerKAT, the CORNISH sources are prefiltered to only contain sources that are expected to have similar morphologies at all frequencies, i.e.\ ultracompact \HII\ regions, planetary nebulae and radio stars.}
    \label{fig:cornishcomp}
\end{figure}

Fig.\ \ref{fig:cornishcomp} shows a scatter plot of the Galactic longitude ($l$) and Galactic latitude ($b$) offsets between corresponding SMGPS and CORNISH/CORNISH-South point sources. Clearly there is an element of minor systematic error in the positions as the points are not symmetric around an ($l$, $b$) offset of (0, 0). The median offset in ($l$, $b$) is (0\farcs16, 0\farcs30) with a standard deviation of 0\farcs5 in both $l$ and $b$.

In addition to CORNISH/CORNISH-South we also conducted an examination of the positional offsets between the \SMGPS\ catalogue and 28 ICRF sources found within the survey area. These 28
sources are relatively bright in the SMGPS (median 1.29 GHz flux density of 1.3 Jy), so the statistical uncertainty in the MeerKAT position determinations is small.  The sources were chosen for the ICRF by virtue of being compact on
milli-arcsecond scales, and generally do not have any structure on the
scale of MeerKAT's resolution of 8\arcsec\ which might affect the position
determination.  Since the ICRF constitutes the most accurately known
set of astronomical positions, they are an ideal additional check of the absolute
astrometry in the MeerKAT images.

We therefore compared our point-source catalog positions with those of the 
ICRF3\footnote{\url{http://hpiers.obspm.fr/icrs-pc/newwww/icrf/index.php}}
sources that were included in our fields, which we list in 
Table~\ref{tab:ICRF}.

\begin{table}
	\begin{center}
		\caption{ICRF sources in the \MKGPS.}
		\label{tab:ICRF}
		\begin{tabular}{l l l}
			\hline\hline
J080125.9$-$333619 & J080644.7$-$351941 & J082804.7$-$373106 \\
J093333.1$-$524019 & J120651.4$-$613856 & J135546.6$-$632642 \\
J151512.6$-$555932 & J163246.7$-$455801 & J171738.6$-$394852 \\
J173657.8$-$340030 & J174317.8$-$305818 & J174423.5$-$311636 \\
J175151.2$-$252400 & J175526.2$-$223210 & J183220.8$-$103511 \\
J184603.7$-$000338 & J185146.7+003532   & J185535.4+025119   \\
J185802.3+031316   & J190539.8+095208   & J192234.6+153010   \\
J192439.4+154043   & J193052.7+153234   & J193450.2+173214   \\
J193510.4+203154   & J193629.3+224625   & J194606.2+230004   \\
J194933.1+242118   \\
\hline
\end{tabular}
\end{center}
\end{table}

We found that the MeerKAT catalog positions differed from the ICRF3
ones by $<1\farcs 5$.  The largest offset was for ICRF
J193629.3+224625, for which the MeerKAT catalog
position was ($-0\farcs8, +1\farcs2$) from the ICRF position in $(l,
b)$.  Over all 28 sources, the mean and RMS deviation between our catalog
positions and the ICRF ones were $-0\farcs21 \pm 0\farcs35$ in $l$ and
$-0\farcs13\ \pm 0\farcs40$ in $b$\@. This is consistent with the results as found from the comparison with CORNISH/CORNISH-South, lending confidence that the positions of SMGPS sources are known to an RMS accuracy of $\sim$ 0\farcs5.

\subsection{Spectral indices}
\label{sec:spi}

The ability to derive spectral indices from images made from
interferometer data depends on the array adequately sampling all of
the relevant size scales across the entire observing band.  This
causes problems in the Galactic Plane with structure on a huge range
of scales. 

In the construction of the ``dirty'' images zeros are substituted
for unsampled visibilities.  Since the region around the origin is
almost never sampled due to the physical constraints of moving
antennas, this results in a dirty image which in the average is zero.
This leaves negative regions balancing positive ones.  Deconvolution
(here CLEAN) is a technique for interpolating over regions of the $uv$
plane which were not sampled.  Deconvolution has its limits.

The largest scales which can be imaged are limited by the $uv$ coverage
of the shortest baselines measured in wavelengths.  With the 2:1
frequency coverage of MeerKAT L band this means that the data at the
bottom of the band can image structures of twice the size of data at
the top of the band.  Alternatively, the portion of the image at the
bottom of the band can recover a significantly higher fraction of the
flux density for resolved sources than at the top end of the
band. Deconvolution helps with this problem but, if the extent of
the feature is significant, will not eliminate it.  A naive spectral
index derived from such data can appear much more negative than reality.

The difference in largest structure sampled across the band can be
greatly reduced by using an ``inner'' taper \citep{Cotton+2020} to equalize
the short baseline coverage across the band.  This leads to better
estimates of in-band spectral index but comes at a cost of filtering
out the largest scale structures which are only visible at the bottom
of the band.  This seems like a poor trade-off for the Galactic Plane.

The CLEAN used for the data presented here is relatively shallow and
used only point components.  No inner $uv$ taper was applied. A deeper CLEAN using multiple scales (or
the equivalent) combined with an inner taper to equalize the $uv$
coverage could result in more accurate spectral indices for the
surviving structures.  The ultimate fix is to include filled aperture
(e.g. large single dish) data.  Such exercises are deferred to future
data releases.

Although the \MKGPS\ data release includes in-band spectral index values,
$\alpha$, determined by fitting a power law
to the brightness in the frequency-resolved planes, these values
of $\alpha$ should only be used with considerable caution. 
As described above, near regions of bright emission the effective zero level in the images can be significantly offset from zero \Jb, usually being negative.  This ``zero offset'' is strongly frequency dependent.

As a consequence, the spectral index fitted to the layers in the
cubes, and present as layer 1 in the ``refit'' cubes, can be
significantly biased.  For a better estimation of the spectral index,
some frequency-dependent estimate of the local ``zero level'' near the
source of interest should be made. For more accurate values, a deeper, multi-resolution
CLEAN using an inner $uv$ taper should 
be carried out.

\begin{figure}
\centering
\includegraphics[width=\linewidth]{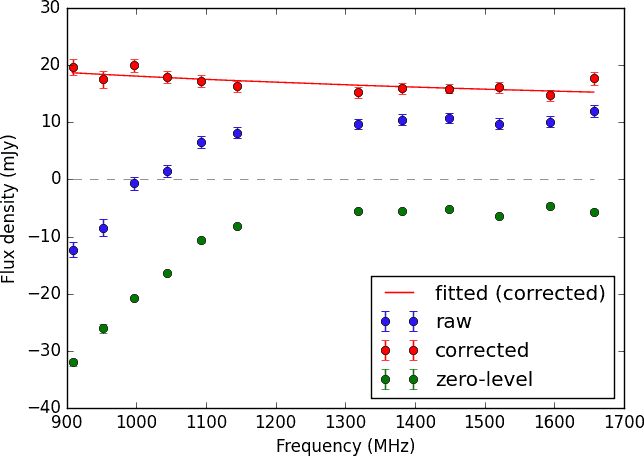}
\caption{An example of the effect of zero levels on the determination of
spectral indices: the spectrum for the candidate PWN around PSR J1208$-$6238
(see \S~\ref{sec:PWNe} and Fig.~\ref{fig:PWNe} for further details on this object).  The candidate PWN
is a relatively compact source, about 23\arcsec\ ($\sim 3$ beamwidths) in radius
and with a total flux density of $\sim$16 mJy.
The blue points show the raw flux density in a region with 
equivalent radius 23\arcsec\ around the candidate PWN.   The 
green points show the flux density due to the 
zero level, which was estimated by taking the mean  brightness from an
approximately annular region just around the putative PWN. The red 
points show the difference of the two, or the corrected flux density, and the red line shows 
the fitted power-law spectrum with $S_\nu = 16.3 \; (\nu/{\rm 1360 \, MHz})^{-0.33}$ mJy. The grey dashed line shows flux density 0 for reference. The 
plotted uncertainties include both statistical and systematic contributions.
}
\label{fig:psrj1208}
\end{figure}

As an example, we show in Fig.\ \ref{fig:psrj1208} the spectrum of some 
emission around PSR J1208$-$6238, which is a new candidate pulsar wind 
nebula discovered in the \MKGPS\  (for further details, 
see Section~\ref{sec:PWNe} and Fig.~\ref{fig:PWNe}a below).  As can be seen in the figure,
the zero levels are strongly frequency-dependent, and therefore the correction for the zero level strongly affects the determination
of the spectral index.

We attempted a correction for the zero levels by fitting, for each 
sub-band independently, a mean zero-level brightness to an approximately annular area selected
to be apparently devoid of real emission around the source, and then subtracted the mean brightness value in this
annular region from the source brightness.  A power-law function ($B_\nu \propto
\nu^\alpha$) fitted by least-squares to the resulting flux densities (brightness integrated over the
source region) results in a good fit, and a credible
spectral index of $-0.33$ which is in the expected range
for a PWN.

The determination of local zero level depends on the dimension of the source. 
For extended sources, the fluctuations of the zero level in the image 
can be relatively high, depending on the brightness of the target source and of 
any other sources around it that were not adequately CLEANed.
The fluctuations in the zero level at any point in the image are often due to several nearby
sources and  therefore depend significantly on the wavelength, and can 
have a significant effect on the slope of the spectrum (as seen in Fig.~\ref{fig:psrj1208}).

As further examples, we selected three SNRs several arcminutes across, namely 
G340.6+0.3 (diameter of $6\farcm2$), G346.6$-$0.2 (8\arcmin), G344.7$-$0.1 (8\arcmin), with known spectral indices
$\alpha=-0.35, -0.50$ and $-0.53$, respectively \citep{Trushkin+1999}. 
From the \MKGPS\ mosaic cubes, for each frequency plane, we compute
the integrated flux density inside a circular region with radius equal to the 
maximum radius of the source.  First, we subtracted the average brightness computed in an annular region
just outside the source. Then we did a linear least-squares fit of
$\log\,S_\nu$ to $\log \nu$ to find
$\alpha$. The nominal \MKGPS\ in-band spectra are quite steep: $\alpha = -0.67, -1.35$ and $-1.65$ for
the three SNRs, respectively.   For all three SNRs, the \MKGPS\ in-band spectra
are notably steeper than those from the literature, indicating that for
these three sources, with diameters between 6\arcmin\ and 8\arcmin,
simple estimation of a constant zero level from a region just outside
the source is not adequate.

\begin{figure}
    \centering
    \includegraphics[width=\linewidth,trim=0 7mm 0 0, clip]{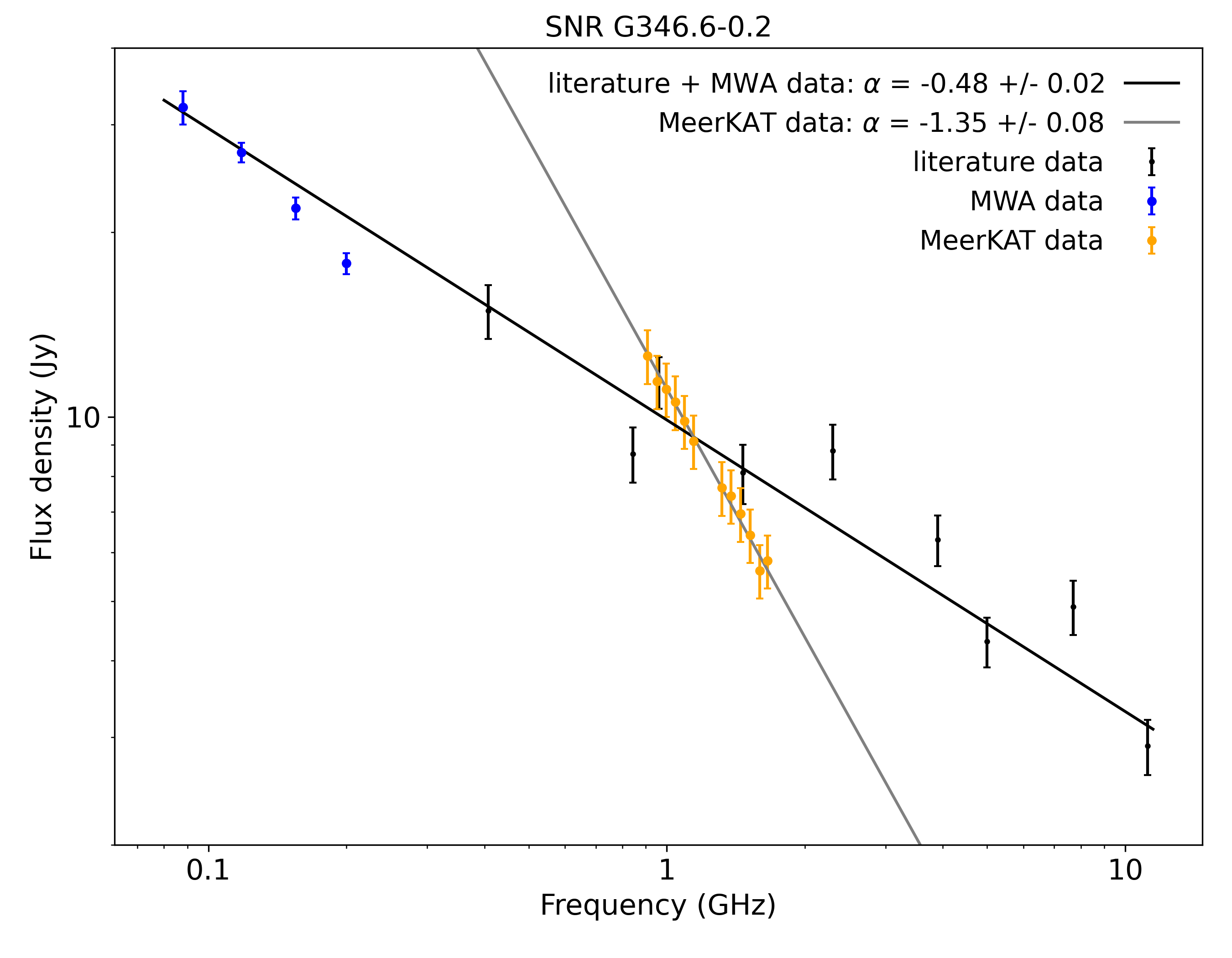}
    \caption{Spectrum of the SNR G346.6$-$0.2. The thick black line is the fit to literature values \citep[][]{Trushkin+1999,Hurley-Walker+2019b}. 
    Yellow points are the \MKGPS\ values, with the thin grey line showing the corresponding fit. 
    The two fits intersect approximately at the central frequency of the MeerKAT band, but have different spectral indices. 
    }
    \label{fig:G346_02_SED}
\end{figure}

Despite these issues, the flux densities computed for the
center of the \SMGPS\ band are in good agreement with the literature values. For example, 
measurements of the SNR G346.6$-$0.2, shown in Fig.~\ref{fig:G346_02_SED}, 
clearly show that, near the center of the \SMGPS\ band,  the \SMGPS\ 
flux density measurements agree
with the spectrum obtained from the literature \citep{Trushkin+1999,Hurley-Walker+2019b}.  
However, the slope corresponding to the \SMGPS\ values is notably
steeper than that determined over a much wider frequency range from
literature values.

For point-like and compact sources, smaller than a few synthesized beam areas, the local 
zero-level brightness does not vary much over the source, 
and should be reliably estimated
from the region immediately around the source.   The spatial and brightness scales 
on which the zero level
varies depends on the complexity of the surrounding field, and there is no
general rule to estimate the spatial scale for which they become problematic,
but in general they can be more reliably estimated for smaller sources.

\begin{figure}
    \centering
    \includegraphics[width=\linewidth]{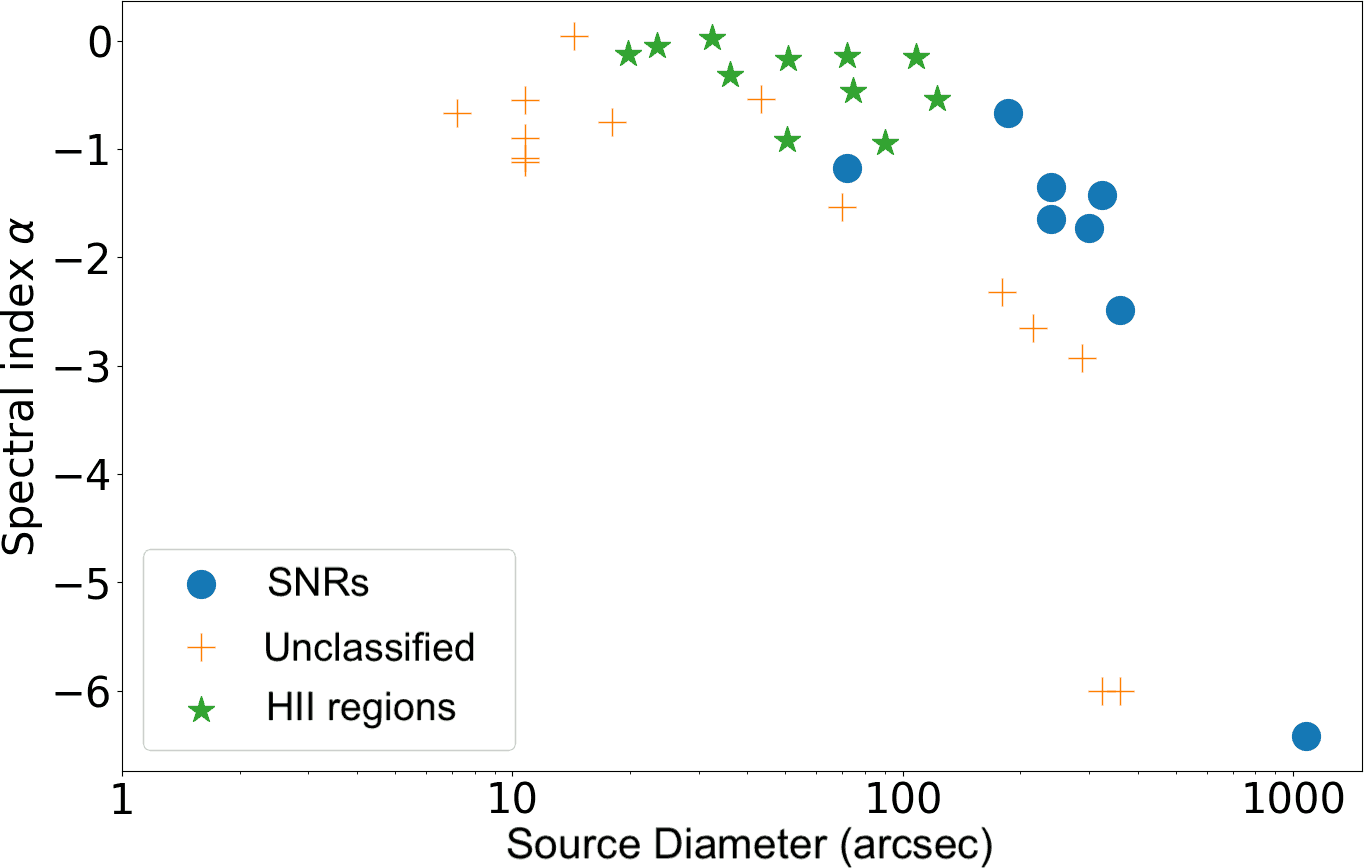}
    \caption{\MKGPS\ nominal in-band spectral indices as a function of 
    the angular diameter of the source for three classes of resolved sources
    with relatively circular morphologies.}
    \label{fig:alphavsize}
\end{figure}

To try to determine for which source size the \MKGPS\ in-band spectral indices may be
reliable, we computed the \MKGPS\ in-band $\alpha$ for a sample of sources with 
angular diameters ranging from 
7\arcsec\ to 1000\arcsec\@. We selected a number of resolved sources of three different classes,
SNRs (eight, including the three mentioned above), \HII\ regions (11) and 
unclassified sources (14), all of them with approximately circular morphology. 
We plot the resulting nominal in-band values of $\alpha$ against the source size
in Fig.~\ref{fig:alphavsize}.  There is a clear correlation between
the \MKGPS\ in-band $\alpha$ and the angular size. For small diameters,
the values of $\alpha $ are broadly within the expected range between $-1$ 
and 0, but as the angular size 
increases, the in-band $\alpha$ values become quite negative, with 
 $\alpha < -3$ determined for sources $> 300\arcsec$.  Such values are unphysical for these source classes, and the \SMGPS\ values are unreliable here.

In conclusion, 
the in-band spectral index values should be interpreted very cautiously.
The effect of the frequency-dependent local zero levels in the images
must be taken into account unless the brightness of the source is much larger
than the zero-level offsets.  For small sources, the zero levels may be reliably
estimated from the region just around the source, but for larger sources, the zero-level offsets are anti-correlated with the actual source
brightness, and the derived in-band spectra can be significantly in
error, most often being too steep. It appears that for the current data release, estimation of $\alpha$ should not be attempted for sources with diameters $\ga 1\arcmin$, and even for smaller sources great care should be taken with estimation of zero levels.

\section{Galactic Science Highlights}
\label{sec:highlights}

In this section we present some of the science highlights from the survey to illustrate the data quality, scientific results and the different source populations discovered in the SMGPS.

\subsection{Radio filaments}
\label{sec:filaments}

\begin{figure*}
	\centering
    \includegraphics[width=\linewidth]{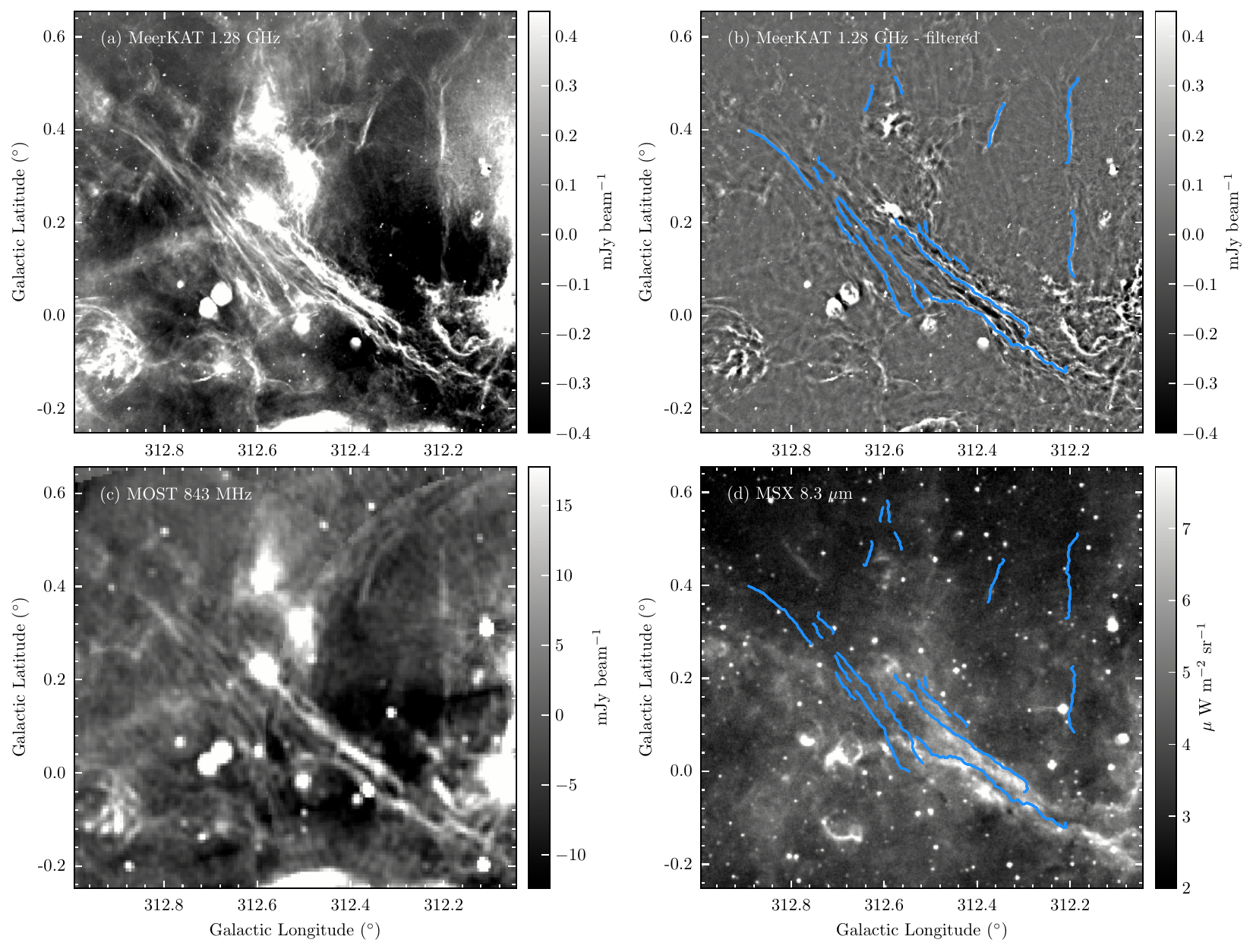}
	\caption{Zoomed-in view of a filamentous region in the $311\degr < l < 314\degr$ SMGPS tile showing \emph{(a)} the 1.28\,GHz MeerKAT image (with 8\arcsec\ angular resolution), \emph{(b)} the high-pass filtered  MeerKAT image (i.e. with removal of contamination from foreground/background emission), \emph{(c)} the 843\,MHz MOST image \protect\citep[][with $46\arcsec$ resolution]{Green+1999}, and \emph{(d)} the 8.3$\mu$m MSX image  \protect\citep[][with $20\arcsec$  resolution]{Price+2001}. The spines of 21 identified filaments (where the spine traces the high-intensity crest/backbone of the filament) are marked in blue in panels \emph{(b)} and \emph{(d)}, and the colourscale in all panels is shown on a linear scale.}
	\label{fig:G312_filaments}
\end{figure*}

Filamentary structures are ubiquitous constituents of the interstellar medium (ISM) across a variety of size scales and environments, 
from low-density dust cirrus \citep[e.g.][]{Low+1984,Bianchi+2017}, 
to neutral atomic \HI\ filaments \citep[e.g.][]{Kalberla+2016,KalberlaKH2020,Soler+2020}, 
and both quiescent and star forming filaments in high density dust and molecular gas \citep[e.g.][]{Andre+2014,Li+2016,Schisano+2020}. 
In the ionised ISM, an intriguing kind of filamentary phenomenon has been identified towards the Galactic Centre \citep[][]{Yusef-ZadehMC1984,Yusef-ZadehHC2004,Heywood+2019,BarkovL2019,Heywood+2022} and Orion \citep{Yusef-Zadeh1990}.
These so-called non-thermal filaments (NTFs) are highly linear with large aspect ratios (with many of the longest aligned perpendicular to the Galactic Plane), are inherently non-thermal in nature, and are highly magnetised with intrinsic B field vectors aligned parallel to their lengths \citep{LangME1999}. 
A consensus is yet to be reached on the means of formation of these NTFs, with many proposed theories including both external and in-situ acceleration of relativistic particles \citep[e.g.][]{Yusef-ZadehMC1984,MorrisS1996,RosnerB1996,BarkovL2019,Yusef-Zadeh+2022D,CoughlinNG2021}. 
Due to a lack of NTFs observed elsewhere in the Galaxy, it has so far been generally agreed that they are structures that occur in the uniquely extreme and energetic environment of the Galactic Centre. 

The SMGPS images reveal a plethora of complex, elongated, filamentary structures. Some are associated with extended sources such as supernova remnants (see Section~\ref{sec:SNR}) and \HII\ regions, while others appear to be isolated.  
We implemented a semi-automated method for their identification in as consistent a way as possible across the survey area. A high-pass filter was applied to the moment zero images for the removal of large scale diffuse emission, in effect increasing the contrast of compact sources and the ridges/spines of narrow structures against the local background \citep[e.g.][]{Yusef-Zadeh+2022A}. The filtered images were thresholded to create a mask of all emission brighter than 3 times the local background RMS brightness. 
The masks were segmented based on their shape by calculation of their principal moments of inertia \citep[using the $J$-plots algorithm;][]{Jaffa+2018}, and only the most elongated structures (with aspect ratios $>4$) were selected.
A final visual inspection of all extracted structures allowed removal of any clear artefacts such as strong sidelobe features. 
We emphasise that our method does not pick up structures connected to diffuse or extended emission, because we are choosing only the brightest, isolated and most elongated structures in a semi-automated fashion.

With this method, we identify a population of radio filaments across the Galactic Plane.
The full catalogue will be presented elsewhere.
As an example, we present here a sub-sample of 21 filaments
from the G312.5 tile ($311\degr < l < 314\degr$).  We show both the unfiltered and filtered images of this region in
Fig.~\ref{fig:G312_filaments}a and b, and list the filament properties in Table~\ref{tab:filament_properties}.

\begin{table*}
    \centering
    \caption{Observed properties of a sub-set of filaments identified in the $311\degr < l < 314\degr$ SMGPS tile.}
    \label{tab:filament_properties}
    \setlength\tabcolsep{9.4pt}
    \begin{tabular}{ccccccccccc}
    \hline \hline
    \multicolumn{2}{c}{Centroid coordinates$^{a}$} & Peak intensity  & Length   & Width$^{b}$ & Aspect ratio & PA$^{c}$ &   \multicolumn{2}{c}{$J$ moments$^{d}$}  & MIR$^{e}$ & Bundle$^{f}$ \\
    $l$ (\degr)     & $b$ (\degr)    & (m\Jb)   & (arcmin) & (arcsec) & & (\degr) & $J_1$ & $J_2$ & assoc. & vs. Isolated \\ \hline
    312.367 &   $-$0.026 &   3.27 &   27.3 &     9.0 &   183.0 &  $-$60.0 & 0.59 & $-$0.76 & Y & Bundle \\
    312.434 &   0.078 &   1.24 &    25.9 &    25.4 &    61.0 &  $-$48.0 & 0.59 & $-$0.81 & Y & Bundle \\
    312.602 &   0.082 &   0.49 &    13.9 &    12.3 &    68.0 &  $-$33.0 & 0.55 & $-$0.82 & Y & Bundle \\
    312.563 &   0.093 &   1.82 &    11.8 &    22.6 &    31.0 &  $-$31.0 & 0.70 & $-$0.78 & Y & Bundle \\
    312.196 &   0.155 &   0.40 &    10.0 &    26.8 &    22.0 &  $-$1.0 & 0.77 & $-$0.84 & N & Isolated \\
    312.434 &   0.112 &   0.41 &     2.7 &     5.8 &    27.0 &  $-$41.0 & 0.60 & $-$0.78 & Y & Bundle \\
    312.495 &   0.156 &   0.94 &     7.2 &    11.8 &    37.0 &  $-$42.0 & 0.63 & $-$0.83 & Y & Bundle \\
    312.571 &   0.163 &   0.37 &     1.9 &     8.3 &    14.0 &  $-$52.0 & 0.75 & $-$0.78 & Y & Bundle \\
    312.656 &   0.204 &   0.39 &     9.6 &    16.3 &    35.0 &  $-$41.0 & 0.55 & $-$0.66 & Y & Bundle \\
    312.685 &   0.182 &   0.39 &     4.8 &    15.4 &    19.0 &  $-$36.0 & 0.45 & $-$0.69 & Y & Bundle \\
    312.511 &   0.191 &   0.24 &     1.5 &     9.9 &     9.0 &  $-$14.0 & 0.60 & $-$0.74 & Y & Bundle \\
    312.690 &   0.220 &   0.48 &     4.2 &    11.9 &    21.0 &  $-$31.0 & 0.17 & $-$0.92 & Y & Bundle \\
    312.819 &   0.343 &   0.38 &    11.9 &    21.2 &    34.0 &  $-$47.0 & 0.72 & $-$0.75 & N & Bundle \\
    312.743 &   0.302 &   0.22 &     2.5 &    16.4 &     9.0 &  $-$34.0 & 0.73 & $-$0.80 & N & Bundle \\
    312.727 &   0.319 &   0.35 &     3.9 &    15.0 &    16.0 &  $-$34.0 & 0.59 & $-$0.87 & N & Bundle \\
    312.201 &   0.419 &   0.57 &    13.0 &    23.7 &    33.0 &  8.0 & 0.82 & $-$0.84 & N & Isolated \\
    312.362 &   0.410 &   1.04 &     6.6 &    24.1 &    16.0 &  19.0 & 0.82 & $-$0.82 & N & Isolated \\
    312.632 &   0.467 &   0.22 &     4.0 &    17.0 &    14.0 &  18.0 & 0.52 & $-$0.81 & N & Isolated \\
    312.568 &   0.495 &   0.29 &     2.6 &    16.3 &    10.0 &  $-$22.0 & 0.57 & $-$0.83 & N & Isolated \\
    312.590 &   0.559 &   0.19 &     3.2 &    14.2 &    13.0 &  $-$7.0 & 0.36 & $-$0.83 & N & Isolated \\
    312.606 &   0.553 &   0.22 &     2.1 &    15.5 &     8.0 &  10.0 & 0.57 & $-$0.72 & N & Isolated \\ \hline
    \end{tabular}
    \begin{flushleft}
        $^a$ Centroid position of the filament spine. \\
        $^b$ Deconvolved FWHM of a Gaussian fitted to the mean transverse intensity profile of the filaments. \\
        $^c$ Mean position angle of the filament spine, measured from Galactic North (where PA$=0^{\circ}$), with positive values in the clockwise direction. \\
        $^d$ Derived from the principal moments of inertia of the structure masks \citep[using the $J$-plots algorithm;][]{Jaffa+2018}. The $J$ moments describe the shape of the object; positive $J_1$ and negative $J_2$ values together denote elongated, filamentary-like structures.\\
        $^e$ A flag noting whether the filament is coincident with $8.3\,\mu$m MSX emission (Y) or not (N).\\
        $^f$ A note describing whether the filament is isolated, or belongs to the bundle of braided filaments described in the text (see Section~\ref{sec:filaments}).
    \end{flushleft}
\end{table*}

The filaments in this region appear with two distinct morphologies: (i) a bundle of filaments oriented at an angle to the Galactic Plane, and (ii) relatively isolated filaments oriented almost perpendicular to it.
Concerning the former, \cite{CohenG2001} noted the presence of ``large-scale braided filamentary structures'' in the Molonglo Galactic Plane Survey data \citep{Green+1999} observed at 843\,MHz with MOST (see Fig.~\ref{fig:G312_filaments}c). It was further noted by \cite{CohenG2001} that these filaments appear to be coincident with similarly braided ``tendrils'' of $8.3\,\mu$m mid-infrared (MIR) emission observed with MSX \citep[see Fig.~\ref{fig:G312_filaments}d;][]{Price+2001}, likely signposting emission from polycyclic aromatic hydrocarbons (PAHs); this strongly suggests that these radio filaments are thermal in nature. 
\cite{CohenG2001} posited that the alternating pattern of MIR-ridge to radio-ridge (see Fig.~\ref{fig:G312_filaments}d)
suggests that the filaments are
limb brightened sheets of emission at the edge of a large-scale bubble with diameter $>1\degr$ centred above $b=0\fdg5$, though this remains uncertain.
With an angular resolution of 46\arcsec, \cite{CohenG2001} were unable to resolve the
filaments, nor distinguish whether the radio emission was as intricate as that seen in the MIR. With the 8\arcsec\ angular resolution of \SMGPS, we give a first estimate for the resolved width of these filaments 
in Table~\ref{tab:filament_properties}, 
and confirm that the radio emission structure is indeed highly complex --- intertwining tendrils of radio emission are indeed interspersed with MIR emission on the south-west side of the bundle, whilst the filaments on the north-east side of the bundle appear unrelated to MIR emission (Fig.~\ref{fig:G312_filaments}d). 
Thus, as they are on the whole likely thermal, these braided filaments are fundamentally different in nature to the NTFs identified towards the Galactic Centre \citep[e.g.][]{Heywood+2019}. 
Concerning the isolated filaments oriented almost perpendicular to the Galactic Plane, we identify three highly elongated filaments to the north-west of the filament bundle, and four shorter filaments to the north-east. Despite the three longest filaments being, at least retrospectively, noticeable in the 843\,MHz MOST image (Fig.~\ref{fig:G312_filaments}c) they were not discussed by \citet{CohenG2001}. 
All seven of these filaments appear to be unrelated to 8.3\,$\mu$m MIR emission (Fig.~\ref{fig:G312_filaments}d), strongly suggesting
they are non-thermal in nature.

Due to the limitations discussed in Section~\ref{sec:spi}, we are at this
point
unable to derive reliable spectral indices for these extended filamentary structures, or
their polarisation properties, since polarized data reduction was not done for the G312.5 tile.
Though the braided, MIR-associated, likely thermal filaments in the bundle are fundamentally different in nature to NTFs, the isolated filaments identified here may be good candidates for the first NTFs identified outside the Galactic Centre region. 
Their highly linear nature and 
the fact that their widths are resolved
(see Table~\ref{tab:filament_properties}) are reminiscent of the Galactic Centre population of NTFs. 
The apparent lack of coincident MIR emission towards these isolated filaments is also suggestive of a non-thermal nature. Unlike many NTFs in the Galactic Centre, the isolated filaments presented here do not appear to have point sources along their lengths, however they do reside in a region filled with bubbles, \HII{} regions and supernova remnants, and therefore may possibly be examples of externally driven NTFs. 
These results raise the possibility that the NTF phenomenon might be more prevalent across the Galaxy than previously appreciated. A forthcoming investigation of filamentary structures in the full SMGPS data release will address this topic in detail.

\subsection{Supernova remnants}
\label{sec:SNR}

Supernova remnants (SNRs) emerge in the aftermath of most supernova explosions (SNe).  SNe are the end stage of massive stars,
and the birthplace of neutron stars and stellar-mass black holes.
They  also inject a substantial amount of
energy into the ISM, as well as being the major contributors to the
chemical enrichment of the ISM.

In a SN explosion, typically several solar masses of material are ejected from the star,
and plough outward through the circumstellar medium, typically the
stellar wind of the progenitor star, or the ISM.  In
so doing, they produce strong shocks which accelerate particles to
relativistic energies and amplify the magnetic field, the combination
of which produces non-thermal radio emission.

Radio emission is therefore one of the important hallmarks of SNRs,
and is most often the way that they are identified. The most complete current catalogue as of 2022 is that of Green\footnote{
See
\url{https://www.mrao.cam.ac.uk/surveys/snrs}.}, which is an updated version of \citet{Green2019} containing 303
SNRs.

Studies of SNRs in the radio are often limited by either resolution or
surface-brightness sensitivity. MeerKAT's high sensitivity and
resolution, alongside superb image fidelity, therefore make it an ideal instrument for studying known SNRs.
Also, its southern location allows a view of the central parts of our
Galaxy, where many SNRs are expected due to the high stellar density.
There is a long-standing tension between the $\approx 300$ currently established Galactic SNRs, and
the $\approx 1000$--2000 that may be expected  \citep{Li+1991, TammannLS1994,
  Gerbrandt+2014,RanasingheL2022}. The large area coverage of the \MKGPS\ allows a more detailed study of many known SNRs, and should enable the discovery of many new SNRs in a good fraction of the
Galaxy to lower surface brightness than previous wide-area surveys.

\subsubsection{Known SNRs}
\label{sec:knownSNR}

\begin{figure*}
\centering
\includegraphics[width=\linewidth, trim=0 18mm 0 2cm,clip]{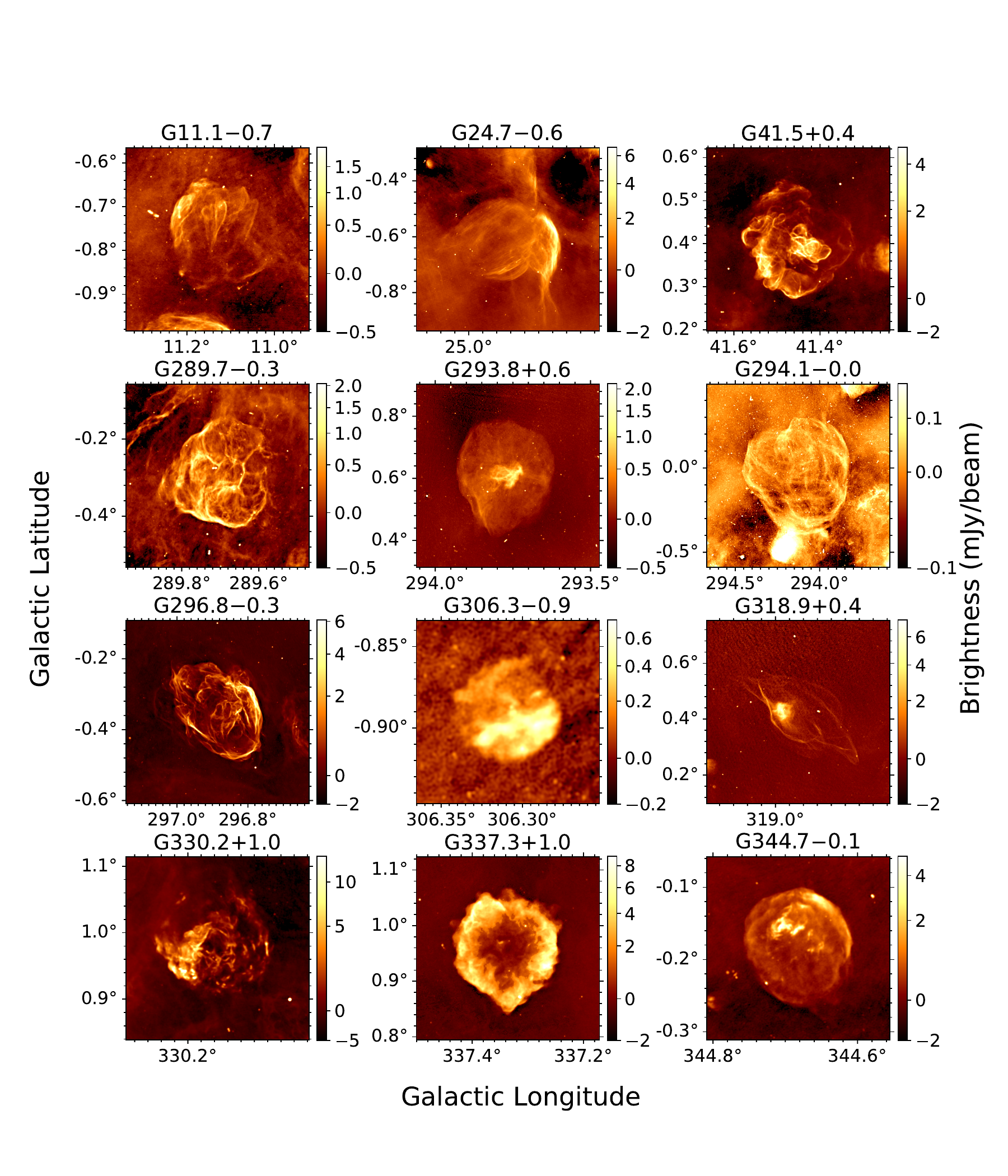}
\caption{\MKGPS\ images for 12 known SNRs. Each panel is centred on the MeerKAT
  emission identified with the SNR and has a width of 3 times the
  identified radius.  The colourscale shows brightness with a square
  root stretch.}
\label{fig:knownSNR}
\end{figure*}

Approximately 200 known SNRs are imaged in the SMGPS. In Fig.~\ref{fig:knownSNR} we show images for 12 of them. These were selected to show examples of the ways in which the SMGPS  provides a markedly improved view over previously available radio images, in a variety of environments. For example, in several of the images the full(er) extent of the SNR becomes apparent, owing to improved surface-brightness sensitivity; in many others, fine scale filamentary features become apparent where previously none were known. A further example is given 
in Appendix \ref{sec:W44}, where we present a full polarization analysis of the SNR W44.

\paragraph*{SNR G11.1$-$0.7.}
The \MKGPS\ image shows a complex structure
$\sim$15\arcmin\ in diameter, brighter to the Galactic NE, with a
number of filaments, but not clearly a shell.  Much lower resolution
and dynamic range images at 327 MHz and 1465 MHz were published in
\citet{Brogan+2004}, which do not show the extent or the filamentary
nature of the emission.

\paragraph*{SNR G24.7$-$0.6.} Centred at $l, b = 24\fdg9, -0\fdg6$, approximately circular in outline with radius $\sim$9\arcmin\ but fading to the Galactic SSE, this SNR
has an area of filamentary emission.  It is somewhat edge brightened, and possibly a shell.  Superposed, and perhaps interacting with this,
is a dominant filamentary bundle running from Galactic NNE to SSW, which overlaps
and curves around the western edge of the possible shell, that is
likely identified as PMN J1838$-$0734 \citep{Griffith+1995}.
The best published radio image seems to be that of \citet[VLA;
resolution $\sim$50\arcsec]{Dubner+1993}, which gives only the vaguest hint of
the possible shell component.

\paragraph*{SNR G41.5+0.4.}
Roughly circular in outline, $\sim$16\arcmin\ in diameter, with an
enhancement both near the centre and towards the Galactic
ESE (identified as [ADD2012] SNR 21 in \citealt{Alves+2012} and
RRF 305 in \citealt{ReichRF1990}). A VLA 332-MHz image was presented in
\citet{Kaplan+2002}, with resolution $\approx$50\arcsec.
The \MKGPS\ image reveals numerous loops and filaments throughout. \citet{Kaplan+2002} had suggested a possible central pulsar wind nebula (PWN), $\sim
3\arcmin \times 1\farcm5$ in extent, but with our improved image fidelity it is possible that this is simply a centrally located brighter complex of loops and filaments.

\paragraph*{SNR G289.7$-$0.3.}
The only previously published image seems to be from
MOST at 843 MHz
\citep{WhiteoakG1996}, with much lower resolution (43\arcsec) and
dynamic range than from \MKGPS.  Although the earlier image reveals the
$\sim$14\arcmin\ extent of the emission and some filamentation, it does not
show the details of the filaments or the unusual interlocking loops
seen in the \MKGPS\ image.

\paragraph*{SNR G293.8+0.6.}
The only previously published image
seems to be that of \citet{WhiteoakG1996}.  The emission extends over a roughly circular
region of $\sim$20\arcmin\ diameter, and shows a prominent central
condensation about 5\arcmin\ in extent, which is quite filamentary and could plausibly be a PWN. The filamentary nature of the central component was not
evident in the previous image.

\paragraph*{SNR G294.1$-$0.0.}
The only previously published image
seems to be that of \citet{WhiteoakG1996}. The emission extends over a
roughly circular area of $\sim$40\arcmin\ in diameter, with extensive
filamentation.  The SNR may be interacting with another, brighter,
extended source, $\sim 6\arcmin \times 12\arcmin$ in extent, which
lies contiguous to the SNR to Galactic SSE, which is identified as
an infrared bubble, likely associated with the \HII\ region [HKS2019]
E140 \citep{Hanaoka+2019}.

\paragraph*{SNR G296.8$-$0.3.}
\citet{GaenslerMG1998} published 1.3-GHz, $\sim$23\arcsec-resolution
images from ATCA of this SNR, which are consistent with, but
of considerably lower resolution and dynamic range than the \MKGPS\ image.  The SNR has an elongated morphology, $\sim 18\arcmin
\times 12\arcmin$ in overall extent, with edge-brightened filamentary
structure, including interior loops.  It is detected in X-rays, and
has a central point-like X-ray source, 2XMMi J115836.1$-$623519
\citep{Sanchez-Ayoso+2012}.  We do not see any compact radio source $\gtrsim 100\,\muJb$
at the location of 2XMMi J115836.1$-$623519.

\paragraph*{SNR G306.3$-$0.9.}
A small source, only $\sim$3\arcmin\ in diameter.  It is roughly
circular in outline, but is not noticeably edge-brightened. It is
divided into two halves along an ESE-WNW line, with the half to the
Galactic S being notably brighter.  The \MKGPS\ image shows much more
detail than the 5.5-GHz one  (ATCA, resolution
  24\arcsec) of \citet{Reynolds+2013}, revealing a somewhat flocculent
structure, with the most prominent features being aligned
approximately ESE-WNW.  The same division into two halves of unequal
brightness is seen in X-rays \citep{Reynolds+2013}, although the
detailed correspondence between radio and X-ray brightness is poor.
The object is also identified in the infrared as [SPK2012]
MWP1G306301$-$008946, with the 24 $\mu$m emission largely lying
along the outline of the radio emission region, and appearing more
filamentary than the radio. It is conceivable that this source could be an \HII\ region.

\paragraph*{SNR G318.9+0.4.}
Unusual, elongated morphology,
$\sim 30\arcmin \times 14\arcmin$ in extent, with elliptical arcs, and
an off-centre bright ``core''.
The best published image is that of
\citet{Whiteoak1993} at 843~MHz from MOST
(resolution $\sim$47\arcsec).  The \MKGPS\ image shows the
arcs much more clearly, as well as showing that the core is relatively
small at only $\sim$5\arcmin\ diameter, and has considerable internal
structure, which appears quite distinct from the arcs, although the arcs
are brighter near it.

\paragraph*{SNR G330.2+1.0.}
Radio emission which is relatively circular in outline and
$\sim$7\arcmin\ in diameter.  It is brighter in a wedge to the Galactic SE, and
has a flocculent structure throughout.  An 843-MHz image from MOST
\citep{WhiteoakG1996} shows the general outline, but it is at much
lower resolution ($\sim$49\arcsec) and does not show the flocculent
structure visible in the \MKGPS\ image.  There is a central X-ray
source, 2XMM J160103.4$-$513353, likely associated
\citep{MayerB2021}, however no compact counterpart with brightness
$>200 \, \muJb$ is visible in the \MKGPS.

\paragraph*{SNR G337.3+1.0.}
Nearly circular, edge-brightened shell structure,
$\sim$12\arcmin\ in diameter, with a possible axis of symmetry perpendicular to the Galactic Plane.  The best published radio images
seem to be from MOST at 843 MHz \citep{KestevenC1987,
  WhiteoakG1996}, which show the general shape but very little
detail. Low surface brightness ``blowouts'' or ``ears'' are prominent in the SMGPS image, especially in the northern sector of the SNR.

\paragraph*{SNR G344.7$-$0.1.}
Shell-like structure, $\sim$9\arcmin\
in diameter, brighter to the Galactic NW, with some bright interior
features, somewhat to the Galactic N of the geometrical centre.
The \citet{Giacani+2011} 1.4-GHz radio image from ATCA and VLA
data, with resolution similar to that of the \MKGPS\ image but RMS noise approximately one order of magnitude higher, does not show the complete and almost circular
outline of the radio emission.

\subsubsection{New candidate SNRs}
\label{sec:SNRcandidate}

\begin{figure*}
\centering
\includegraphics[width=\linewidth]{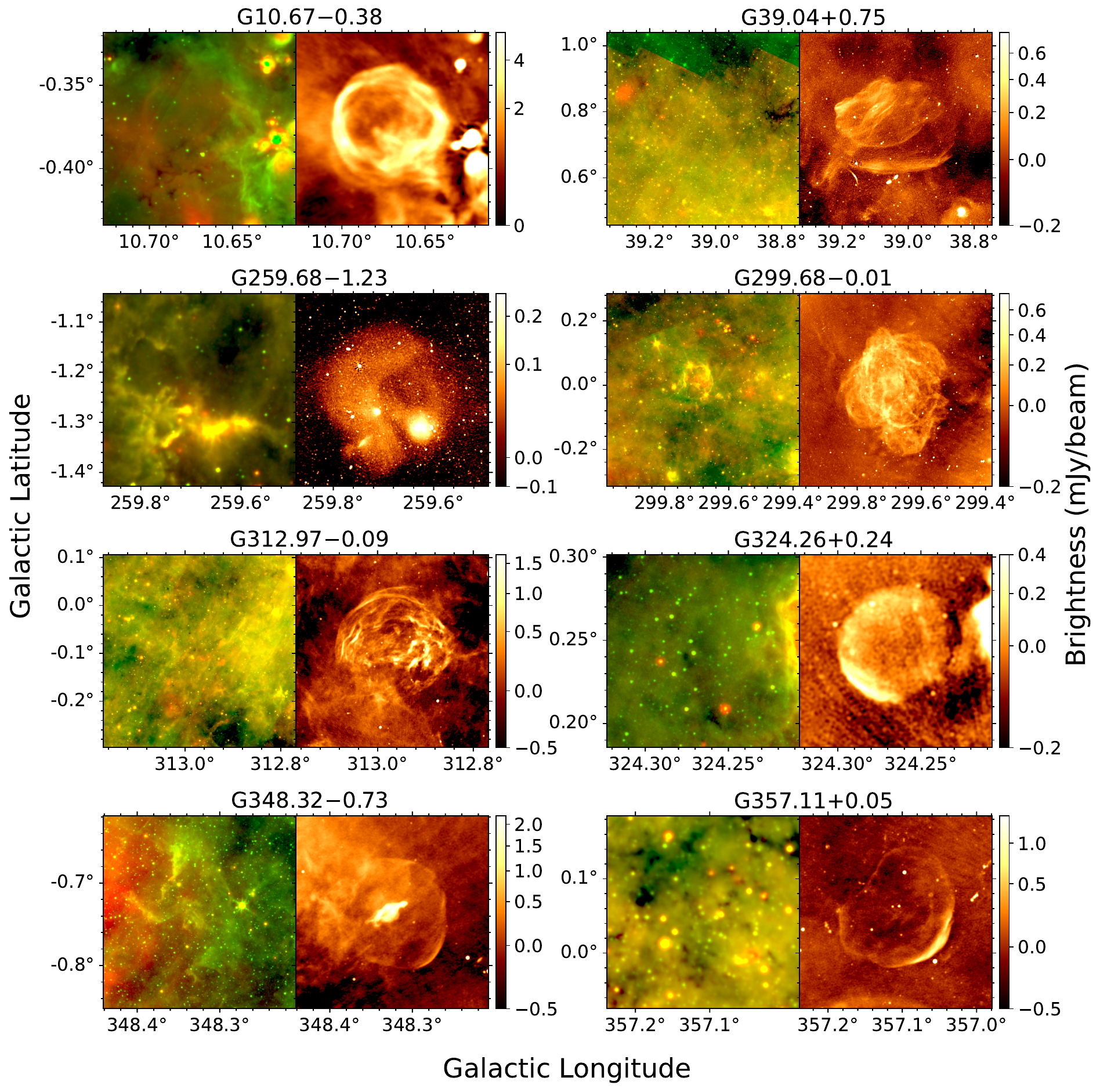}
\caption{SNR candidates from \MKGPS.  \emph{Left} panels: mid infra-red (MIR)
  images using data from the \emph{Spitzer} GLIMPSE 8.0\,$\mu$m (green) and MIPSGAL
  24\,$\mu$m (red) surveys for sources with $|b|<1\fdg0$ or WISE 12\,$\mu$m
  (green) and 22\,$\mu$m (red) data for sources further from the
  Galactic Plane.  \emph{Right} panels: \MKGPS\ 1.3 GHz images. The colourscale shows radio
  brightness, with a square-root stretch.   Each panel is centred
  on the MeerKAT emission identified with the SNR candidate and has a
  width of 3 times the candidate's radius.}
\label{fig:candidateSNR}
\end{figure*}

The \SMGPS\ images surely contain
many as-yet unknown SNRs, and indeed are an excellent tool
for identifying new candidate SNRs.  We did so, using a method
of SNR detection similar to that of
\citet{Anderson+2017} and \citet{Dokara+2021}.  The vast majority of
discrete extended sources $\ga 1\arcmin$  in the \MKGPS\ images are either SNRs (whether
previously known ones or new candidates) or \HII\ regions.

We visually examined \MKGPS\ images overlaid with the positions of
\HII\ regions from the WISE Catalogue of Galactic \HII\ Regions
\citep[][hereafter the ``WISE Catalogue’']{Anderson+2014} to create a
catalogue of extended MeerKAT radio continuum sources that cannot be
explained as being previously-known \HII\ regions.  For each
identified source we noted its centroid and the circular radius
necessary to contain the emission.
For each catalogue source we then examined mid-infrared (MIR) data from
\emph{Spitzer} GLIMPSE \citep{Benjamin+2003, Churchwell+2009} at 8.0\,$\mu$m
and MIPSGAL \citep{Carey+2009} at 24\,$\mu$m for sources at
$|b|<1\fdg0$ or WISE \citep{Wright+2010} 12\,$\mu$m\ and 22\,$\mu$m\ data for
sources further from the Galactic Plane.  We removed sources from the
catalogue with associated $\sim10\,\mu$m emission surrounding
$\sim20\,\mu$m emission of a similar or complementary morphology
to that of the MeerKAT radio continuum, as this morphology is
associated with thermally-emitting objects \citep{Anderson+2014}.
This process removes  \HII\ regions that are not listed in the WISE
Catalogue; the remaining catalogue should contain only known SNRs and SNR
candidates.  

Finally, we identified previously known SNRs in the catalogue using the
2022 compilation of Green\footnote{\url{https://www.mrao.cam.ac.uk/surveys/snrs}.},
and previously-identified SNR candidates using the
results from \citet{Helfand+2006}, \citet{GreenRM2014},
\citet{Anderson+2017}, \citet{Hurley-Walker+2019}, and
\citet{Dokara+2021}.
The remaining catalogue sources are SNR candidates newly-identified in the SMGPS data.

The full catalogue of $\sim 100$ new SNR candidates will be presented elsewhere. Here, in Fig.~\ref{fig:candidateSNR}, we show
eight candidate SNRs from this catalogue with a variety of morphologies and in a variety of environments. We designate them
by the Galactic coordinates of their estimated centre.

\paragraph*{G10.67$-$0.38.}
Relatively circular shell structure, $\sim$2\farcm5 in
diameter.  The region is somewhat confused, and
G10.67$-$0.38 may form part of the $\sim10\arcmin$-diameter radio
source GPA 010.68$-$0.37 identified in the
Green Bank Galactic Plane survey \citep{Langston+2000}. The MIR panel of G10.67$-$0.38 in
Fig.~\ref{fig:candidateSNR} shows a complex of \HII\ regions to the
Galactic W.  These are distinguishable from SNRs by their bright
MIR emission that is spatially correlated with that of the MeerKAT
radio continuum.  Although this SNR candidate has faint associated
24\,$\mu$m emission, it lacks the 8.0\,$\mu$m emission that would be present for an \HII\ region.

\paragraph*{G39.04+0.75.}
Unusual structure extending over
a region $\sim$18\arcmin\ in diameter, but broadly reminiscent of some SNRs as observed with MeerKAT (e.g. see G11.1$-$0.7 in Fig.~\ref{fig:knownSNR}).  It is quite strongly
filamentary, and there is bright condensation to the Galactic
NE. \citet{Dokara+2021} identified this bright condensation as
SNR candidate G039.203+0.811 but did not discern the large region of
emission to its SW seen in the \MKGPS\ image.  A feature elongated along the
Galactic E-W direction lies largely separated from the main
structure by $\sim$2\arcmin\ to the Galactic S, and it is not
clear whether they are in fact physically related.

\paragraph*{G259.68$-$1.23.}
Diffuse relatively circular structure, $\sim$12\arcmin\ in
diameter, with several nearby and overlapping \HII\ regions in its southern half.

\paragraph*{G299.68$-$0.01.}
Roughly circular structure, 
$\sim$20\arcmin\ in diameter, not noticeably edge-brightened but
very filamentary and with blowouts/extensions to the S and SE. Broadly speaking, the morphology of this SNR candidate has parallels to those of SNRs G41.5+0.4 and G296.8$-$0.3 as seen in Fig.~\ref{fig:knownSNR}. There is a small \HII\ region, IRAS 12202$-$6222,
slightly to the Galactic NE of the centre, but no extended
infrared emission covering the bulk of the radio source.

\paragraph*{G312.97$-$0.09.}
Approximately circular region
$\sim$16\arcmin\ in diameter with unusual flocculent and somewhat
filamentary structure. Based on MOST 843\,MHz and ATCA 20\,cm images, \citet{Roberts+1999} referred to the G313.0--0.1 ``diffuse shell'' as a ``potential SNR'', but the MeerKAT
image reveals its intricate morphology for the first time.

\paragraph*{G324.26+0.24.}
Approximately circular,
edge-brightened structure, bright to the Galactic SE,
$\sim$3\farcm5 in diameter.  There is no associated IR emission. This source was also identified as an SNR candidate in a recent ASKAP image \citep{Ball+2023}.

\paragraph*{G348.32$-$0.73.}
Unusual structure suggesting a composite
SNR with a shell and a central PWN. It has an
elongated central bright region $\sim 2\arcmin \times 1\arcmin$
in extent, surrounded by a faint, possibly shell-like region
$\sim$7\arcmin\ in diameter, edge brightened to the Galactic SW.
The radio source GPA 348.30$-$0.72 from
the Green Bank 8 GHz and 14 GHz survey \citep{Langston+2000}
likely corresponds to the central bright region. 

\paragraph*{G357.11+0.05.}
Possibly partial shell, relatively circular in outline, $\sim$8\arcmin\ in
diameter, bright to the Galactic SW and incomplete to the NE\@.  The
shell appears ``indented'' to the Galactic NE, suggesting possible
interaction in this direction.  There is little emission in the
interior detected in the MeerKAT image.

\subsection{Youthful pulsars and their environments}
\label{sec:PWNe}

\begin{figure*}
\centering
\includegraphics[width=\linewidth]{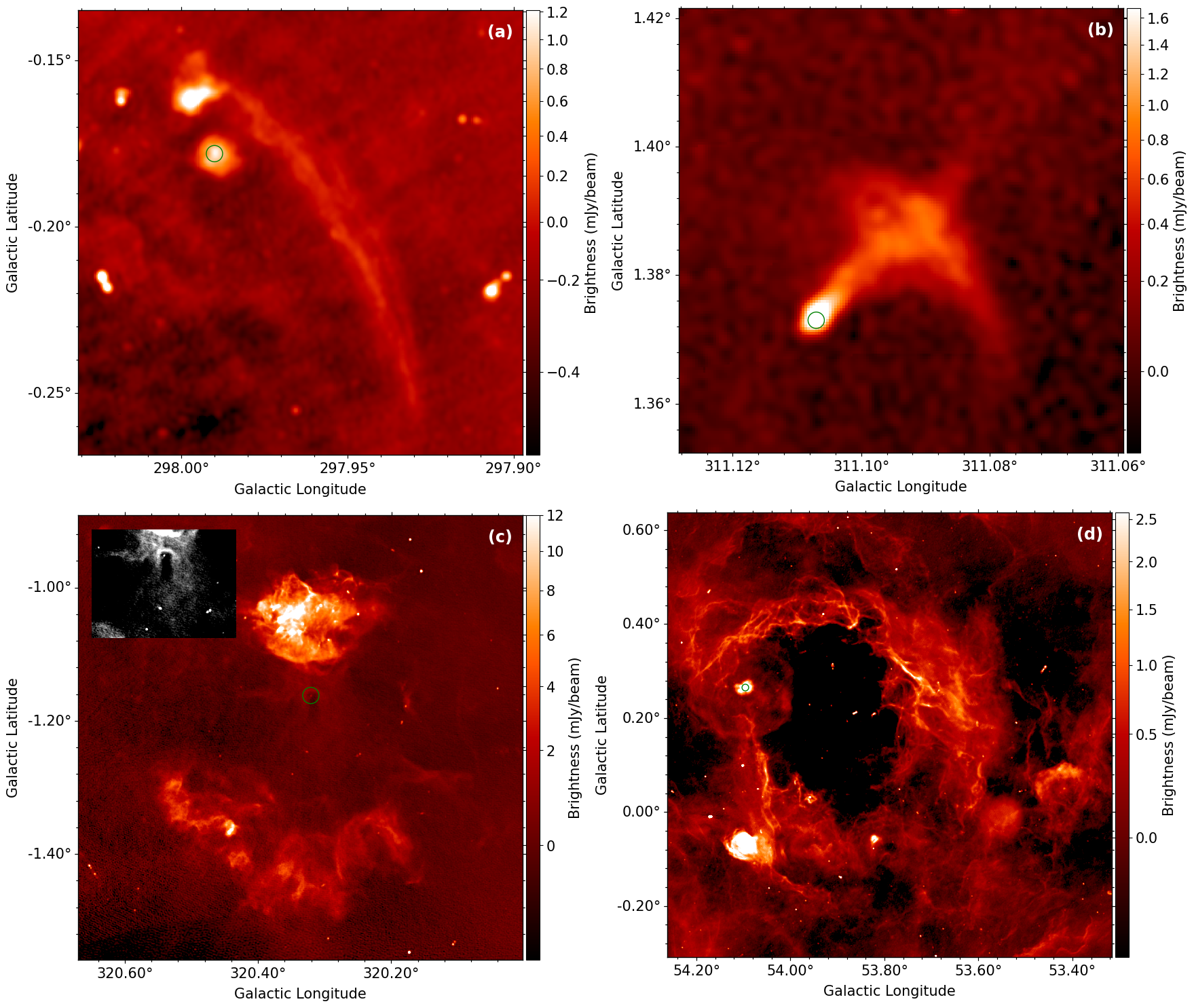}
\caption{Extended emission from SMGPS images surrounding four energetic and youthful pulsars. The colourscale shows the brightness and the green circles indicate the positions of the pulsars. \emph{(a):} Gamma-ray (radio-quiet) PSR J1208$-$6238 with associated PWN-like diffuse emission. The long arc of emission across the image is presumably unassociated. \emph{(b):} Gamma-ray (radio-quiet) PSR J1358$-$6025 with coincident and trailing emission that may be a bow-shock nebula. \emph{(c):} PSR J1513$-$5908 with its surrounding PWN and SNR G320.4$-$1.2. The $14\arcmin \times 10\arcmin$ inset shows the point-like pulsar within a radio cavity.
\emph{(d):} PSR J1930+1852 within its bright G54.1+0.3 PWN, towards upper left. This complex is surrounded by a circular shell that may be the corresponding SNR. Towards lower right is the unrelated SNR G53.41+0.03, with a positionally coincident central compact source.
}
\label{fig:PWNe}
\end{figure*}

Young and energetic pulsars are often associated with pulsar wind nebulae (PWNe) as they inject an energetic particle wind into the surrounding medium 
\citep[e.g.][]{Slane2017}. PWNe typically manifest in radio images as extended, diffuse sources with relatively flat spectra and a variety of morphologies 
depending on how the pulsar wind interacts with its dynamic environment. 
The unprecedented surface brightness sensitivity of MeerKAT therefore provides
an  opportunity to discover previously unknown PWNe. 
In order to do so, we visually inspected 
\SMGPS\ images in the vicinity of a subset of known pulsars. We applied two different criteria in selecting the sample.
First, we selected pulsars 
from the ATNF pulsar catalogue \citep{Manchester+2005}\footnote{Version 1.64; \url{https://www.atnf.csiro.au/research/pulsar/psrcat/expert.html}} with large spindown flux ($\propto \dot{E} / d^2 > 10^{35}$ erg s$^{-1}$ kpc$^{-2}$), where $\dot{E}$ is the spindown luminosity and $d$ is the 
distance. 
Seventy-seven pulsars satisfied this criteria. We also selected another 30 pulsars with high inferred surface magnetic field strength ($B_{\rm 
s}> 3 \times 10^{13}$ G). 
Below we summarize interesting findings for the fields surrounding four of these pulsars.

\subsubsection{PSR J1208--6238}

PSR J1208$-$6238 was discovered in a blind gamma-ray search with  \emph{Fermi}-LAT \citep{ClarkPW2016}. Despite a sensitive search with the Parkes telescope, this remains a radio-quiet pulsar.
 
The pulsar has a very high magnetic field strength, $B_{\rm s}=3.8 \times 10^{13}$ G, and a very small characteristic age, $\tau_c = 2.1$ kyr. There is no prior association with either an SNR or a PWN, although \citet{Bamba+2020} detected a possible X-ray PWN with $4.4\,\sigma$ significance.
The \MKGPS\ image in Fig.~\ref{fig:PWNe}a clearly shows diffuse emission from a compact source (diameter $\approx 45\arcsec$) at the position of the pulsar. 

The fitted spectrum for this source is shown in Fig.~\ref{fig:psrj1208}.
Given the radio-quiet nature of the pulsar and the relatively flat spectrum ($\alpha = -0.3$) of the extended emission, the radio
emission in Fig.~\ref{fig:PWNe}a is most probably from the PWN associated with PSR J1208$-$6238.

\subsubsection{PSR J1358--6025}

PSR J1358$-$6025 was discovered by  Einstein@Home\footnote{\url{https://einsteinathome.org/gammaraypulsar/FGRP1_discoveries.html}}
in \emph{Fermi}-LAT data, and has $\tau_c = 318$ kyr.
The \SMGPS\ image in Fig.~\ref{fig:PWNe}b shows a 
structure reminiscent of a bow-shock PWN.

Using the method summarized in Section~\ref{sec:spi} and Fig.~\ref{fig:psrj1208}, we attempted to measure the radio spectrum for this nebula. This is complicated because the source is located near the edge of a data cube, and has some sub-bands blanked due to
primary-beam effects. Nevertheless, we obtain $\alpha_{\rm head} = -0.4\pm0.1$ and
$\alpha_{\rm tail} = -0.9\pm0.2$.
These values should be regarded as preliminary, however they are 
consistent with the flatter-spectrum ``head'' being generated by 
electrons freshly accelerated by the pulsar. This appears to be a bow-shock PWN associated with PSR J1358$-$6025.

\subsubsection{PSR J1513--5908}

PSR J1513$-$5908 (B1509$-$58) is a young and very energetic pulsar, powering a spectacular PWN visible from radio to TeV energies. The intricate morphology of the system includes apparent interaction between the PWN and the SNR G320.4$-$1.2 within which it is embedded (see \citealt{Gaensler+2002} and \citealt{Romani+2023}).

Fig.~\ref{fig:PWNe}c shows the SNR G320.4$-$1.2 field as observed with MeerKAT. Three things are notable in this image.

\begin{itemize}

\item There is a cavity of extent $\approx 2\farcm5 \times 1\arcmin$, with the pulsar near its northern tip, within which no radio emission is detectable (see the inset in the panel). This corresponds very well to the brighter portion of the X-ray jet trailing the pulsar \citep{Romani+2023}. \citet{Gaensler+2002} had noted a region of reduced radio emission trailing the pulsar in ATCA 1.4 GHz data, but the MeerKAT image indicates that this is a bounded cavity.

\item Surrounding this cavity, there is a large region of very low surface brightness radio emission, previously undetected. Given the complexity of the field, it is unclear what portion of this emission may correspond to the PWN. To the north, this faint emission connects, at least in projection, to the bright SNR emission.

\item The northern portion of the cavity is bounded by a relatively bright radio arc. This had been previously faintly detected in linearly polarized emission \citep{Gaensler+2002}, but it is now clearly detected in Stokes $I$. The feature has remarkable correspondence to the X-ray emission \citep{Romani+2023}.

\end{itemize}

\subsubsection{PSR J1930+1852}

\begin{figure*}
	\centering
	\includegraphics[width=\textwidth]{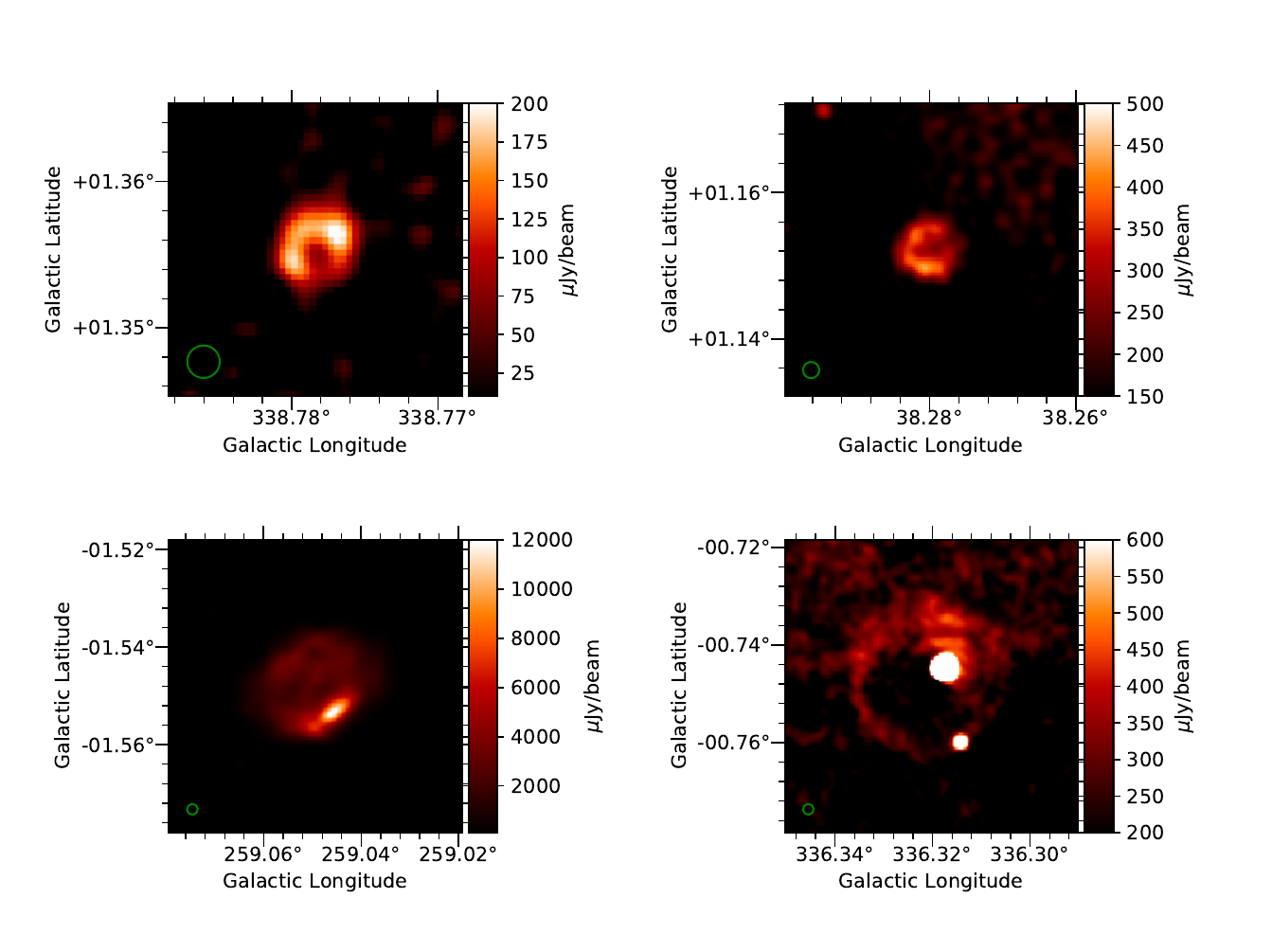}
    \caption{
    Four examples of new possible PN candidates from the SMGPS. In the lower-right panel the bright point sources are likely unrelated field sources. The green circle in the lower-left corner of each panel indicates the synthesized beam.}
	\label{fig:new_PN}
\end{figure*}

PSR J1930+1852 is one of the most energetic pulsars in the Galaxy \citep{Camilo+2002} and powers a prominent PWN from radio to X-ray wavelengths \citep{Lang+2010}. However, it is not clearly associated with an SNR shell, despite repeated searches. \citet{Lang+2010} report on the discovery of a shell with the VLA at 1.4 GHz, but \citet{Driessen+2018} revisit the question of whether this feature is the SNR, and on the basis of additional observations at lower frequencies with WSRT and LOFAR conclude that there is no detected shell surrounding the PWN G54.1+0.3.

The SMGPS image of this region (Fig.~\ref{fig:PWNe}d) shows that there is a clear circular feature, $\approx 14\arcmin$ in diameter, surrounding the pulsar and its PWN. While it sits within a complex region of overlapping features, it is morphologically distinct from its surroundings. Comparison with MIR images clearly shows the previously identified surrounding \HII\ region, but there is no detectable IR emission from the circular feature. We regard this as a candidate SNR shell associated with PSR J1930+1852.

Fig.~\ref{fig:PWNe}d also shows, towards the bottom right, the best image to date of the young SNR G53.41+0.03 confirmed by \citet{Driessen+2018}. Curiously, we detect a point-like source near the center of the SNR shell, which may be deserving of further investigation.

\subsection{Evolved stars}

At the end of their life, stars shed their outer layers to form circumstellar envelopes. During this process, when the exposed inner layers are hot enough, the circumstellar envelopes become ionised and radio emission arises. Stars over a large mass range undergo this phase: in the following we describe our findings for low- (Section \ref{sec:pne}) and high-mass (Section \ref{sec:high}) evolved stars.

\subsubsection{Planetary nebulae}
\label{sec:pne}

\begin{table*}
	\caption{Sub-sample of six MBs detected in SMGPS tiles.}
	\begin{tabular}{lcccc}
		\hline
        \hline
		
		Name & RA & Dec & 24 $\mu$m morphology$^{a}$ & Previous radio detection  \\
		&    &     &   & \\
		\hline
		
		MGE G020.4513$-$00.9867 & 18:31:57.1  & $-$11:32:46 & 2b &  -- \\
		MGE G023.3894$-$00.8753 & 18:37:02.8  & $-$08:53:15 & 2a  & -- \\
		MGE G028.7440+00.7076  & 18:41:16.0    & $-$03:24:11 & 2b  & 5 GHz \citep{Ingallinera+2014a}\\
		MGE G343.6641+00.9584 & 16:55:46.9    & $-$41:48:42 &  2b & 2 GHz \citep{Ingallinera+2019}\\
		MGE G354.1474$-$01.1141 & 17:35:27.6  & $-$34:29 19 &  2a & -- \\
		MGE G356.8235+00.0139 & 17:37:47.6    & $-$31:37:32 & 1a  & -- \\
		\hline
	\end{tabular}
    \begin{flushleft}
        $^a$ Morphology is from \citet{Mizuno+2010}: 1a -- object with clear central component; 2a -- ring with regular brightness distribution and no central source; 2b -- ring with irregular brightness distribution and no central source. \\
    \end{flushleft}
	\label{tab:MBs}
\end{table*}

Planetary nebulae (PNe) represent the last evolutionary stage of low- and intermediate-mass stars (zero-age main sequence mass below 8~\Msun)\@. 
Despite decades of study, several open questions remain. The number of known Galactic PNe is 
$\sim\!3000$, one order 
of magnitude lower than expected \citep{Sabin+2014}. If there really are
fewer PNe than expected, it 
would impact 
low-mass star evolution models. However, this could be simply due to an observational bias. Indeed, a possible explanation is that the main probe for the discovery of PNe, H$\alpha$ emission, is strongly affected
by extinction and confusion at very low Galactic latitude, where most PNe are expected \citep[e.g.][]{ZijlstraP1991}. Unlike H$\alpha$, 
radio emission is largely unaffected by Galactic dust and, since PNe are radio emitters, radio observations are important
for discovering new, low-latitude PNe. For example, \citet{Irabor+2018} used the CORNISH survey to find 90 new, compact, young PNe. However, even if a potential PN is detected in radio,
it may be hard to confirm its identity. In regions where optical and infrared observations cannot be used, radio morphology provides a powerful means for identifying new PN candidates. \citet{Ingallinera+2016} showed that a typical feature of resolved PNe is that they usually appear as small rings or disks ($\lesssim 1$\arcmin) in radio images, isolated from other nearby sources. This morphology is typical of PNe and only rarely mimicked by other Galactic sources. 
From a visual inspection of the \SMGPS\ tiles, we extracted 176 previously unidentified sources that show the typical appearance of a PNe. In Fig.\ \ref{fig:new_PN} we show some new candidates.

\begin{figure}
	\centering
	\includegraphics[width=\linewidth]{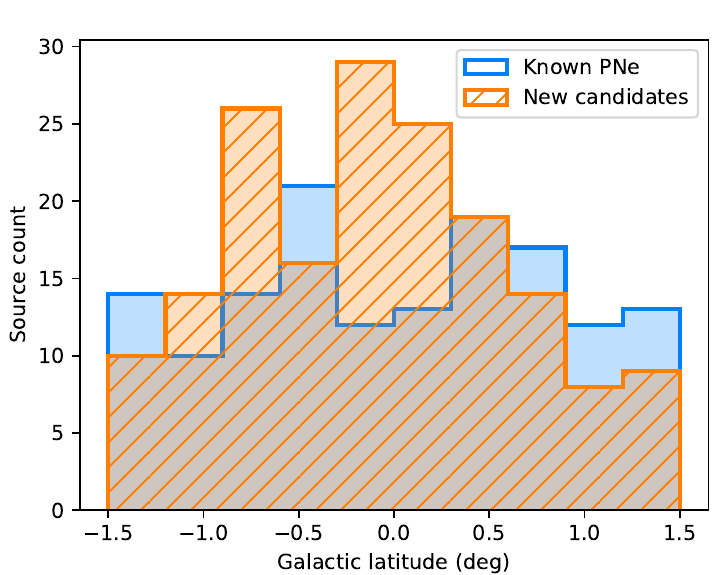}
	\caption{Galactic latitude distribution of previously identified or candidate PNe \citep[blue histogram --- 145 in total within the SMGPS survey area, from the HASH catalogue;][]{ParkerBF2016}, and of 176 SMGPS PN candidates. All sources from both samples have diameter larger than 30\arcsec.}
	\label{fig:PN_dist}
\end{figure}

The main limitation to identifying PNe by their radio morphology is that it requires the source to be resolved enough to determine the morphology. Given the SMGPS resolution, this implies that we are not able to identify PNe much less extended than 30\arcsec\ across.
But even with this important limitation, some results can already be obtained. Notably, the distribution of these 176 new possible PNe,
shown in Fig.\ \ref{fig:PN_dist},
is strongly peaked at $b=0\degr$. By comparison, the latitudinal distribution of known PNe with a diameter larger than 30\arcsec\ extracted from the HASH catalogue \citep{ParkerBF2016} is significantly flatter than in \SMGPS, where both surveys overlap.

This suggests that the \MKGPS\ may be recovering many   previously missing low-latitude PNe. 
This is possible thanks to three fundamental features of the survey: high sensitivity, good resolution, and large survey area. No previous survey matches \SMGPS\ in this regard. 
We note that the number of new possible PNe in Fig.\ \ref{fig:PN_dist} is larger than the number of previously identified PNe, with the same minimum dimension, in the same sky area. In other words, the \SMGPS\ 
may allow us to double the number of known PNe larger than 30\arcsec.

\subsubsection{Mid-infrared bubbles}
\label{sec:mbs}

\begin{figure*}
    \centering
	\includegraphics[width=\textwidth]{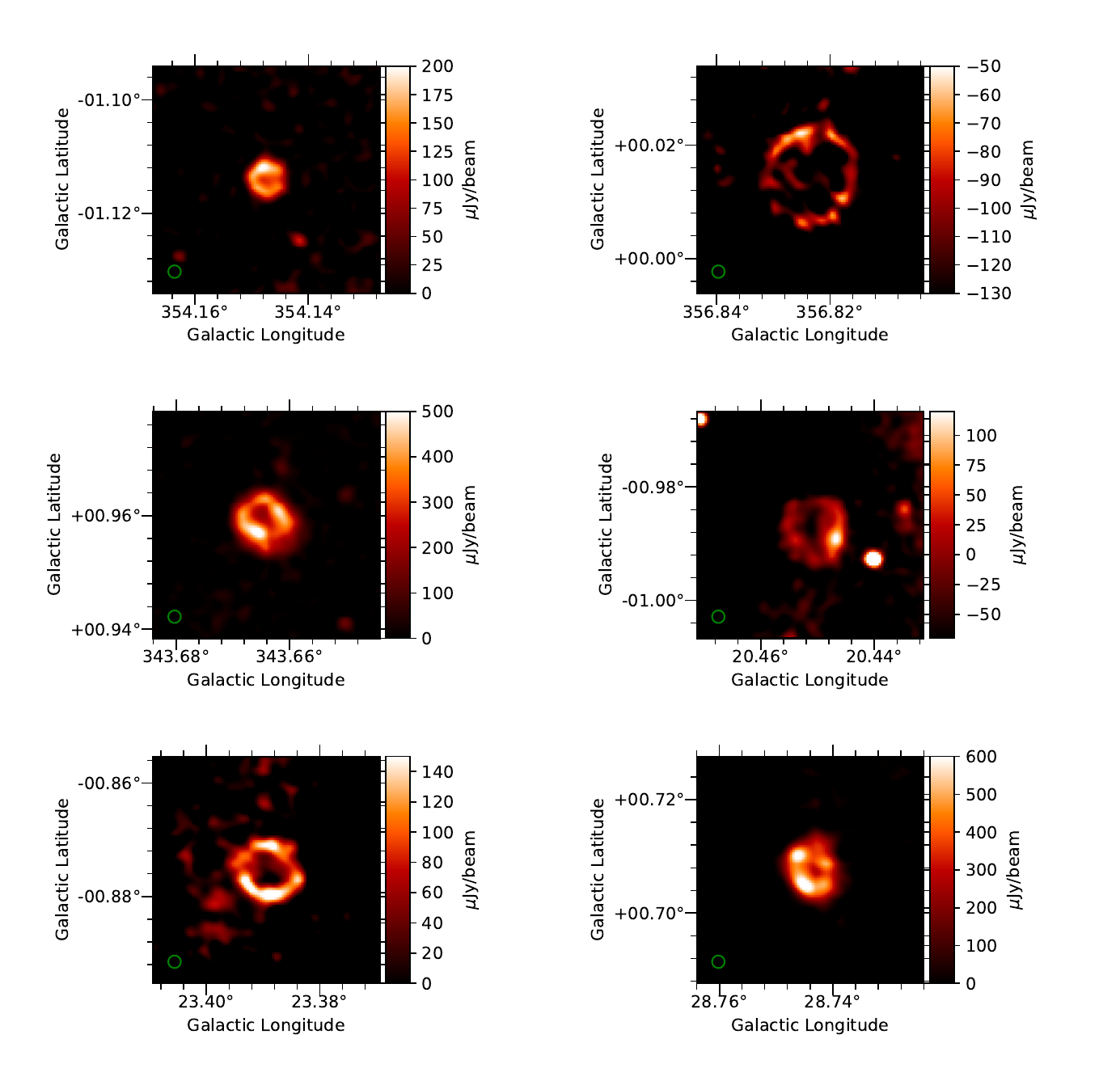}
	\caption{
    1.3-GHz \SMGPS\ images of six infrared bubbles from the Mizuno sample. From left to right and top to bottom: MGE G354.1474$-$01.1141, MGE G356.8235+00.0139, MGE G343.6641+00.9584, MGE G020.4513$-$00.9867, MGE G023.3894$-$00.8753 and MGE G028.7440+00.7076.  All the images are $2\arcmin \times 2\arcmin$ and centered at the bubble's position. The green circle in the lower-left corner of each panel indicates the synthesized beam. Note that the brightness levels for MGE G356.8235+00.0139 are all negative, a striking indication that this source sits in a ``bowl'' of emission. As for all images in this paper unless indicated, the brightness has not been corrected for any variations in the zero level; see Section~\ref{sec:spi}.}
	\label{fig:bubbles_radio}
\end{figure*}

\begin{figure*}
	\centering
    \includegraphics[width=\textwidth]{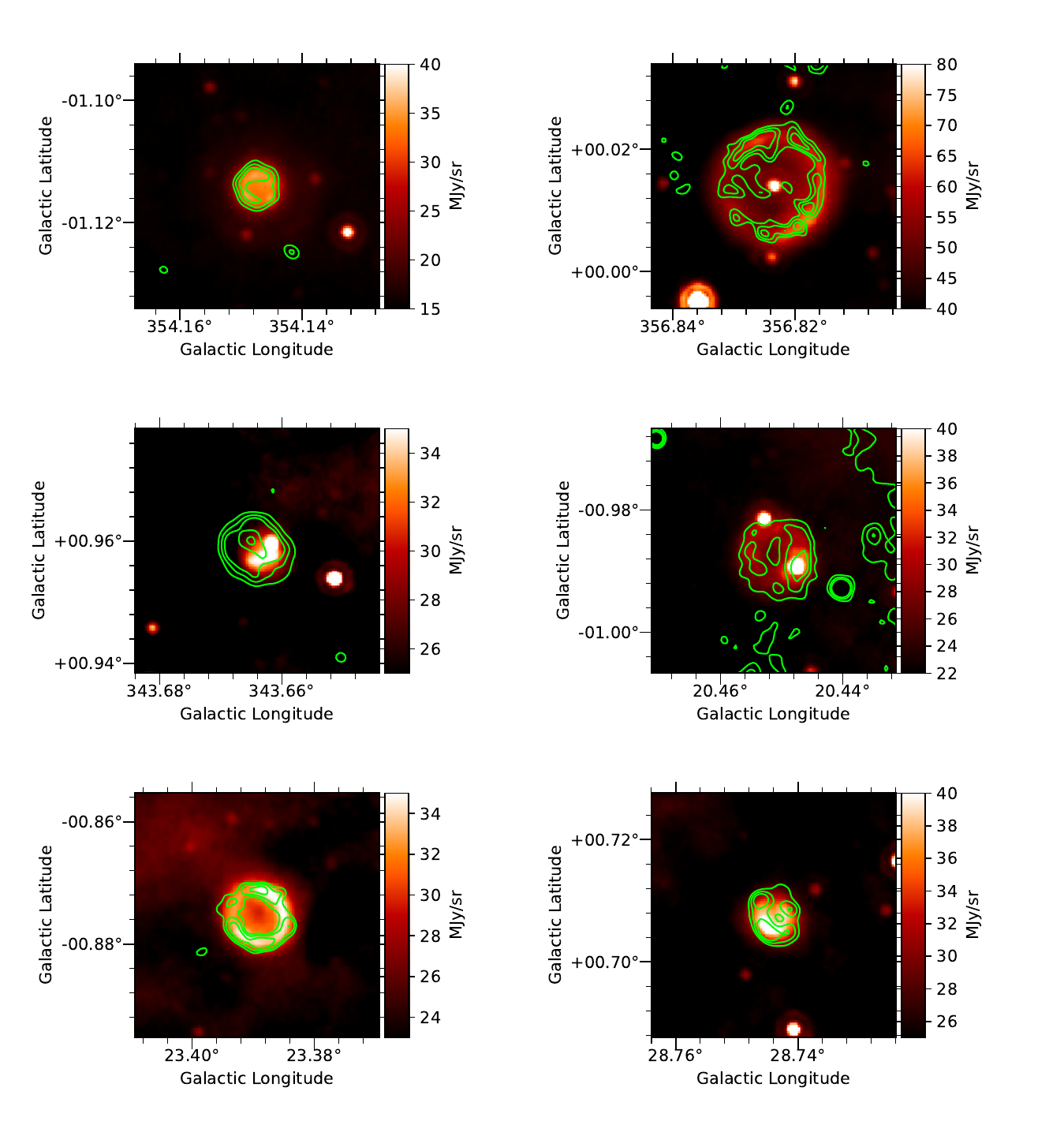}
    \caption{
    Mid-infrared morphology (24 $\mu$m MIPSGAL) of the same bubbles of Fig.~\ref{fig:bubbles_radio}. SMGPS 1.3 GHz images are superimposed as green contours. All the images are $2\arcmin \times 2\arcmin$ and  centered at the bubble's position.}
	\label{fig:bubbles_24microon}
\end{figure*}

The MIPSGAL Legacy Survey \citep{Carey+2009}, covering the inner Galactic Plane at 24 $\mu$m, allowed the discovery of 428 compact mid-IR bubbles (hereafter ``MBs'', \citealt{Mizuno+2010}). MBs have varying morphologies (rings, disks or shells), are compact ($ \leq 1 \arcmin $) and are widespread  through the entire Galactic Plane, with an approximately uniform distribution of about 1.5 bubbles per square degree. Most of them seem linked to PNe \citep{Nowak+2014}, and are part of the HASH catalogue. 
Indeed, the HASH catalogue designates 54 MBs as true PNe, seven as likely, 83 as possible and 189 as candidate.
Among the remaining MBs, 63 are massive stars or candidates, three are SNRs and 29 are still unclassified.

MBs usually show emission only at 24 $\mu$m and seldom have counterparts in 3.6 $\mu$m or 8.0 $\mu$m GLIMPSE images. This is probably the reason why this population had been missed by previous visible or near-IR surveys. 
\citet{Ingallinera+2016} investigated the radio morphology of
18 previously-known MBs using high-resolution 5-GHz VLA observations, and noted a possible correlation between
an MB's radio morphology and its classification: an association with a massive star is almost certain   
when the radio morphology at 5~GHz indicates the presence of a central object, otherwise the source is more likely a PN.
In this context it is therefore worthwhile to  
determine how many MBs are detected in the \SMGPS\@.

From a visual inspection of the \SMGPS\ tiles, we find that 146 out of the 244 MBs in the \SMGPS\ survey area (60\%) are detected, and have various radio morphologies.  According to the SIMBAD database\footnote{\url{http://simbad.u-strasbg.fr/simbad}}, 113 of the detected MBs (77\%) are new radio detections, and 137 (94\%) are new detections in the 1--2 GHz range. 
In Fig.\ \ref{fig:bubbles_radio}, we show radio images of six example MBs detected in \SMGPS,
and in Fig.\ \ref{fig:bubbles_24microon} 
we show mid-IR images with SMGPS contours for the same objects. Table \ref{tab:MBs} lists coordinates, MIR morphology, and status of previous radio detection for the six MBs.

It is evident that even the small sample displayed in Fig.\ \ref{fig:bubbles_24microon} includes different radio and MIR morphologies. In sources MGE G354.1474$-$01.1141 and MGE G356.8235+00.0139 the MIR nebula is slightly more extended than the radio nebula, as is observed in some LBV candidates. The lack of an evident central source in the radio images may be ascribed to insufficient sensitivity,
inadequate to detect the stellar wind associated with these LBV candidates. By contrast, in MGE G343.6641+00.9584, identified as a possible PN candidate by \citet{Ingallinera+2019}, the 24$\mu$m nebula lies well within the ionized nebula. This hints at the possibility of classifying MBs based on a comparison of MIR and radio morphology, but that possibility needs to be assessed with the analysis of the complete sample of 146 MBs detected in SMGPS.

\subsubsection{Massive stars}
\label{sec:high}

\begin{figure*}
    \includegraphics[height=6.75cm,clip]{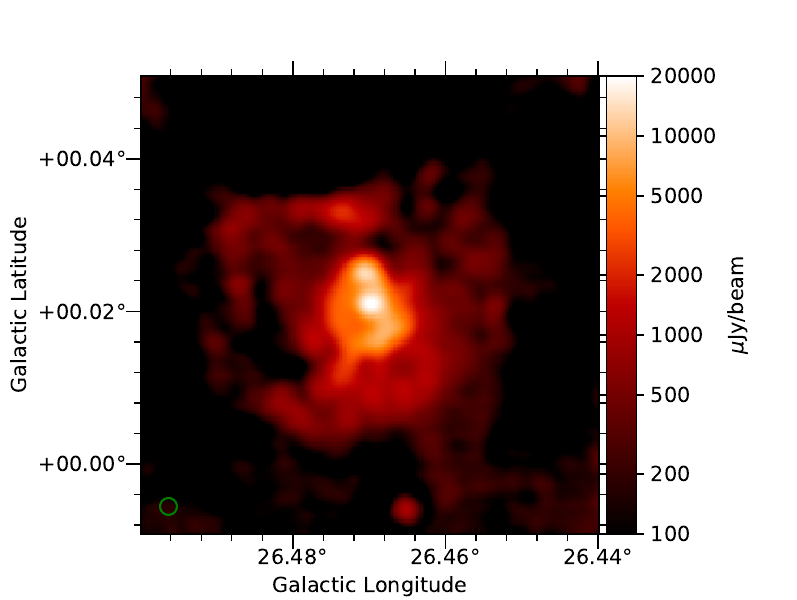}
    \includegraphics[height=6.75cm,clip]{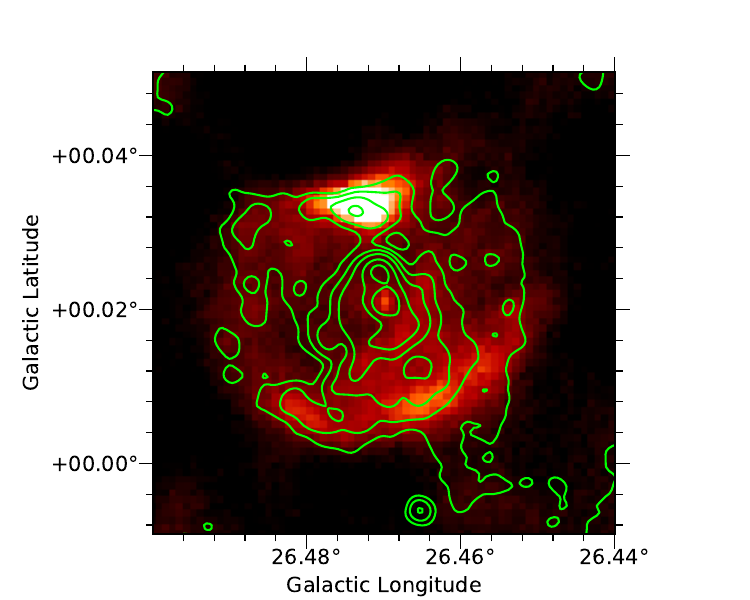}
    \caption{
    LBV candidate G026.47+0.02. \emph{Left:} SMGPS 1.3 GHz image with logarithmic colour scale at angular resolution 8\arcsec\ (green circle in the lower-left corner). \emph{Right:} 70 $\mu$m Hi-GAL \citep{Molinari+2010} image with linear colour scale at angular resolution 6\arcsec, with the radio image superimposed as green contours.}
	\label{fig:G26}
\end{figure*}

The evolution of disk galaxies is strongly influenced by their populations of massive stars, that via their copious and energetic stellar winds provide both processed material and energy to the interstellar medium, triggering the formation of new generations of stars.
Mass-loss is a key factor in the evolution of a massive star ($M \geq$  8 \Msun) towards its endpoint, which is essentially determined by the initial mass of the star and  the total mass lost during its lifetime.
However, many aspects of the evolution of massive stars are still not clear and, in particular, how  mass-loss properties change over time is not well constrained. 
A very massive O-type star on the main sequence, with 60--100 \Msun\ and a mass-loss rate of $\sim 10^{-6} \,\Msun$\,yr$^{-1}$, evolves into a Wolf-Rayet star, whose mass usually does not exceed 30 \Msun\ \citep{Langer+1994}. This implies that severe mass-loss must occur in the post-main-sequence evolution, through strong winds and/or eruptions. Extreme instability in the post-MS evolution of massive stars has been observed in so-called Luminous Blue Variables (LBVs)\@.
LBVs are among the most luminous and massive stars,
and  are characterized by variability and strong mass loss. They can 
also undergo giant eruptions, during which they lose a significant fraction of their envelopes \citep{Humphreys+1994}. The presence of extended stellar ejecta around several LBVs and LBV candidates indicates that eruptions are  common in these objects \citep{WeisB2020}. Because of these characteristics they represent an important phase of massive star evolution, and it is important to derive observational constraints 
on LBVs for a full understanding of the evolution of the most massive and luminous stars.
LBVs are very rare in our Galaxy, and all the information we have on their properties relies on a very small sample of 
19 confirmed LBVs and 42 candidates \citep{Richardson2018}.
Radio observations are crucial to  
determine the mass-loss properties of LBVs:
the current mass-loss rate can be measured from the radio emission of the central object's wind, and from the analysis of the stellar ejecta.
Together with mid-IR observations, the total mass (gas + dust) and the occurrence of multiple eruptions can be determined
\citep{Umana+2011a}.

In Section \ref{sec:mbs} we introduced the mid-infrared bubble sample as an MIR catalogue in which to search for low- and high-mass evolved stars. When a central object is detected in a MB, optical and infrared spectroscopy can provide clues to its nature. Spectroscopic studies indicate that the majority of MBs showing a central MIR object are indeed massive evolved stars, with a significant fraction being LBV candidates \citep{GvaramadzeKF2010, Wachter+2010, Flagey+2014, Nowak+2014, Silva+2017}.
These spectroscopic studies have resulted in a robust classification of 30\% of the total MB sample.
MGE G356.8235+00.0139 in Fig. \ref{fig:bubbles_24microon} shows a central MIR source, making it a potential evolved massive star candidate. 

The new radio detections of MBs in the SMGPS will allow detailed studies focusing on the classification of these radio/MIR nebulae, and leading to a better understanding of their origin in the framework of massive star evolution.

\paragraph{Gal 026.47+0.02.}

To illustrate the potential of the SMGPS for studying and determining the mass-loss properties of massive evolved stars, we extracted from the SMGPS tiles the image of Gal 026.47+0.02, hereafter G26,
which was proposed as an LBV candidate by \citet{Clark+2003}.

G26 was observed in radio and MIR by  \citet{Umana+2012}, who
performed a detailed comparison of 
the dust and ionised components of the stellar ejecta. The morphology of both components is consistent with a series of nested toroidal shells, with a common axis. 
The inner torus is ionized and its radio emission traces a bipolar component.  At its  center there is a compact component, related to the wind of the central star. The outermost shell of material is traced only  by thermal dust emission.
Between these two main structures, some diffuse radio emission was 
evident in the VLA maps. Besides the morphology of the circumstellar material, the current mass-loss rate and the properties of the associated ejecta were also derived, most notably the total mass.

MeerKAT's high sensitivity, particularly to extended emission, shows a radio shell $\sim 1\farcm8 \times 2\farcm7$ in size, more extended and better defined than previously known. This shell nicely matches the structure observed at 70 $\mu$m, both visible in Fig.\ \ref{fig:G26}, indicating the presence of an ionized gas component cospatial with the dusty component.
All the major structures visible at 70 $\mu$m are readily identifiable at 1.3 GHz. 
The greater dimension of the radio nebula implies a greater mass of ionised gas than previously appreciated. 
This result reinforces the problem, noted by \citet{Umana+2012}, of reconciling the source of UV photons required 
to explain the inner component of the radio nebula with the spectral classification of the central source given by \citet{Clark+2003}. It additionally provides evidence that episodes of strong mass loss occurred in G26 also during a hotter phase of its evolution \citep{Umana+2012}.

LBVs have been proposed as direct progenitors of core-collapse SNe \citep{Miller+2010,Pastorello+2018,Taddia+2020}, and
the morphology of some SNRs has been modeled
as resulting from the interaction of the SN blast waves with non-uniform pre-existing circumstellar environments (CSEs), including the typical ejecta observed around LBVs \citep{Ustamujic+2021,Chiotellis+2021}. One of the critical parameters for these models is the mass content of the CSE, for which radio observations provide excellent diagnostics.

Several LBVs and LBV candidates have been detected in the \SMGPS. Analysis following that outlined here for G26 has the potential to place improved constraints on the gas content in their CSEs.

\subsection{Radio stars}
\label{sec:stars}

\begin{figure*}
\centering
\includegraphics[width=\textwidth]{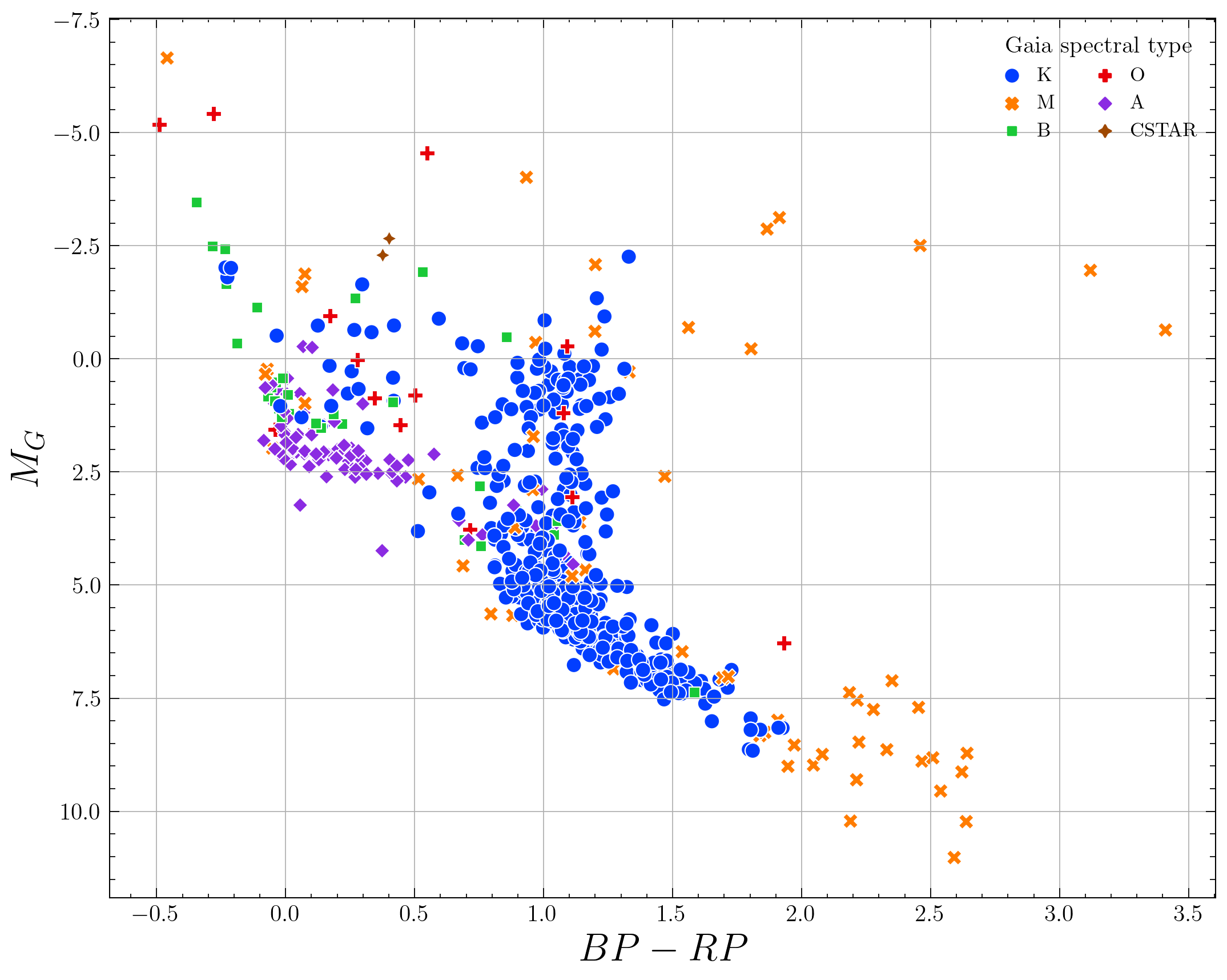}
\caption{\label{fig:colmag}Color--magnitude diagram for 624 dereddened \textit{Gaia} counterpart stars to SMGPS. Here, $BP - RP$ represents the dereddened intrinsic color of the star, while $M_G$ denotes the absolute magnitude after extinction correction. Each source is color-coded based on spectral type information obtained from \textit{Gaia}.}
\end{figure*}

While stars emit a negligible fraction of their luminosity in radio ($L_\mathrm{rad}/L_\mathrm{bol}\approx10^{-11}$ for the quiet Sun), certain phenomena, occurring during  different phases of stellar evolution, can be particularly well observed only at 
radio frequencies. Radio emission from stars can originate from the thermal radiation of stellar winds and/or circumstellar nebulae \citep[e.g.\ WR stars and LBVs,][]{Umana+2011b}, from 
maser mechanisms \citep[e.g.\ AGB and post-AGB stars,][]{Ingallinera+2022} or from non-thermal radiation caused by the coherent or incoherent interaction between charged particles and magnetic fields \citep[e.g.\ magnetic chemically peculiar stars and flare stars,][]{Trigilio+2018}. Currently there are no reliable statistics on radio emission for each class of radio stars, with the existing information collected through targeted observations that favoured certain stellar types. New-generation wide-area sensitive radio surveys such as the SMGPS have the potential to detect many hundreds to thousands of radio stars without prior selection bias.

Among the usual methods exploited to discriminate Galactic from extragalactic sources, and in particular radio stars from unresolved 
galaxies, e.g.\ spectral indices, 
variability, polarization, we focused, due to the nature of our data, on 
two: cross-matching with an optical catalogue, in our case \textit{Gaia},
and circular polarization.

\subsubsection{Cross matching with \textit{Gaia}}

\begin{table*}
    \centering
    \caption{Circularly polarised sources in the SMGPS G336.5 tile.}    
    \label{tab:stokesV}
    \setlength\tabcolsep{14.0pt}
    \begin{tabular}{ccccccc}
    \hline\hline
    Gal. Long. & Gal. Lat. & Stokes $V$  & Stokes $I$ & $V/I$ & SIMBAD match & SMGPS$-$ATNF offset\\
    (dd mm ss.s) & (dd mm ss.s) & (mJy) & (mJy) & (\%) &  & (arcsec)\\
    \hline
    334 56 36.3& $-$00 15 58.7& $-$0.209 &   1.20 &  $-$17\% & -- & --\\
    335 31 45.9& +00 38 10.6 & +0.059  &  0.51 &  +11\% & PSR J1626$-$4807 & 1.0\\
    335 45 57.6& +00 27 39.8 &  +0.125 &   0.65 &  +19\% & PSR J1628$-$4804& 0.2\\
    336 24 13.9& +00 33 46.8 &  +1.724 &   7.62 &  +22\% &  PSR B1626$-$47& 3.2\\
    336 51 50.9& +00 31 47.5 &  $-$0.086 &   0.12 &  $-$72\% & --&--\\
    337 09 59.1& $-$00 15 33.2 &  +0.124 &   0.48 &  +26\% & --&--\\
    \hline
    \end{tabular}
\end{table*}

To assess the stellar population among point-like sources in the \SMGPS, we performed a cross-match of all the \SMGPS\ catalogue point sources and the \cite{Gaia+2022} DR3 using a 2\arcsec\ search radius. We used the Bayesian cross-matching approach first proposed by \cite{budavari2008} and modified by \cite{salvato2018} that allows for the incorporation of multiple parameters, such as magnitude and color, as priors. Radio stars detected in the optical are usually red and relatively nearby; hence, we considered \SMGPS\ and \textit{Gaia} positional uncertainties, Gaia color\footnote{\textit{Gaia} color refers to $G_{BP} - G_{RP}$, the difference between the \textit{Gaia} Blue and Red filter magnitudes.}, and proper motion as priors \citep[see also][for the use of proper motions to detect radio stars]{Driessen+2023}. These priors and the separation between \SMGPS\ and \textit{Gaia} sources were used to estimate the probability of association, $p_{\text{any}}$, between sources in both surveys. This is the likelihood that a given pair of astronomical objects are related based on the cross-matching procedure. In order to determine a suitable threshold on $p_{\text{any}}$ for genuine associations, we first created a shifted \SMGPS\ catalogue by randomly modifying the position of each actual \SMGPS\ source by $2^{\prime}$, while ensuring a separation of at least $0\farcm5$ compared to each original \SMGPS\ source. The shifted \SMGPS\ catalogue was cross-matched with \textit{Gaia} using a 2\arcsec\ radius to obtain a shifted (unphysical) cross-matched sample. Comparing the sets of $p_{\text{any}}$ for both the real and shifted samples enabled us to calculate reliability ($1-p_{\mathrm{false\_positive}}$) and completeness statistics for actual associations. We selected for further analysis the 52,576 SMGPS sources with $p_{\text{any}} \geq 0.77$, for which the sample completeness and the reliability are both 77\%.

For the 52,576 sources, we further considered probabilistic photo-geometric distance estimates from the \textit{Gaia} DR3 distance catalogue \citep{bailer2021}. We selected stellar candidates within 3 kpc and with proper motion $\geq 5$ mas yr$^{-1}$. This limited our sample to 831 \textit{Gaia} and SMGPS sources, which we consider to be good candidates for radio stars. To investigate the intrinsic properties of these stars, we dereddened the 624 stars with measured extinction given in the \textit{Gaia} catalogue, and computed their absolute magnitude. The resulting color--magnitude diagram is shown in Fig.~\ref{fig:colmag}. This shows different stellar objects at different evolutionary stages, as previously noted by \cite{Gudel:2002} and \cite{Yu:2021}.

We also searched for known stellar sources amongst the 831 \SMGPS-\textit{Gaia} optical counterpart candidates by querying the SIMBAD database \citep{Wenger:2000}, identifying 124 matches. Typical objects represented include massive stars (O, B, WR systems), binaries (eclipsing, spectroscopic, RS CVn types), long period variables, super-giants and YSOs. A full report on this cross-matching procedure, statistical analysis and results will be presented elsewhere.

With the identification of 831 radio stellar candidates enabled by \SMGPS, the prospects for discovering and studying a statistically significant sample of radio stars are promising, and MeerKAT is poised to contribute significantly to this endeavour.

\subsubsection{Circular polarization}

Radio stars can be highly circularly polarised ($\sim 10\%$--100\%, depending on the emission type), while extragalactic sources are usually only very weakly circularly polarised, making detections in  Stokes $V$ images relatively likely to be stars, although only some of the stellar emission mechanisms listed earlier will be detected. For a previous application of this method, see \citet{Pritchard2021}. Here we inspected the Stokes $V$ images corresponding to two SMGPS tiles for which we did polarisation calibration, G056.5 and G336.5.

In the G056.5 tile we found an interesting source at RA=19:39:56.63, decl=+21:39:05.0\@. 
With a flux density of 73 $\mu$Jy it is faint, but 100\% circularly polarised. There are no \textit{Gaia} matches, but a 2MASS YSO candidate is offset by 1\farcs3.
For a source this faint we expect a statistical positional uncertainty $\approx 1\farcs3$ (see Section~\ref{sec:astrometry} and \citealt{Fomalont1999}),
so this is plausibly a potential match.
This source's high circular polarisation 
suggests a coherent emission mechanism. The most likely such mechanism for stars is electron-cyclotron maser emission \citep{Melrose1982}. Such emission is usually limited in time, with 
timescales ranging from minutes to days \citep[e.g.][]{Das2023}. Variability of this source during the multi-hour span of the relevant SMGPS observations could be probed by inspection of the raw visibilities.

In the more crowded G336.5 tile, we identified six circularly polarised sources, listed in Table \ref{tab:stokesV}.
Three of them are classified as pulsars on SIMBAD. The nominal offset between the position of PSR~B1626$-$47 (with uncertainty of $0\farcs9$ in the ATNF pulsar catalogue) and the corresponding SMGPS 
source is 3\farcs2. However this is based on an old timing solution; more recent measurements with the Parkes telescope yield a timing position that is $1\farcs3\ \pm 1\farcs0$ away from the SMGPS source (M. Lower, priv. comm.). For the other two pulsars, the catalogued positional uncertainties are larger than the nominal offsets listed in Table \ref{tab:stokesV}.
The other three sources have no known matches. However, the first and last source in Table~\ref{tab:stokesV} have a percentage of circular polarization compatible with pulsars. With $|V|/I = 72\%$, the remaining source is perhaps less likely to be a pulsar, but like the 100\% circularly polarized source in the G056.5 tile, its emission mechanism is also likely to be coherent and perhaps variable.

\subsection{\HII\ regions}

\begin{figure}
	\centering
	\includegraphics[width=\linewidth]{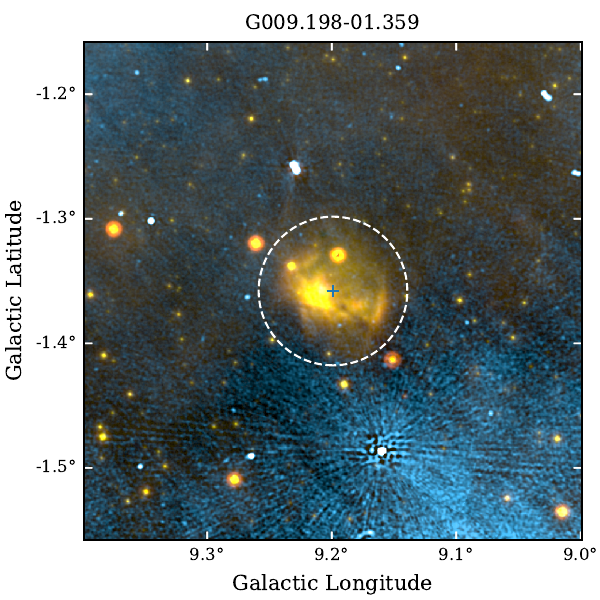}
 	\includegraphics[width=\linewidth]{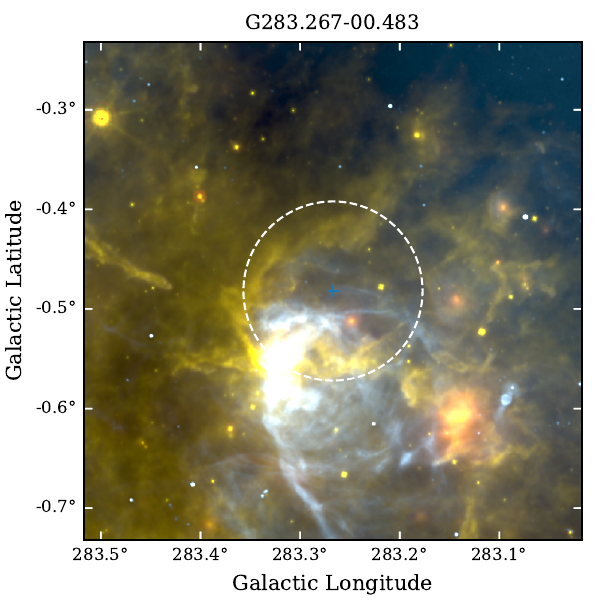}
	\centering
	\caption{WISE and SMGPS 3-colour images of example radio-quiet \HII\ regions from the \citet{Anderson+2014} WISE \HII\ region catalogue. WISE 12 $\mu$m and 22 $\mu$m are coded as yellow and red, with SMGPS 1.3 GHz coded as blue. Crosses and dashed circles mark the central positions and radii of the WISE \HII\ region candidates. \emph{Top:} An example of a radio-quiet \HII\ region candidate that is not detected by SMGPS. \emph{Bottom:} A radio-quiet \HII\ region candidate where the greater sensitivity and fidelity of SMGPS reveals faint radio emission within the infrared cavity.}
	\label{fig:largehiis}
\end{figure}

An unambiguous signpost of a site of high-mass star formation is the presence of an ionized hydrogen (\HII) region, usually identified by the strong free-free emission in the radio spectrum or the mid-infrared morphology \citep{Anderson+2014}. 
\HII~regions are typically classified by their size  and emission measures into hyper-compact (HC\HII), ultra-compact (UC\HII), compact (C\HII), and classical \HII\ regions \citep[e.g.][]{Kurtz2005}. Some of the larger giant \HII~regions are produced by the combined ionizing power of a cluster of OB stars \citep[e.g.][]{Chibueze+2013,Hindson+2013,Davies+2012}. 

Radio
observations have been instrumental in compiling large catalogues of \HII\ regions, particularly of ultracompact, compact and classical  
ones \citep{Urquhart+2013, Djordjevic+2019, Kalcheva+2018, Anderson+2011,Wenger+2021}. More recently, detailed multi-wavelength mid-IR imaging from the WISE mission has been used to 
confirm several thousand candidate \HII\ regions from their 12 $\mu$m and 22 $\mu$m morphologies \citep{Anderson+2014}. Nevertheless, a substantial fraction of the WISE
sample remain  
unconfirmed due to the observational limitations of existing radio surveys or dedicated followups (approximately 42\% of the WISE sample\footnote{As of the latest v2.3 catalogue: \url{http://astro.phys.wvu.edu/wise}} are classed as ``radio quiet''). 

Moreover, unlike their evolved cousins, the youngest and smallest HC\HII\ regions remain rare and elusive. The largest sample of these objects discovered to date have been in ``blind'' GHz surveys, utilising 1.4--5 GHz spectral indices and 1.4 GHz ``dropouts'' \citep{Yang+2019}, but this technique is necessarily limited by the lack of sensitive and high angular resolution 1.4 GHz survey data. Furthermore, the Yang et al.\ search was based upon the CORNISH survey \citep{Hoare2012}, whose footprint is within Quadrant I of the Galaxy. \SMGPS\ thus offers the potential to more than double the sample of known HC\HII\ regions by virtue of its improved sensitivity and matching coverage of the Quadrant IV CORNISH-South survey \citep{Irabor+2023}.

Here we explore the potential of \SMGPS\ to improve the radio classifications of WISE \HII\ regions. \citet{Umana+2021} showed that radio-quiet \HII\ regions from the WISE catalogue tend to be fainter and more compact than the general population, suggesting that their ``radio quietness'' is due to the sensitivity limitations of the radio data used for classification. With deeper ASKAP 912 MHz SCORPIO observations, \citet{Umana+2021} detected 45\% of the WISE sources previously classified as radio quiet. Similar results are reported by \citet{Armentrout+2021} who undertook deep 9 GHz JVLA observations of a subset of the WISE sample.

Both \citet{Umana+2021} and \citet{Armentrout+2021} focused on smaller radio-quiet \HII\ regions, indeed the largest angular scale recoverable by \citet{Armentrout+2021} was 70\arcsec. However there are many radio-quiet \HII\ region candidates in the WISE sample with much larger angular diameters, up to several 10s of arcminutes or even degree scales. These regions are not well imaged by radio interferometers with limited $uv$ coverage, which includes many of the original snapshot imaging surveys used for classification (e.g.~CORNISH) and also a number of dedicated follow-up studies \citep{Armentrout+2021, Wenger+2021}.

SMGPS on the other hand has dense coverage of the inner $uv$ plane and is also at a lower frequency than the dedicated radio follow-up studies, hence has excellent recovery of emission at much larger angular scales. Moreover, even with  conservative flux-density detection limits we expect the SMGPS to detect the majority of classical \HII\ regions within the Milky Way. An isothermal, optically thin, homogenous \HII\ region powered by a B0.5 star would be detectable to a distance of 18 kpc in the SMGPS, using the standard equations and predicted Lyman fluxes of \citet{Carpenter+1990} and \citet{Sternberg+2003}. This pessimistically assumes a uniform flux distribution, whereas in practice \HII\ regions are often limb-brightened.  Hence \SMGPS\ is ideally placed to resolve the question of whether large angular diameter WISE radio-quiet candidates are true \HII\ regions.

In order to test our hypothesis we  examined the WISE radio-quiet \HII\ region candidates with diameters $\ge$ 200\arcsec\ \citep[i.e.\ the largest angular scale of the observations of][]{Wenger+2021} in the \SMGPS\ moment zero images. Within the \SMGPS\ footprint there are 74 radio-quiet WISE \HII\ region candidates with diameters larger than 200\arcsec, excluding those \HII\ regions that are larger than a single \SMGPS\ tile and those that are only partially covered by \SMGPS\ observations. We excluded three of these regions from further analysis due to the presence of nearby bright confusing radio sources that result in image artefacts.

Out of the 71 WISE radio-quiet candidate \HII\ regions in our sample we see significant radio emission towards only 22 objects, of which an example is given in the bottom panel of Fig.~\ref{fig:largehiis}. G283.267$-$0.483 is a radio-quiet WISE \HII\ region candidate found to the N of the radio-bright WISE \HII\ region G283.322$-$0.557. SMGPS shows that the mid-IR cavity is indeed associated with low surface brightness radio emission (shown in faint blue in Fig.~\ref{fig:largehiis}).	However, the majority of the large WISE radio-quiet candidate \HII\ regions are not associated with 1.3 GHz emission (as in the top panel of Fig.~\ref{fig:largehiis}) or are associated with diffuse extended emission that is morphologically unrelated to the WISE mid-infrared. We conclude that, given the lack of SMGPS 1.3 GHz radio emission, the majority of WISE radio-quiet objects with diameters $\ge$ 200\arcsec\ are \emph{not} \HII\ regions. This is exemplified by G9.198$-$1.359 in Fig.\ \ref{fig:largehiis}. Despite the presence of the confusing bright radio source TXS 1808$-$217 which elevates the local noise, there is no discernible 1.3 GHz emission at the location of this WISE candidate above a 3$\sigma$ limit of $\sim$0.1 mJy\,beam$^{-1}$, an order of magnitude below what is expected for even a distant \HII\ region. A detailed study of these objects is beyond the scope of this paper, but we  speculate that these mid-infrared radio-quiet nebulae may instead be intermediate-mass star forming regions \citep[e.g.][]{Arce+2011}.

\section{The \HI\ Universe behind the \SMGPS} 
\label{sec:gps-hi}

As evident from the previous sections, the main scientific goals of the \SMGPS\ are studies of Galactic objects and the inner Galaxy itself. The \SMGPS, however, opens a further avenue for exploration, namely the study of the large-scale structures (LSS) in the Universe through the 21\,cm spectral line emission of gas-rich galaxies. While it might seem far-fetched that a narrow strip along the Galactic Plane (GP) 
could add significantly to our view of the Local Universe, many dynamically important structures such as the Local Void \citep[LV;][]{Tully2008,Tully+2019}, the Great Attractor \citep[GA;][]{Dressler+1987,Woudt+2004}, and the Vela supercluster \citep[VSCL;][]{Kraan-Korteweg+2017} are bisected by the southern GP and remain poorly mapped. The dispute about the volume within which the CMB dipole and the residual bulkflows arise continues unabated \citep[e.g.][]{Springob+2016,Kraan-Korteweg+2017}.

One of the few efficient tools that allow an unhindered view through the dust-obscured Zone of Avoidance (ZoA) are `blind' systematic \HI\ surveys. Many such surveys have been carried out since the early 1990s, 
mostly with single dish telescopes. The most relevant to the southern ZoA is the Parkes MultiBeam survey, HIZOA \citep[$|b| < 5\degr; v_{\rm Heliocentric} < 12,000$\,\kms,][]{Staveley-Smith+2016}. While it revealed unprecedented insight into hidden structures (such as the GA), the survey is quite shallow with its 883 detections. Moreover, the survey becomes increasingly incomplete for latitudes below $|b| \la 1\fdg5$ due to the increase in Galactic continuum radiation \citep[see Fig.~7 in][]{Staveley-Smith+2016}.

In this regard, the SMGPS provides an ideal complement: it fully encompasses the incompleteness strip of HIZOA. We therefore embarked on a search of extragalactic \HI\ signals hidden in the SMGPS. Given the significant improvement in  sensitivity and resolution of the
\SMGPS, it has high potential to shed new light onto currently hidden LSS. The survey is, for instance, sensitive to $M^*_{\rm HI}$ galaxies in the volume out to 
recession velocity $v = 25,000$\,\kms, reaching $\log M_{\rm HI} \ga 8.0$\,\Msun\ at the GA distance (compared to 9.5\,\Msun\ in HIZOA). 

In the following, we give a brief summary of the \HI\ data reduction and the resulting data products (a more detailed description will be provided in Rajohnson et al., in prep.), followed by a few highlights of initial science results.

\subsection{Data reduction and source finding}
\label{reds}

\begin{figure}
		\centering
		\includegraphics[width=0.9\linewidth]{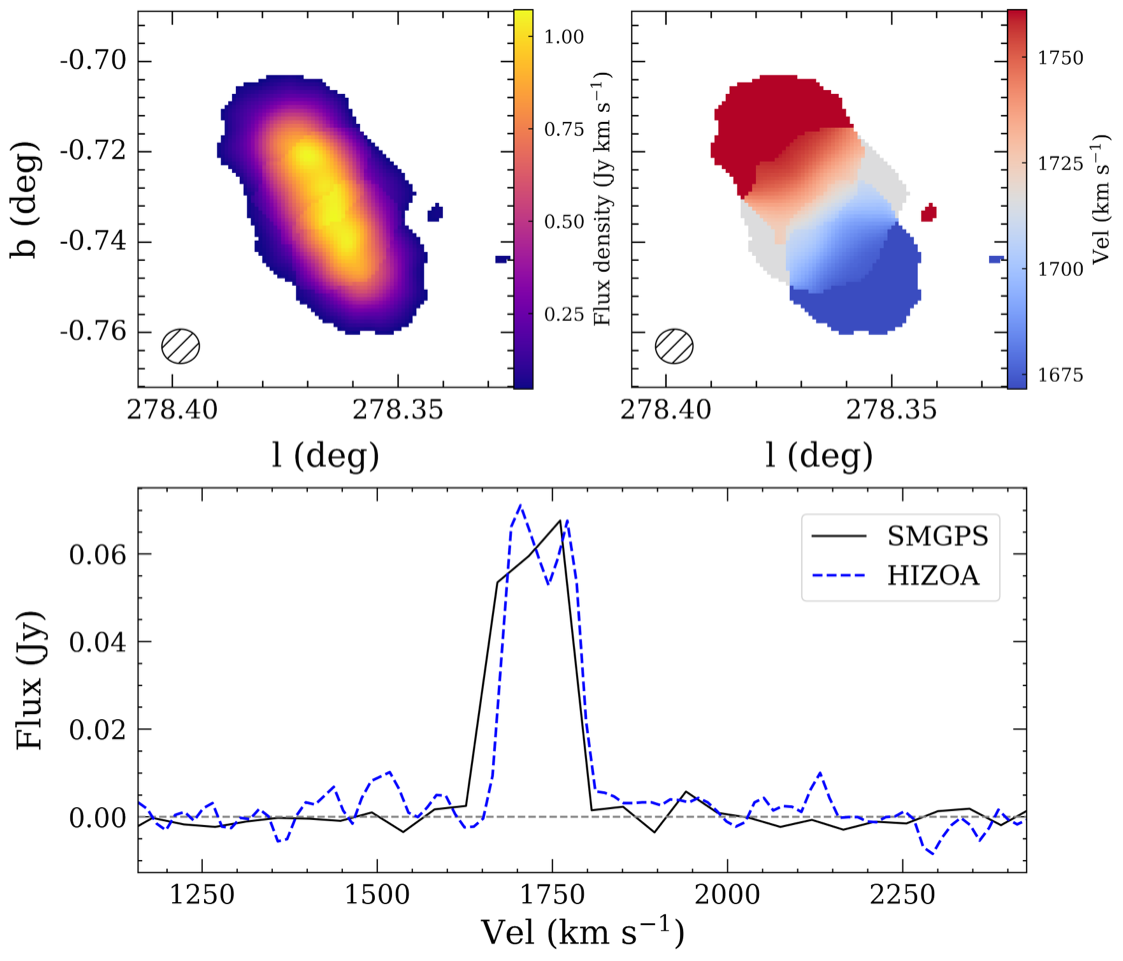}
		\centering
		\caption{SMGPS observations of HIZOA\,J0947$-$54. The \emph{top} panels display the moment maps (total \HI\ intensity; rotation) with the FWHM restoring beam size in the lower left corner. The colour bars indicate the flux density and velocity scales. The \emph{bottom} panel shows the global \HI\ profile from SMGPS vs that from HIZOA.}
		\label{fig:hizoa_detection}
	\end{figure}

 Both parallel-hand polarizations
 of the raw (visibility) data were downloaded from the SARAO Archive for the frequency range 1308--1430\,MHz. The lower frequency boundary was set to select the predominantly RFI-free part of the spectrum which coincides closely with the volume relevant to (residual) bulkflows ($z < 0.08$). The correlator mode used, with 4096 channels across the full L band, gives a frequency resolution of 209.0 kHz, corresponding to a channel width of 44.1\,\kms\ at redshift $z = 0$.

We used the container-based CARACal pipeline \citep{Jozsa+2020,Makhathini2018} which includes various open-source radio interferometry software packages such as CASA, Cubical, WSClean and Montage. The reduction steps included RFI flagging, cross-calibration, self-calibration and continuum subtraction. For the \HI\ imaging of the individual fields, we worked with $\mbox{robust}=0$ weighting, a pixel size of $3\arcsec$ and further convolved the images with a circular Gaussian of FWHM 15\arcsec, leading to a final restoring beam
of FWHM $\sim 30\arcsec \times 27\arcsec \; (\pm2\arcsec)$. This angular resolution is sufficient for our science goals, i.e.\ sufficient for nearby quite extended galaxies as well as more distant (higher column-density) galaxies. 

For the science analysis, 22 contiguous, primary-beam corrected MeerKAT pointings were combined into \HI\ mosaics, each covering an area along the GP of $4\fdg0 \times 3\fdg5$, spaced by $\Delta l = 3\degr$.  The mosaics have a median RMS noise level of $\sim 0.45$\,m\Jb, rising from $\sim$0.3 m\Jb\ at latitudes
far from the Galactic Centre to $\sim$0.6 m\Jb\ towards it. The increase is primarily due to the Galactic continuum emission (see Fig.~\ref{fig:svy_area}). Some bright, extended and diffuse continuum sources result in patches of increased RMS values which can locally affect the \HI\ detectability.

The automated source finding and parameterization software SoFiA-2 \citep{Westmeier+2021} was used for the identification of galaxy candidates, with parameter settings optimised to the slight noise variations at these low latitudes, followed by a visual verification process based on moment maps (total intensity and velocity), the global \HI\ spectrum and its S/N.

For a comparison of the data quality with HIZOA we used galaxies in common with it (hardly any other counterparts exist this deep in the ZoA). As an example we show in Fig.~\ref{fig:hizoa_detection} the \HI\ data product for HIZOA\,J0947$-$54. The top panel demonstrates the exquisite detail provided by the moment maps (total intensity and rotation) while the bottom panel shows the excellent agreement of the \HI\ profile (and flux) between the SMGPS and HIZOA.
 
A more quantitative assessment of the integrated flux, systemic velocity and linewidths between SMGPS and  HIZOA sources found no deviation from linearity for any of the \HI\ parameters, with errors well within expectations.

\subsection {Extragalactic science}

Here we give a brief synopsis of some of the exciting new insights gained so far from our investigations of \HI\ signals of extragalactic large-scale structures hidden in the SMGPS.

\subsubsection{The Vela Supercluster (VSCL)}
\label{VSCL}

\begin{figure}
	\centering
     \includegraphics[width=0.48\textwidth]{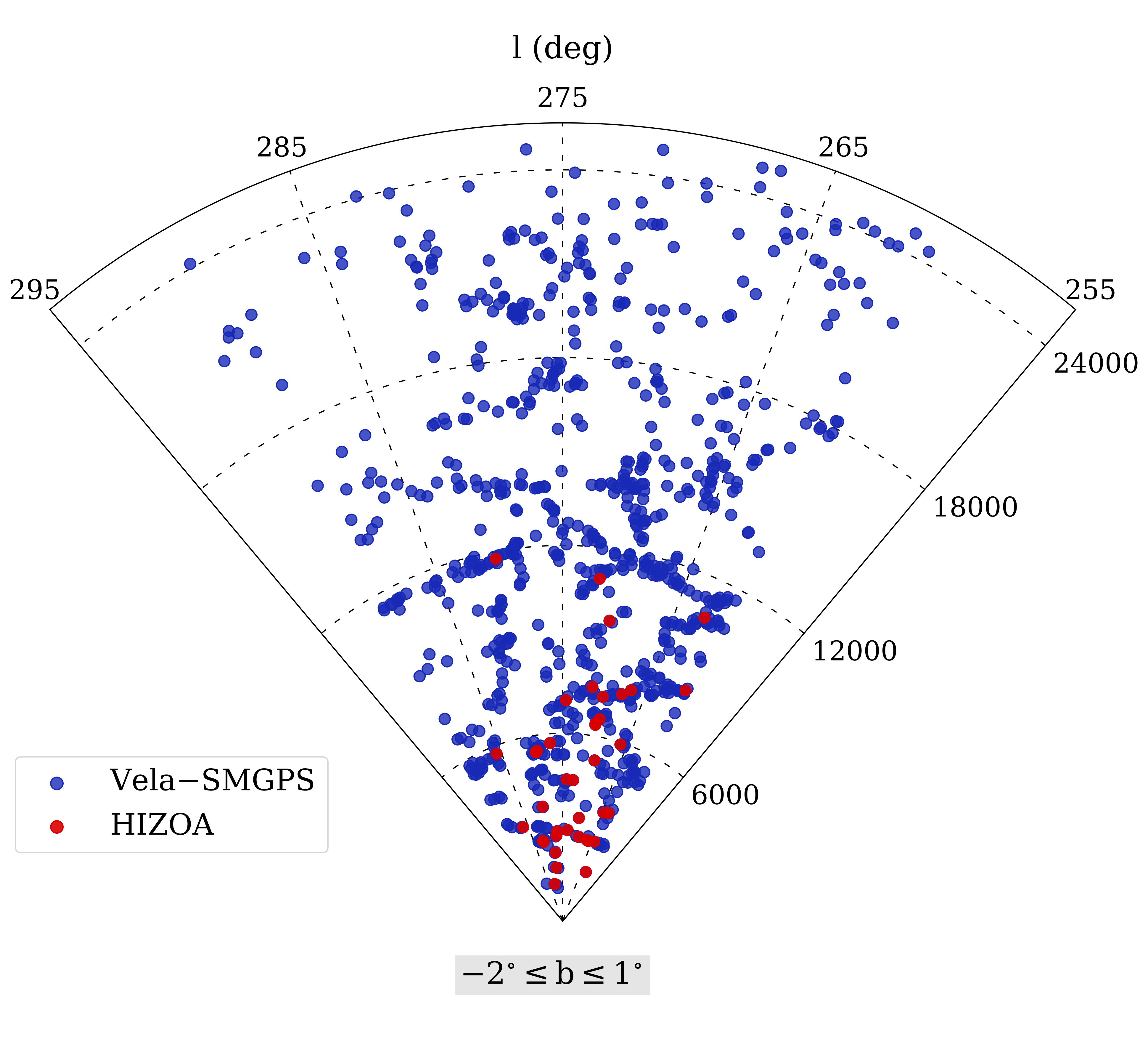}
	\centering
	\caption{Redshift wedge out to $v
    < 25,000$\,\kms\ of the galaxies detected in the VSCL-SMGPS (blue dots) for the latitude range $-2^{\circ} \leq b \leq 1^{\circ}$ (red dots mark HIZOA detections).}
	\label{fig:vscl_wedge}
	\end{figure}

Optical spectroscopy of galaxies at the outer Galactic latitude border of the ZoA suggested the existence of a massive and very extended supercluster in Vela that seems to consists of two --- possibly merging --- walls \citep[at $v \sim$\,18,500 and  21,000\,\kms;][]{Kraan-Korteweg+2017}. Their location coincides with the general direction and distance of the residual bulkflow \citep[e.g.][]{Springob+2016}. However, little is known about the
supercluster's structure at latitudes below $|b| \la 7\degr$.
We therefore extracted \HI\ data from the SMGPS strip for the longitude range $260\degr$--$290\degr$. It comprises 10 slightly overlapping \HI\ mosaics constructed from 157 individual MeerKAT pointings. With an average RMS = 0.39\,m\Jb, 843 \HI\ detections were identified. In agreement with expectations, typical spiral galaxies were identified out to the volume edge of $v = 25,000$\,\kms.  Only 39 had known counterparts in HIZOA.
 
The redshift wedge of the \HI\ detections displayed in Fig.~\ref{fig:vscl_wedge} reveals various new structures. With regard to the  primary goal of this investigation we can report that the number of detections at the VSCL distance is indeed enhanced compared to simulations, with the highest density  enhancements coinciding with the location of the VSCL walls. The appearance of the walls bears a striking resemblance to what was found on either side of the ZoA in the VSCL discovery paper \citep[see Fig.~3 of][]{Kraan-Korteweg+2017}. Additional details will be provided elsewhere.

\subsubsection{The Great Attractor (GA)}
\label{GA}

The discovery in 1987 of a then mysterious great attractor was based on some of the earliest whole-sky peculiar velocity field data \citep{Dressler+1987}. Subsequent dedicated galaxy surveys close to the ZoA demonstrated that the large-scale inflow to the GA is due to a big wall-like supercluster centered on the Norma cluster \cite[at $l,b,v\sim325\degr,-7\degr,4900$\,\kms;][]{Woudt+2004,Kraan-Korteweg+1996}. The scarcity of data in the ZoA remained a conundrum, 
however, in particular the inner ZoA ($|b| \la 1\fdg5$).

\begin{figure} 
    	\centering 
        \includegraphics[width=0.48\textwidth]{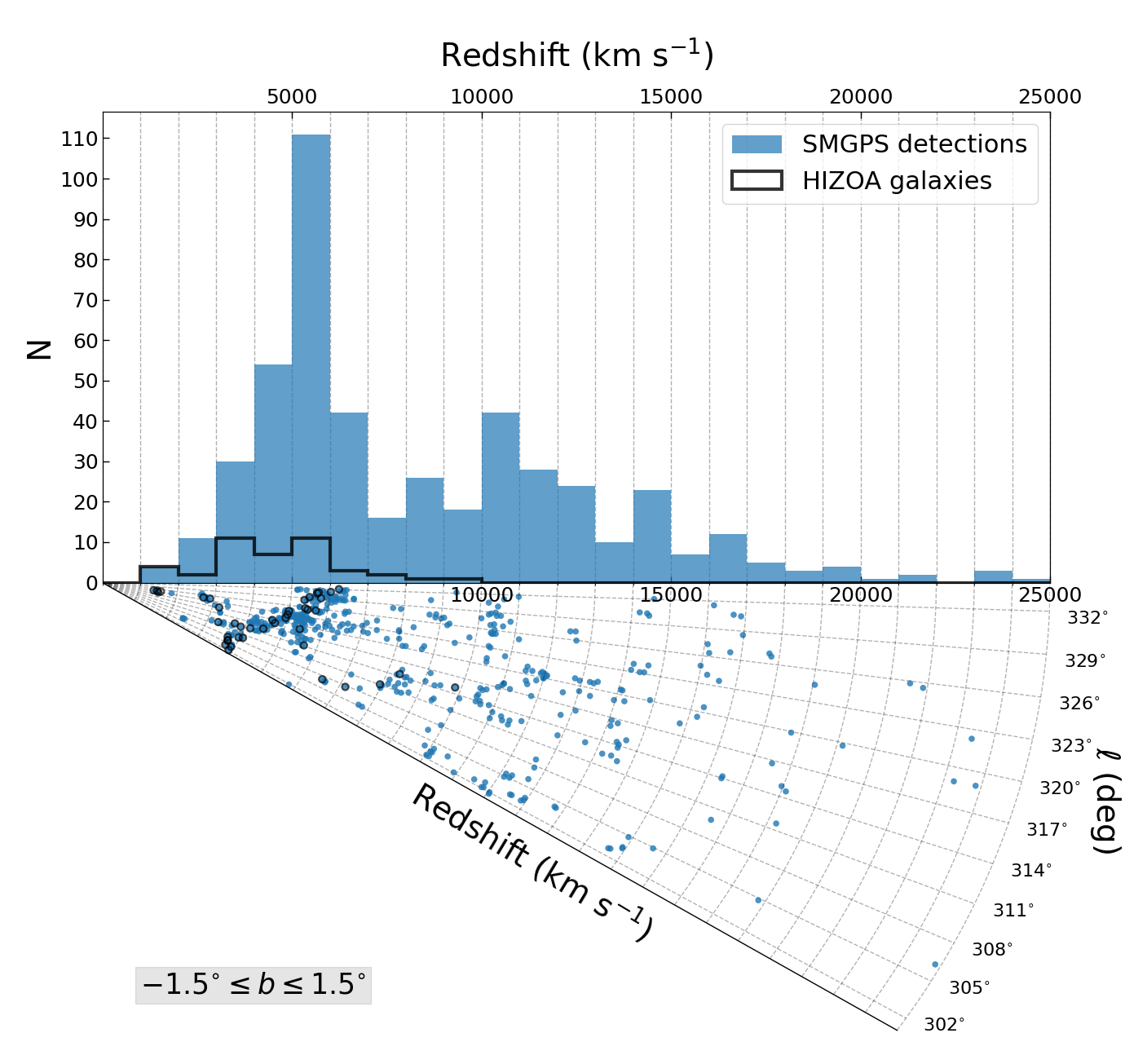} 
    	\centering
    	\caption{Velocity histogram and corresponding redshift wedge out to 25,000\,\kms\ of the \HI\ detections in the GA-SMGPS search. Blue dots represent SMGPS detections and black contours mark HIZOA galaxies.}
    \label{fig:GA_hist-wedge}
\end{figure}

It was therefore an obvious endeavour to assess the SMGPS \HI\ data for signatures of the GA Wall extension across the ZoA. The GA-SMGPS survey encloses the longitude range $302\degr < l < 332\degr$. Although the GA is relatively nearby, \HI\ signals were searched for out to the full volume bounded by  $v  = 25,000$\,\kms. The resulting mean RMS sensitivity in the mosaics (0.47\,m\Jb) is slightly higher than that of SMGPS for the VSCL, in line with expectations of the increasing background continuum.

In total, 477 galaxy candidates were identified, of which 42 had counterparts in HIZOA. The spatial distribution of the uncovered galaxies is quite distinct from that of the VSCL-SMGPS region. To get a better insight into the uncovered LSS, Fig.~\ref{fig:GA_hist-wedge} displays the redshift distribution of the \HI\ detections as a histogram and wedge diagram. A brief inspection of these plots immediately reveals the striking peak around the GA velocity range (3500--6500\,\kms). Being offset by 7\degr\ from the Norma cluster --- the core of the GA --- the prominence of the GA Wall is astounding: a major fraction of the \HI\ detections ($N = 214$; 45\%) lie within the narrow GA velocity range of $\Delta v = 3000$\,\kms. A comparison to simulations substantiates that the observed number density around the GA distance is a factor of 3--5 higher than that in simulations, inferring the GA Wall to be very dense. Further details are given in \citet{Steyn+2023}.

\subsubsection{The Local Void (LV)}
\label{LV}

Another enigma of the ZoA is the Local Void \citep{Tully2008, Kraan-Korteweg+2008, Tully+2019}. Being one of the largest voids in the local Universe, its proximity makes it an ideal structure for in-depth studies of the void properties {\em and} the few galaxies that live there. From the point of view of cosmic flows, underdensities are as important as overdensities because of the opposite forces they exert (repulsion vs attraction). Moreover, voids provide us with a unique opportunity to study galaxies at an early stage in their evolution. 

The LV-SMGPS survey region was optimised to contain the full extent of the LV, inclusive of its borders in velocity space. With an on-sky extent of the LV of $\sim 60\degr$, we focused on the longitude strip $55\degr > l > 329\degr$ ($\Delta l = 85\degr$); in view of the surmised outer edge of the LV ($\sim 6000$\,\kms), we restricted the velocity range to $v < 7500$\,\kms. We also used larger $uv$ tapers (30\arcsec\ to 120\arcsec)
when reducing the LV data set,
in addition to the standard 15\arcsec\ taper, to facilitate the detection of nearby extended low \HI\ column density sources. The resulting mosaics were inspected visually for \HI\ signals, and with the {SoFiA} algorithm. 

\begin{figure} 
     \centering
    \includegraphics[width=0.48\textwidth]{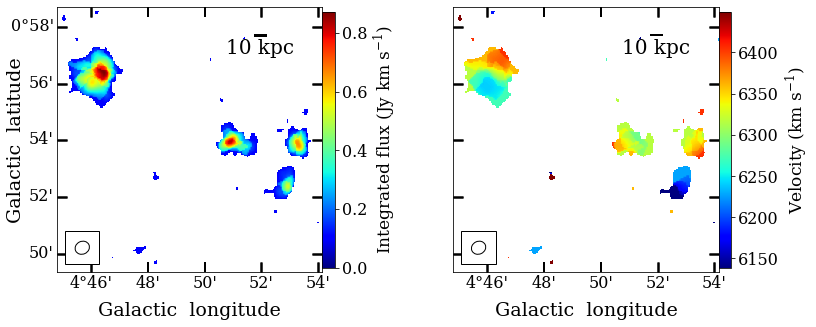} 
     \centering
   \caption{\HI\ intensity (moment-0) and velocity (moment-1) maps of the SMGPS data of HIZOA\,J1753$-$24A, which resolves the HIZOA source into a small group of 4 galaxies.}
    \label{fig:LV_group}
\end{figure}

In total, 291 galaxies were detected. Of the 39 HIZOA galaxies within the explored volume, all were recovered.  We found a considerable number of galaxies to have small companions, or to form part of small groups. As an example Fig.~\ref{fig:LV_group} displays the SMGPS data of HIZOA\,J1753$-$24A which appears to consist of four small galaxies in an approximately linear configuration. Some other groups in or at the edge of the Void also seem to lie along intra-void filaments --- the underdense environment possibly improving their chances of individual survival rather than being merged into one larger galaxy \citep{Kreckel+2012}.

The huge extent of the LV is the most notable LSS in the LV-SMGPS. The diameter for the LV determined in the inner ZoA is 58\,Mpc, consistent with determinations obtained beyond the ZoA \citep[e.g.][]{Tully+2019}. Only a few galaxies ($N=17$) were found to reside deep within the LV itself, all of them of low \HI\ mass. Further details, including population study as a function of environment, are given in \citet{Kurapati+2023}.

\section{Summary and Conclusions}
\label{sec:conclusions}

We present the largest, most sensitive and highest angular resolution $\sim$1 GHz survey of the Galactic Plane to date, the SARAO MeerKAT Galactic Plane Survey (SMGPS). The SMGPS covers almost half of the Galactic Plane in two discrete blocks of 251\degr\ $\le l \le$ 358\degr\ and 2\degr\ $\le l \le$ 61\degr, each covering Galactic latitudes  $|b| \le 1\fdg5$. The angular resolution of SMGPS is 8\arcsec\ and the RMS sensitivity is $\sim$10--20 \muJb\ in regions unaffected by confusion from nearby bright sources (over an order of magnitude deeper than previous Galactic Plane surveys). The new parameter space explored by SMGPS offers tremendous scope for the detection of new radio populations within and without the Milky Way.

In this paper we describe the first data products from the SMGPS comprising raw $uv$ data and reduced $3\degr \times 
3\degr$ data cubes and zeroth moment images. We explore the limitations in the reduced data and provide a thorough assessment of the survey calibration. A selection of scientific highlights are presented to illustrate the power of the SMGPS for future discoveries. We summarise these highlights below:

\begin{itemize}

\item The much deeper sensitivity and improved image fidelity of SMGPS have revealed a new population of non-thermal radio filaments in the Galactic Plane, which previously had only been known
near the Galactic Centre.

\item A number of new candidate supernova remnants (SNRs) have been identified using a radio/mid-IR matching procedure, demonstrating the sensitivity of SMGPS to low surface-brightness and extended objects. SMGPS also reveals previously unknown complex structure within known SNRs, identifying new shells and filaments.   

\item A number of new pulsar wind nebulae are discovered
around a sample of young energetic pulsars, illustrating the potential of SMGPS for future discoveries.

\item Candidate planetary nebulae (PNe) were identified from their distinct radio morphology. This method is not affected by extinction and thus can uniquely identify PNe along the Galactic mid-Plane that are missed by H$\alpha$ surveys --- indeed our initial census doubles the number of candidate PNe in the survey area.

\item We show the ability of combined SMGPS and mid-IR observations to classify rare Luminous Blue Variable candidates, which may be the progenitors of supernovae. In the case of the LBV candidate Gal 026.47+0.02, SMGPS reveals that the ionised gas component is more extended and has a greater mass than previously thought.

\item Via a Bayesian cross-matching technique we have identified approximately 800 stars with likely 1.3 GHz radio emission, across the entire HR diagram. In SMGPS tiles with polarization calibration we additionally identify a sample of highly circularly polarised stars, of which one is $\sim$100\% polarised.

\item Using the high sensitivity of the SMGPS to extended low surface brightness emission we show that many large radio-quiet \HII\ region candidates from WISE have no detectable radio continuum emission and so are unlikely to be true \HII\ regions. We speculate that these mid-IR bright objects are instead intermediate-mass star forming regions. 

\item Reprocessing of the raw SMGPS visibility data with a custom pipeline to detect \HI\  emission from background galaxies reveals several hundred \HI\ galaxies, most of which were previously undetected in HIZOA. The detections reveal new structures within the Vela Supercluster, a much greater density of galaxies along the Great Attractor Wall, and the presence of small galaxy groups in the Local Void.

\end{itemize}

The SMGPS dataset is both rich and complex, revealing structures across the Milky Way that have many different emission mechanisms and which cover a large range in angular scales. SMGPS offers tremendous potential for new scientific discoveries and is an ideal pathfinder survey to guide future exploration.

\section*{Acknowledgements}

The MeerKAT telescope is operated by the South African Radio Astronomy Observatory, which is a facility of the National Research Foundation, an agency of the Department of Science and Innovation.
The National Radio Astronomy Observatory is a facility of the National Science Foundation operated under cooperative agreement by Associated Universities, Inc.
The Centre for Astrophysics Research at the University of Hertfordshire kindly provided access to their HPC facilities for data processing and storage. MAT and GMW gratefully acknowledge the support of the Science \& Technology Facilities Council through grant awards ST/R000905/1 and ST/W00125X/1. 
LDA is supported by NSF award AST2307176.
DAHB acknowledges research support by the National Research Foundation.
DE is supported by an SAAO Prize PhD Scholarship.
This work was supported in part by the Italian Ministry of Foreign Affairs and International Cooperation, grant number ZA23GR03.
This research would not have been possible without NASA's Astrophysics Data System Bibliographic Services; the SIMBAD database  operated at CDS, Strasbourg, France; and the Astropy community-developed core Python package.
We acknowledge the Einstein@Home project for providing the position of PSR J1358$-$6025.
We thank Dave Green for bringing to our attention the Roberts et al. (1999) reference.

\section*{Data Availability}

All SMGPS DR1 data products are available through the DOI 
\url{https://doi.org/10.48479/3wfd-e270}. When using DR1 products, this paper should be cited, and the MeerKAT telescope acknowledgement included. The raw visibilities are hosted on the SARAO Data Archive (\url{https://archive.sarao.ac.za/}) under project code SSV-20180721-FC-01.



\typeout{} 
\bibliographystyle{mnras}
\bibliography{smgps-ffix}

\begin{thebibliography}{}
\makeatletter
\relax
\def\mn@urlcharsother{\let\do\@makeother \do\$\do\&\do\#\do\^\do\_\do\%\do\~}
\def\mn@doi{\begingroup\mn@urlcharsother \@ifnextchar [ {\mn@doi@}
  {\mn@doi@[]}}
\def\mn@doi@[#1]#2{\def\@tempa{#1}\ifx\@tempa\@empty \href
  {http://dx.doi.org/#2} {doi:#2}\else \href {http://dx.doi.org/#2} {#1}\fi
  \endgroup}
\def\mn@eprint#1#2{\mn@eprint@#1:#2::\@nil}
\def\mn@eprint@arXiv#1{\href {http://arxiv.org/abs/#1} {{\tt arXiv:#1}}}
\def\mn@eprint@dblp#1{\href {http://dblp.uni-trier.de/rec/bibtex/#1.xml}
  {dblp:#1}}
\def\mn@eprint@#1:#2:#3:#4\@nil{\def\@tempa {#1}\def\@tempb {#2}\def\@tempc
  {#3}\ifx \@tempc \@empty \let \@tempc \@tempb \let \@tempb \@tempa \fi \ifx
  \@tempb \@empty \def\@tempb {arXiv}\fi \@ifundefined
  {mn@eprint@\@tempb}{\@tempb:\@tempc}{\expandafter \expandafter \csname
  mn@eprint@\@tempb\endcsname \expandafter{\@tempc}}}

\bibitem[\protect\citeauthoryear{{Alves}, {Davies}, {Dickinson}, {Calabretta},
  {Davis}  \& {Staveley-Smith}}{{Alves} et~al.}{2012}]{Alves+2012}
{Alves} M. I.~R.,  {Davies} R.~D.,  {Dickinson} C.,  {Calabretta} M.,  {Davis}
  R.,   {Staveley-Smith} L.,  2012, \mn@doi [\mnras]
  {10.1111/j.1365-2966.2012.20796.x}, \href
  {https://ui.adsabs.harvard.edu/abs/2012MNRAS.422.2429A} {422, 2429}

\bibitem[\protect\citeauthoryear{{Anderson}, {Bania}, {Balser}  \&
  {Rood}}{{Anderson} et~al.}{2011}]{Anderson+2011}
{Anderson} L.~D.,  {Bania} T.~M.,  {Balser} D.~S.,   {Rood} R.~T.,  2011,
  \mn@doi [\apjs] {10.1088/0067-0049/194/2/32}, \href
  {https://ui.adsabs.harvard.edu/abs/2011ApJS..194...32A} {194, 32}

\bibitem[\protect\citeauthoryear{{Anderson}, {Bania}, {Balser}, {Cunningham},
  {Wenger}, {Johnstone}  \& {Armentrout}}{{Anderson}
  et~al.}{2014}]{Anderson+2014}
{Anderson} L.~D.,  {Bania} T.~M.,  {Balser} D.~S.,  {Cunningham} V.,  {Wenger}
  T.~V.,  {Johnstone} B.~M.,   {Armentrout} W.~P.,  2014, \mn@doi [\apjs]
  {10.1088/0067-0049/212/1/1}, \href
  {http://adsabs.harvard.edu/abs/2014ApJS..212....1A} {212, 1}

\bibitem[\protect\citeauthoryear{{Anderson} et~al.,}{{Anderson}
  et~al.}{2017}]{Anderson+2017}
{Anderson} L.~D.,  et~al., 2017, \mn@doi [\aap] {10.1051/0004-6361/201731019},
  \href {http://adsabs.harvard.edu/abs/2017A%26A...605A..58A} {605, A58}

\bibitem[\protect\citeauthoryear{{Andr{\'e}}, {Di Francesco}, {Ward-Thompson},
  {Inutsuka}, {Pudritz}  \& {Pineda}}{{Andr{\'e}} et~al.}{2014}]{Andre+2014}
{Andr{\'e}} P.,  {Di Francesco} J.,  {Ward-Thompson} D.,  {Inutsuka} S.~I.,
  {Pudritz} R.~E.,   {Pineda} J.~E.,  2014, in {Beuther} H.,  {Klessen} R.~S.,
  {Dullemond} C.~P.,   {Henning} T.,  eds, Protostars and Planets VI. p.~27
  (\mn@eprint {arXiv} {1312.6232}),
  \mn@doi{10.2458/azu\_uapress\_9780816531240-ch002}

\bibitem[\protect\citeauthoryear{{Anglada}, {Villuendas}, {Estalella},
  {Beltr{\'a}n}, {Rodr{\'\i}guez}, {Torrelles}  \& {Curiel}}{{Anglada}
  et~al.}{1998}]{Anglada+1998}
{Anglada} G.,  {Villuendas} E.,  {Estalella} R.,  {Beltr{\'a}n} M.~T.,
  {Rodr{\'\i}guez} L.~F.,  {Torrelles} J.~M.,   {Curiel} S.,  1998, \mn@doi
  [\aj] {10.1086/300637}, \href
  {https://ui.adsabs.harvard.edu/abs/1998AJ....116.2953A} {116, 2953}

\bibitem[\protect\citeauthoryear{{Arce}, {Borkin}, {Goodman}, {Pineda}  \&
  {Beaumont}}{{Arce} et~al.}{2011}]{Arce+2011}
{Arce} H.~G.,  {Borkin} M.~A.,  {Goodman} A.~A.,  {Pineda} J.~E.,   {Beaumont}
  C.~N.,  2011, \mn@doi [\apj] {10.1088/0004-637X/742/2/105}, \href
  {https://ui.adsabs.harvard.edu/abs/2011ApJ...742..105A} {742, 105}

\bibitem[\protect\citeauthoryear{{Armentrout}, {Anderson}, {Wenger}, {Balser}
  \& {Bania}}{{Armentrout} et~al.}{2021}]{Armentrout+2021}
{Armentrout} W.~P.,  {Anderson} L.~D.,  {Wenger} T.~V.,  {Balser} D.~S.,
  {Bania} T.~M.,  2021, \mn@doi [\apjs] {10.3847/1538-4365/abd5c0}, \href
  {https://ui.adsabs.harvard.edu/abs/2021ApJS..253...23A} {253, 23}

\bibitem[\protect\citeauthoryear{{Bailer-Jones}, {Rybizki}, {Fouesneau},
  {Demleitner}  \& {Andrae}}{{Bailer-Jones} et~al.}{2021}]{bailer2021}
{Bailer-Jones} C.~A.~L.,  {Rybizki} J.,  {Fouesneau} M.,  {Demleitner} M.,
  {Andrae} R.,  2021, \mn@doi [\aj] {10.3847/1538-3881/abd806}, \href
  {https://ui.adsabs.harvard.edu/abs/2021AJ....161..147B} {161, 147}

\bibitem[\protect\citeauthoryear{{Ball} et~al.,}{{Ball}
  et~al.}{2023}]{Ball+2023}
{Ball} B.~D.,  et~al., 2023, \mn@doi [\mnras] {10.1093/mnras/stad1953}, \href
  {https://ui.adsabs.harvard.edu/abs/2023MNRAS.524.1396B} {524, 1396}

\bibitem[\protect\citeauthoryear{{Bamba}, {Watanabe}, {Mori}, {Shibata},
  {Terada}, {Sano}  \& {Filipovi{\'c}}}{{Bamba} et~al.}{2020}]{Bamba+2020}
{Bamba} A.,  {Watanabe} E.,  {Mori} K.,  {Shibata} S.,  {Terada} Y.,  {Sano}
  H.,   {Filipovi{\'c}} M.~D.,  2020, \apss, 365, 178

\bibitem[\protect\citeauthoryear{{Barkov} \& {Lyutikov}}{{Barkov} \&
  {Lyutikov}}{2019}]{BarkovL2019}
{Barkov} M.~V.,  {Lyutikov} M.,  2019, \mn@doi [\mnras]
  {10.1093/mnrasl/slz124}, \href
  {https://ui.adsabs.harvard.edu/abs/2019MNRAS.489L..28B} {489, L28}

\bibitem[\protect\citeauthoryear{{Benjamin} et~al.,}{{Benjamin}
  et~al.}{2003}]{Benjamin+2003}
{Benjamin} R.~A.,  et~al., 2003, \mn@doi [\pasp] {10.1086/376696}, \href
  {http://adsabs.harvard.edu/abs/2003PASP..115..953B} {115, 953}

\bibitem[\protect\citeauthoryear{{Beuther} et~al.,}{{Beuther}
  et~al.}{2016}]{Beuther+2016}
{Beuther} H.,  et~al., 2016, \mn@doi [\aap] {10.1051/0004-6361/201629143},
  \href {https://ui.adsabs.harvard.edu/abs/2016A&A...595A..32B} {595, A32}

\bibitem[\protect\citeauthoryear{{Bianchi} et~al.,}{{Bianchi}
  et~al.}{2017}]{Bianchi+2017}
{Bianchi} S.,  et~al., 2017, \mn@doi [\aap] {10.1051/0004-6361/201629013},
  \href {https://ui.adsabs.harvard.edu/abs/2017A&A...597A.130B} {597, A130}

\bibitem[\protect\citeauthoryear{{Bihr} et~al.,}{{Bihr}
  et~al.}{2016}]{Bihr+2016}
{Bihr} S.,  et~al., 2016, \mn@doi [\aap] {10.1051/0004-6361/201527697}, \href
  {https://ui.adsabs.harvard.edu/abs/2016A&A...588A..97B} {588, A97}

\bibitem[\protect\citeauthoryear{{Brogan}, {Devine}, {Lazio}, {Kassim}, {Tam},
  {Brisken}, {Dyer}  \& {Roberts}}{{Brogan} et~al.}{2004}]{Brogan+2004}
{Brogan} C.~L.,  {Devine} K.~E.,  {Lazio} T.~J.,  {Kassim} N.~E.,  {Tam} C.~R.,
   {Brisken} W.~F.,  {Dyer} K.~K.,   {Roberts} M.~S.~E.,  2004, \mn@doi [\aj]
  {10.1086/379856}, \href
  {https://ui.adsabs.harvard.edu/abs/2004AJ....127..355B} {127, 355}

\bibitem[\protect\citeauthoryear{{Brunthaler} et~al.,}{{Brunthaler}
  et~al.}{2021}]{Brunthaler+2021}
{Brunthaler} A.,  et~al., 2021, \mn@doi [\aap] {10.1051/0004-6361/202039856},
  \href {https://ui.adsabs.harvard.edu/abs/2021A&A...651A..85B} {651, A85}

\bibitem[\protect\citeauthoryear{{Budav{\'a}ri} \& {Szalay}}{{Budav{\'a}ri} \&
  {Szalay}}{2008}]{budavari2008}
{Budav{\'a}ri} T.,  {Szalay} A.~S.,  2008, \mn@doi [\apj] {10.1086/587156},
  \href {https://ui.adsabs.harvard.edu/abs/2008ApJ...679..301B} {679, 301}

\bibitem[\protect\citeauthoryear{{Calabretta}, {Staveley-Smith}  \&
  {Barnes}}{{Calabretta} et~al.}{2014}]{CalabrettaSB2014}
{Calabretta} M.~R.,  {Staveley-Smith} L.,   {Barnes} D.~G.,  2014, \mn@doi
  [\pasa] {10.1017/pasa.2013.36}, \href
  {https://ui.adsabs.harvard.edu/abs/2014PASA...31....7C} {31, e007}

\bibitem[\protect\citeauthoryear{{Camilo}, {Lorimer}, {Bhat}, {Gotthelf},
  {Halpern}, {Wang}, {Lu}  \& {Mirabal}}{{Camilo} et~al.}{2002}]{Camilo+2002}
{Camilo} F.,  {Lorimer} D.~R.,  {Bhat} N.~D.~R.,  {Gotthelf} E.~V.,  {Halpern}
  J.~P.,  {Wang} Q.~D.,  {Lu} F.~J.,   {Mirabal} N.,  2002, \mn@doi [\apjl]
  {10.1086/342351}, \href
  {https://ui.adsabs.harvard.edu/abs/2002ApJ...574L..71C} {574, L71}

\bibitem[\protect\citeauthoryear{{Camilo} et~al.,}{{Camilo}
  et~al.}{2018}]{Camilo+2018}
{Camilo} F.,  et~al., 2018, \mn@doi [\apj] {10.3847/1538-4357/aab35a}, \href
  {https://ui.adsabs.harvard.edu/abs/2018ApJ...856..180C} {856, 180}

\bibitem[\protect\citeauthoryear{{Carey} et~al.,}{{Carey}
  et~al.}{2009}]{Carey+2009}
{Carey} S.~J.,  et~al., 2009, \mn@doi [\pasp] {10.1086/596581}, \href
  {http://adsabs.harvard.edu/abs/2009PASP..121...76C} {121, 76}

\bibitem[\protect\citeauthoryear{{Carpenter}, {Snell}  \&
  {Schloerb}}{{Carpenter} et~al.}{1990}]{Carpenter+1990}
{Carpenter} J.~M.,  {Snell} R.~L.,   {Schloerb} F.~P.,  1990, \mn@doi [\apj]
  {10.1086/169251}, \href
  {https://ui.adsabs.harvard.edu/abs/1990ApJ...362..147C} {362, 147}

\bibitem[\protect\citeauthoryear{{Charlot} et~al.,}{{Charlot}
  et~al.}{2020}]{Charlot+2020}
{Charlot} P.,  et~al., 2020, \mn@doi [\aap] {10.1051/0004-6361/202038368},
  \href {https://ui.adsabs.harvard.edu/abs/2020A&A...644A.159C} {644, A159}

\bibitem[\protect\citeauthoryear{{Chibueze} et~al.,}{{Chibueze}
  et~al.}{2013}]{Chibueze+2013}
{Chibueze} J.~O.,  et~al., 2013, \mn@doi [\apj] {10.1088/0004-637X/762/1/17},
  \href {https://ui.adsabs.harvard.edu/abs/2013ApJ...762...17C} {762, 17}

\bibitem[\protect\citeauthoryear{{Chiotellis}, {Boumis}  \&
  {Spetsieri}}{{Chiotellis} et~al.}{2021}]{Chiotellis+2021}
{Chiotellis} A.,  {Boumis} P.,   {Spetsieri} Z.~T.,  2021, \mn@doi [\mnras]
  {10.1093/mnras/staa3573}, \href
  {https://ui.adsabs.harvard.edu/abs/2021MNRAS.502..176C} {502, 176}

\bibitem[\protect\citeauthoryear{{Churchwell} et~al.,}{{Churchwell}
  et~al.}{2009}]{Churchwell+2009}
{Churchwell} E.,  et~al., 2009, \mn@doi [\pasp] {10.1086/597811}, \href
  {http://adsabs.harvard.edu/abs/2009PASP..121..213C} {121, 213}

\bibitem[\protect\citeauthoryear{{Clark}, {Egan}, {Crowther}, {Mizuno},
  {Larionov}  \& {Arkharov}}{{Clark} et~al.}{2003}]{Clark+2003}
{Clark} J.~S.,  {Egan} M.~P.,  {Crowther} P.~A.,  {Mizuno} D.~R.,  {Larionov}
  V.~M.,   {Arkharov} A.,  2003, \mn@doi [\aap] {10.1051/0004-6361:20031372},
  \href {https://ui.adsabs.harvard.edu/abs/2003A&A...412..185C} {412, 185}

\bibitem[\protect\citeauthoryear{{Clark} et~al.,}{{Clark}
  et~al.}{2016}]{ClarkPW2016}
{Clark} C.~J.,  et~al., 2016, \mn@doi [\apjl] {10.3847/2041-8205/832/1/L15},
  \href {https://ui.adsabs.harvard.edu/abs/2016ApJ...832L..15C} {832, L15}

\bibitem[\protect\citeauthoryear{{Cohen} \& {Green}}{{Cohen} \&
  {Green}}{2001}]{CohenG2001}
{Cohen} M.,  {Green} A.~J.,  2001, \mn@doi [\mnras]
  {10.1046/j.1365-8711.2001.04421.x}, \href
  {https://ui.adsabs.harvard.edu/abs/2001MNRAS.325..531C} {325, 531}

\bibitem[\protect\citeauthoryear{{Condon}}{{Condon}}{1984}]{Condon+1984}
{Condon} J.~J.,  1984, \mn@doi [\apj] {10.1086/162705}, \href
  {https://ui.adsabs.harvard.edu/abs/1984ApJ...287..461C} {287, 461}

\bibitem[\protect\citeauthoryear{{Cotton}}{{Cotton}}{2008}]{Cotton2008}
{Cotton} W.~D.,  2008, \mn@doi [PASP] {10.1086/586754}, 120, 439

\bibitem[\protect\citeauthoryear{Cotton}{Cotton}{2019}]{Cotton2019}
Cotton W.~D.,  2019, Obit Development Memo Series, 63, 1

\bibitem[\protect\citeauthoryear{Cotton et~al.,}{Cotton
  et~al.}{2020}]{Cotton+2020}
Cotton W.~D.,  et~al., 2020, \mn@doi [MNRAS] {10.1093/mnras/staa1240}, 495,
  1271

\bibitem[\protect\citeauthoryear{{Coughlin}, {Nixon}  \& {Ginsburg}}{{Coughlin}
  et~al.}{2021}]{CoughlinNG2021}
{Coughlin} E.~R.,  {Nixon} C.~J.,   {Ginsburg} A.,  2021, \mn@doi [\mnras]
  {10.1093/mnras/staa3771}, \href
  {https://ui.adsabs.harvard.edu/abs/2021MNRAS.501.1868C} {501, 1868}

\bibitem[\protect\citeauthoryear{{Das} \& {Chandra}}{{Das} \&
  {Chandra}}{2023}]{Das2023}
{Das} B.,  {Chandra} P.,  2023, \mn@doi [\apj] {10.3847/1538-4357/acf929},
  \href {https://ui.adsabs.harvard.edu/abs/2023ApJ...957...53D} {957, 53}

\bibitem[\protect\citeauthoryear{{Davies} et~al.,}{{Davies}
  et~al.}{2012}]{Davies+2012}
{Davies} B.,  et~al., 2012, \mn@doi [\mnras]
  {10.1111/j.1365-2966.2011.19736.x}, \href
  {https://ui.adsabs.harvard.edu/abs/2012MNRAS.419.1871D} {419, 1871}

\bibitem[\protect\citeauthoryear{{Dennett-Thorpe} \& {de
  Bruyn}}{{Dennett-Thorpe} \& {de Bruyn}}{2002}]{dennett-thorpe2002}
{Dennett-Thorpe} J.,  {de Bruyn} A.~G.,  2002, \mn@doi [\nat]
  {10.1038/415057a}, \href
  {https://ui.adsabs.harvard.edu/abs/2002Natur.415...57D} {415, 57}

\bibitem[\protect\citeauthoryear{{Djordjevic}, {Thompson}, {Urquhart}  \&
  {Forbrich}}{{Djordjevic} et~al.}{2019}]{Djordjevic+2019}
{Djordjevic} J.~O.,  {Thompson} M.~A.,  {Urquhart} J.~S.,   {Forbrich} J.,
  2019, \mn@doi [\mnras] {10.1093/mnras/stz1262}, \href
  {https://ui.adsabs.harvard.edu/abs/2019MNRAS.487.1057D} {487, 1057}

\bibitem[\protect\citeauthoryear{{Dokara} et~al.,}{{Dokara}
  et~al.}{2021}]{Dokara+2021}
{Dokara} R.,  et~al., 2021, \mn@doi [\aap] {10.1051/0004-6361/202039873}, \href
  {https://ui.adsabs.harvard.edu/abs/2021A&A...651A..86D} {651, A86}

\bibitem[\protect\citeauthoryear{{Dressler}, {Faber}, {Burstein}, {Davies},
  {Lynden-Bell}, {Terlevich}  \& {Wegner}}{{Dressler}
  et~al.}{1987}]{Dressler+1987}
{Dressler} A.,  {Faber} S.~M.,  {Burstein} D.,  {Davies} R.~L.,  {Lynden-Bell}
  D.,  {Terlevich} R.~J.,   {Wegner} G.,  1987, \mn@doi [\apjl]
  {10.1086/184827}, \href
  {https://ui.adsabs.harvard.edu/abs/1987ApJ...313L..37D} {313, L37}

\bibitem[\protect\citeauthoryear{{Driessen}, {Dom{\v{c}}ek}, {Vink}, {Hessels},
  {Arias}  \& {Gelfand}}{{Driessen} et~al.}{2018}]{Driessen+2018}
{Driessen} L.~N.,  {Dom{\v{c}}ek} V.,  {Vink} J.,  {Hessels} J. W.~T.,  {Arias}
  M.,   {Gelfand} J.~D.,  2018, \mn@doi [\apj] {10.3847/1538-4357/aac32e},
  \href {https://ui.adsabs.harvard.edu/abs/2018ApJ...860..133D} {860, 133}

\bibitem[\protect\citeauthoryear{{Driessen}, {Heald}, {Duchesne}, {Murphy},
  {Lenc}, {Leung}  \& {Moss}}{{Driessen} et~al.}{2023}]{Driessen+2023}
{Driessen} L.~N.,  {Heald} G.,  {Duchesne} S.~W.,  {Murphy} T.,  {Lenc} E.,
  {Leung} J.~K.,   {Moss} V.~A.,  2023, \mn@doi [\pasa] {10.1017/pasa.2023.26},
  \href {https://ui.adsabs.harvard.edu/abs/2023PASA...40...36D} {40, e036}

\bibitem[\protect\citeauthoryear{{Dubner}, {Moffett}, {Goss}  \&
  {Winkler}}{{Dubner} et~al.}{1993}]{Dubner+1993}
{Dubner} G.~M.,  {Moffett} D.~A.,  {Goss} W.~M.,   {Winkler} P.~F.,  1993,
  \mn@doi [\aj] {10.1086/116603}, \href
  {https://ui.adsabs.harvard.edu/abs/1993AJ....105.2251D} {105, 2251}

\bibitem[\protect\citeauthoryear{{Duchesne} et~al.,}{{Duchesne}
  et~al.}{2023}]{Duchesne+2023a}
{Duchesne} S.~W.,  et~al., 2023, \mn@doi [\pasa] {10.1017/pasa.2023.31}, \href
  {https://ui.adsabs.harvard.edu/abs/2023PASA...40...34D} {40, e034}

\bibitem[\protect\citeauthoryear{{Flagey}, {Noriega-Crespo}, {Petric}  \&
  {Geballe}}{{Flagey} et~al.}{2014}]{Flagey+2014}
{Flagey} N.,  {Noriega-Crespo} A.,  {Petric} A.,   {Geballe} T.~R.,  2014,
  \mn@doi [\aj] {10.1088/0004-6256/148/2/34}, \href
  {https://ui.adsabs.harvard.edu/abs/2014AJ....148...34F} {148, 34}

\bibitem[\protect\citeauthoryear{{Fomalont}}{{Fomalont}}{1999}]{Fomalont1999}
{Fomalont} E.~B.,  1999, in {Taylor} G.~B.,  {Carilli} C.~L.,   {Perley} R.~A.,
   eds,  Astronomical Society of the Pacific Conference Series Vol. 180,
  Synthesis Imaging in Radio Astronomy II. p.~301

\bibitem[\protect\citeauthoryear{{Fragkou}, {Parker}, {Boji{\v{c}}i{\'c}}  \&
  {Aksaker}}{{Fragkou} et~al.}{2018}]{Fragkou+2018}
{Fragkou} V.,  {Parker} Q.~A.,  {Boji{\v{c}}i{\'c}} I.~S.,   {Aksaker} N.,
  2018, \mn@doi [\mnras] {10.1093/mnras/sty1977}, \href
  {https://ui.adsabs.harvard.edu/abs/2018MNRAS.480.2916F} {480, 2916}

\bibitem[\protect\citeauthoryear{{Gaensler}, {Manchester}  \&
  {Green}}{{Gaensler} et~al.}{1998}]{GaenslerMG1998}
{Gaensler} B.~M.,  {Manchester} R.~N.,   {Green} A.~J.,  1998, \mn@doi [\mnras]
  {10.1046/j.1365-8711.1998.01387.x}, \href
  {https://ui.adsabs.harvard.edu/abs/1998MNRAS.296..813G} {296, 813}

\bibitem[\protect\citeauthoryear{{Gaensler}, {Arons}, {Kaspi}, {Pivovaroff},
  {Kawai}  \& {Tamura}}{{Gaensler} et~al.}{2002}]{Gaensler+2002}
{Gaensler} B.~M.,  {Arons} J.,  {Kaspi} V.~M.,  {Pivovaroff} M.~J.,  {Kawai}
  N.,   {Tamura} K.,  2002, \mn@doi [\apj] {10.1086/339354}, \href
  {https://ui.adsabs.harvard.edu/abs/2002ApJ...569..878G} {569, 878}

\bibitem[\protect\citeauthoryear{{Gaia Collaboration}}{{Gaia
  Collaboration}}{2022}]{Gaia+2022}
{Gaia Collaboration} 2022, \mn@doi [arXiv e-prints]
  {10.48550/arXiv.2208.00211}, p. arXiv:2208.00211

\bibitem[\protect\citeauthoryear{{Gerbrandt}, {Foster}, {Kothes},
  {Geisb{\"u}sch}  \& {Tung}}{{Gerbrandt} et~al.}{2014}]{Gerbrandt+2014}
{Gerbrandt} S.,  {Foster} T.~J.,  {Kothes} R.,  {Geisb{\"u}sch} J.,   {Tung}
  A.,  2014, \mn@doi [\aap] {10.1051/0004-6361/201423679}, \href
  {https://ui.adsabs.harvard.edu/abs/2014A&A...566A..76G} {566, A76}

\bibitem[\protect\citeauthoryear{{Giacani}, {Smith}, {Dubner}  \&
  {Loiseau}}{{Giacani} et~al.}{2011}]{Giacani+2011}
{Giacani} E.,  {Smith} M.~J.~S.,  {Dubner} G.,   {Loiseau} N.,  2011, \mn@doi
  [\aap] {10.1051/0004-6361/201116768}, \href
  {https://ui.adsabs.harvard.edu/abs/2011A&A...531A.138G} {531, A138}

\bibitem[\protect\citeauthoryear{{Green}}{{Green}}{2019}]{Green2019}
{Green} D.~A.,  2019, \mn@doi [Journal of Astrophysics and Astronomy]
  {10.1007/s12036-019-9601-6}, \href
  {https://ui.adsabs.harvard.edu/abs/2019JApA...40...36G} {40, 36}

\bibitem[\protect\citeauthoryear{{Green}, {Cram}, {Large}  \& {Ye}}{{Green}
  et~al.}{1999}]{Green+1999}
{Green} A.~J.,  {Cram} L.~E.,  {Large} M.~I.,   {Ye} T.,  1999, \mn@doi [\apjs]
  {10.1086/313208}, \href
  {https://ui.adsabs.harvard.edu/abs/1999ApJS..122..207G} {122, 207}

\bibitem[\protect\citeauthoryear{{Green} et~al.,}{{Green}
  et~al.}{2009}]{Green+2009}
{Green} J.~A.,  et~al., 2009, \mn@doi [\mnras]
  {10.1111/j.1365-2966.2008.14091.x}, \href
  {https://ui.adsabs.harvard.edu/abs/2009MNRAS.392..783G} {392, 783}

\bibitem[\protect\citeauthoryear{{Green}, {Reeves}  \& {Murphy}}{{Green}
  et~al.}{2014}]{GreenRM2014}
{Green} A.~J.,  {Reeves} S.~N.,   {Murphy} T.,  2014, \mn@doi [\pasa]
  {10.1017/pasa.2014.37}, \href
  {http://adsabs.harvard.edu/abs/2014PASA...31...42G} {31, e042}

\bibitem[\protect\citeauthoryear{{Griffith}, {Wright}, {Burke}  \&
  {Ekers}}{{Griffith} et~al.}{1995}]{Griffith+1995}
{Griffith} M.~R.,  {Wright} A.~E.,  {Burke} B.~F.,   {Ekers} R.~D.,  1995,
  \mn@doi [\apjs] {10.1086/192146}, \href
  {https://ui.adsabs.harvard.edu/abs/1995ApJS...97..347G} {97, 347}

\bibitem[\protect\citeauthoryear{{Grindlay} et~al.,}{{Grindlay}
  et~al.}{2005}]{Grindlay+2005}
{Grindlay} J.~E.,  et~al., 2005, \mn@doi [\apj] {10.1086/498106}, \href
  {https://ui.adsabs.harvard.edu/abs/2005ApJ...635..920G} {635, 920}

\bibitem[\protect\citeauthoryear{{G{\"u}del}}{{G{\"u}del}}{2002}]{Gudel:2002}
{G{\"u}del} M.,  2002, \mn@doi [Annual Review of Astronomy and Astrophysics]
  {10.1146/annurev.astro.40.060401.093806}, \href
  {https://ui.adsabs.harvard.edu/abs/2002ARA&A..40..217G} {40, 217}

\bibitem[\protect\citeauthoryear{{Gvaramadze}, {Kniazev}  \&
  {Fabrika}}{{Gvaramadze} et~al.}{2010}]{GvaramadzeKF2010}
{Gvaramadze} V.~V.,  {Kniazev} A.~Y.,   {Fabrika} S.,  2010, \mn@doi [\mnras]
  {10.1111/j.1365-2966.2010.16496.x}, \href
  {https://ui.adsabs.harvard.edu/abs/2010MNRAS.405.1047G} {405, 1047}

\bibitem[\protect\citeauthoryear{{Hale} et~al.,}{{Hale}
  et~al.}{2021}]{Hale+2021}
{Hale} C.~L.,  et~al., 2021, \mn@doi [\pasa] {10.1017/pasa.2021.47}, \href
  {https://ui.adsabs.harvard.edu/abs/2021PASA...38...58H} {38, e058}

\bibitem[\protect\citeauthoryear{{Hanaoka} et~al.,}{{Hanaoka}
  et~al.}{2019}]{Hanaoka+2019}
{Hanaoka} M.,  et~al., 2019, \mn@doi [\pasj] {10.1093/pasj/psy126}, \href
  {https://ui.adsabs.harvard.edu/abs/2019PASJ...71....6H} {71, 6}

\bibitem[\protect\citeauthoryear{{Hancock}, {Murphy}, {Gaensler}, {Hopkins}  \&
  {Curran}}{{Hancock} et~al.}{2012}]{Hancock+2012}
{Hancock} P.~J.,  {Murphy} T.,  {Gaensler} B.~M.,  {Hopkins} A.,   {Curran}
  J.~R.,  2012, {Aegean: Compact source finding in radio images}, Astrophysics
  Source Code Library, record ascl:1212.009 (\mn@eprint {ascl} {1212.009})

\bibitem[\protect\citeauthoryear{{Helfand}, {Becker}, {White}, {Fallon}  \&
  {Tuttle}}{{Helfand} et~al.}{2006}]{Helfand+2006}
{Helfand} D.~J.,  {Becker} R.~H.,  {White} R.~L.,  {Fallon} A.,   {Tuttle} S.,
  2006, \mn@doi [\aj] {10.1086/503253}, \href
  {http://adsabs.harvard.edu/abs/2006AJ....131.2525H} {131, 2525}

\bibitem[\protect\citeauthoryear{{Heywood} et~al.,}{{Heywood}
  et~al.}{2019}]{Heywood+2019}
{Heywood} I.,  et~al., 2019, \mn@doi [\nat] {10.1038/s41586-019-1532-5}, \href
  {https://ui.adsabs.harvard.edu/abs/2019Natur.573..235H} {573, 235}

\bibitem[\protect\citeauthoryear{{Heywood} et~al.,}{{Heywood}
  et~al.}{2022}]{Heywood+2022}
{Heywood} I.,  et~al., 2022, \mn@doi [\apj] {10.3847/1538-4357/ac449a}, \href
  {https://ui.adsabs.harvard.edu/abs/2022ApJ...925..165H} {925, 165}

\bibitem[\protect\citeauthoryear{{Hindson}, {Thompson}, {Urquhart}, {Faimali},
  {Johnston-Hollitt}, {Clark}  \& {Davies}}{{Hindson}
  et~al.}{2013}]{Hindson+2013}
{Hindson} L.,  {Thompson} M.~A.,  {Urquhart} J.~S.,  {Faimali} A.,
  {Johnston-Hollitt} M.,  {Clark} J.~S.,   {Davies} B.,  2013, \mn@doi [\mnras]
  {10.1093/mnras/stt1405}, \href
  {https://ui.adsabs.harvard.edu/abs/2013MNRAS.435.2003H} {435, 2003}

\bibitem[\protect\citeauthoryear{{Hoare} et~al.,}{{Hoare}
  et~al.}{2012}]{Hoare2012}
{Hoare} M.~G.,  et~al., 2012, \mn@doi [\pasp] {10.1086/668058}, \href
  {https://ui.adsabs.harvard.edu/abs/2012PASP..124..939H} {124, 939}

\bibitem[\protect\citeauthoryear{{Humphreys} \& {Davidson}}{{Humphreys} \&
  {Davidson}}{1994}]{Humphreys+1994}
{Humphreys} R.~M.,  {Davidson} K.,  1994, \mn@doi [\pasp] {10.1086/133478},
  \href {https://ui.adsabs.harvard.edu/abs/1994PASP..106.1025H} {106, 1025}

\bibitem[\protect\citeauthoryear{{Hurley-Walker} et~al.,}{{Hurley-Walker}
  et~al.}{2019a}]{Hurley-Walker+2019b}
{Hurley-Walker} N.,  et~al., 2019a, \mn@doi [\pasa] {10.1017/pasa.2019.37},
  \href {https://ui.adsabs.harvard.edu/abs/2019PASA...36...47H} {36, e047}

\bibitem[\protect\citeauthoryear{{Hurley-Walker} et~al.,}{{Hurley-Walker}
  et~al.}{2019b}]{Hurley-Walker+2019}
{Hurley-Walker} N.,  et~al., 2019b, \mn@doi [\pasa] {10.1017/pasa.2019.33},
  \href {https://ui.adsabs.harvard.edu/abs/2019PASA...36...48H} {36, e048}

\bibitem[\protect\citeauthoryear{{Ingallinera} et~al.,}{{Ingallinera}
  et~al.}{2014}]{Ingallinera+2014a}
{Ingallinera} A.,  et~al., 2014, \mn@doi [\mnras] {10.1093/mnras/stt2157},
  \href {https://ui.adsabs.harvard.edu/abs/2014MNRAS.437.3626I} {437, 3626}

\bibitem[\protect\citeauthoryear{{Ingallinera} et~al.,}{{Ingallinera}
  et~al.}{2016}]{Ingallinera+2016}
{Ingallinera} A.,  et~al., 2016, \mn@doi [\mnras] {10.1093/mnras/stw2053},
  \href {https://ui.adsabs.harvard.edu/abs/2016MNRAS.463..723I} {463, 723}

\bibitem[\protect\citeauthoryear{{Ingallinera} et~al.,}{{Ingallinera}
  et~al.}{2019}]{Ingallinera+2019}
{Ingallinera} A.,  et~al., 2019, \mn@doi [\mnras] {10.1093/mnras/stz2982},
  \href {https://ui.adsabs.harvard.edu/abs/2019MNRAS.490.5063I} {490, 5063}

\bibitem[\protect\citeauthoryear{{Ingallinera} et~al.,}{{Ingallinera}
  et~al.}{2022}]{Ingallinera+2022}
{Ingallinera} A.,  et~al., 2022, \mn@doi [\mnras] {10.1093/mnrasl/slac017},
  \href {https://ui.adsabs.harvard.edu/abs/2022MNRAS.512L..21I} {512, L21}

\bibitem[\protect\citeauthoryear{{Irabor} et~al.,}{{Irabor}
  et~al.}{2018}]{Irabor+2018}
{Irabor} T.,  et~al., 2018, \mn@doi [\mnras] {10.1093/mnras/sty1929}, \href
  {https://ui.adsabs.harvard.edu/abs/2018MNRAS.480.2423I} {480, 2423}

\bibitem[\protect\citeauthoryear{{Irabor} et~al.,}{{Irabor}
  et~al.}{2023}]{Irabor+2023}
{Irabor} T.,  et~al., 2023, \mn@doi [\mnras] {10.1093/mnras/stad005}, \href
  {https://ui.adsabs.harvard.edu/abs/2023MNRAS.520.1073I} {520, 1073}

\bibitem[\protect\citeauthoryear{{Jaffa}, {Whitworth}, {Clarke}  \&
  {Howard}}{{Jaffa} et~al.}{2018}]{Jaffa+2018}
{Jaffa} S.~E.,  {Whitworth} A.~P.,  {Clarke} S.~D.,   {Howard} A.~D.~P.,  2018,
  \mn@doi [\mnras] {10.1093/mnras/sty696}, \href
  {https://ui.adsabs.harvard.edu/abs/2018MNRAS.477.1940J} {477, 1940}

\bibitem[\protect\citeauthoryear{{Jonas} \& {MeerKAT Team}}{{Jonas} \& {MeerKAT
  Team}}{2016}]{Jonas2016}
{Jonas} J.,  {MeerKAT Team} 2016, in MeerKAT Science: On the Pathway to the
  SKA. p.~1, \mn@doi{10.22323/1.277.0001}

\bibitem[\protect\citeauthoryear{{J{\'o}zsa} et~al.,}{{J{\'o}zsa}
  et~al.}{2020}]{Jozsa+2020}
{J{\'o}zsa} G. I.~G.,  et~al., 2020, {CARACal: Containerized Automated Radio
  Astronomy Calibration pipeline} (\mn@eprint {ascl} {2006.014})

\bibitem[\protect\citeauthoryear{{Kalberla}, {Kerp}, {Haud}, {Winkel}, {Ben
  Bekhti}, {Fl{\"o}er}  \& {Lenz}}{{Kalberla} et~al.}{2016}]{Kalberla+2016}
{Kalberla} P.~M.~W.,  {Kerp} J.,  {Haud} U.,  {Winkel} B.,  {Ben Bekhti} N.,
  {Fl{\"o}er} L.,   {Lenz} D.,  2016, \mn@doi [\apj]
  {10.3847/0004-637X/821/2/117}, \href
  {https://ui.adsabs.harvard.edu/abs/2016ApJ...821..117K} {821, 117}

\bibitem[\protect\citeauthoryear{{Kalberla}, {Kerp}  \& {Haud}}{{Kalberla}
  et~al.}{2020}]{KalberlaKH2020}
{Kalberla} P.~M.~W.,  {Kerp} J.,   {Haud} U.,  2020, \mn@doi [\aap]
  {10.1051/0004-6361/202037602}, \href
  {https://ui.adsabs.harvard.edu/abs/2020A&A...639A..26K} {639, A26}

\bibitem[\protect\citeauthoryear{{Kalcheva}, {Hoare}, {Urquhart}, {Kurtz},
  {Lumsden}, {Purcell}  \& {Zijlstra}}{{Kalcheva} et~al.}{2018}]{Kalcheva+2018}
{Kalcheva} I.~E.,  {Hoare} M.~G.,  {Urquhart} J.~S.,  {Kurtz} S.,  {Lumsden}
  S.~L.,  {Purcell} C.~R.,   {Zijlstra} A.~A.,  2018, \mn@doi [\aap]
  {10.1051/0004-6361/201832734}, \href
  {https://ui.adsabs.harvard.edu/abs/2018A&A...615A.103K} {615, A103}

\bibitem[\protect\citeauthoryear{{Kaplan}, {Kulkarni}, {Frail}  \& {van
  Kerkwijk}}{{Kaplan} et~al.}{2002}]{Kaplan+2002}
{Kaplan} D.~L.,  {Kulkarni} S.~R.,  {Frail} D.~A.,   {van Kerkwijk} M.~H.,
  2002, \mn@doi [\apj] {10.1086/338039}, \href
  {https://ui.adsabs.harvard.edu/abs/2002ApJ...566..378K} {566, 378}

\bibitem[\protect\citeauthoryear{{Kesteven} \& {Caswell}}{{Kesteven} \&
  {Caswell}}{1987}]{KestevenC1987}
{Kesteven} M.~J.,  {Caswell} J.~L.,  1987, \aap, \href
  {https://ui.adsabs.harvard.edu/abs/1987A&A...183..118K} {183, 118}

\bibitem[\protect\citeauthoryear{{Knowles} et~al.,}{{Knowles}
  et~al.}{2022}]{Knowles+2022}
{Knowles} K.,  et~al., 2022, \mn@doi [\aap] {10.1051/0004-6361/202141488},
  \href {https://ui.adsabs.harvard.edu/abs/2022A&A...657A..56K} {657, A56}

\bibitem[\protect\citeauthoryear{{Kraan-Korteweg}, {Woudt}, {Cayatte},
  {Fairall}, {Balkowski}  \& {Henning}}{{Kraan-Korteweg}
  et~al.}{1996}]{Kraan-Korteweg+1996}
{Kraan-Korteweg} R.~C.,  {Woudt} P.~A.,  {Cayatte} V.,  {Fairall} A.~P.,
  {Balkowski} C.,   {Henning} P.~A.,  1996, \mn@doi [\nat] {10.1038/379519a0},
  \href {http://adsabs.harvard.edu/abs/1996Natur.379..519K} {379, 519}

\bibitem[\protect\citeauthoryear{{Kraan-Korteweg}, {Shafi}, {Koribalski},
  {Staveley-Smith}, {Buckland}, {Henning}  \& {Fairall}}{{Kraan-Korteweg}
  et~al.}{2008}]{Kraan-Korteweg+2008}
{Kraan-Korteweg} R.~C.,  {Shafi} N.,  {Koribalski} B.~S.,  {Staveley-Smith} L.,
   {Buckland} P.,  {Henning} P.~A.,   {Fairall} A.~P.,  2008, in Galaxies in
  the Local Volume. p.~13 (\mn@eprint {arXiv} {0710.1795}), \mn@doi{J47-52578}

\bibitem[\protect\citeauthoryear{{Kraan-Korteweg}, {Cluver}, {Bilicki},
  {Jarrett}, {Colless}, {Elagali}, {B{\"o}hringer}  \& {Chon}}{{Kraan-Korteweg}
  et~al.}{2017}]{Kraan-Korteweg+2017}
{Kraan-Korteweg} R.~C.,  {Cluver} M.~E.,  {Bilicki} M.,  {Jarrett} T.~H.,
  {Colless} M.,  {Elagali} A.,  {B{\"o}hringer} H.,   {Chon} G.,  2017, \mn@doi
  [\mnras] {10.1093/mnrasl/slw229}, \href
  {https://ui.adsabs.harvard.edu/abs/2017MNRAS.466L..29K} {466, L29}

\bibitem[\protect\citeauthoryear{{Kreckel}, {Platen}, {Arag{\'o}n-Calvo}, {van
  Gorkom}, {van de Weygaert}, {van der Hulst}  \& {Beygu}}{{Kreckel}
  et~al.}{2012}]{Kreckel+2012}
{Kreckel} K.,  {Platen} E.,  {Arag{\'o}n-Calvo} M.~A.,  {van Gorkom} J.~H.,
  {van de Weygaert} R.,  {van der Hulst} J.~M.,   {Beygu} B.,  2012, \mn@doi
  [\aj] {10.1088/0004-6256/144/1/16}, \href
  {https://ui.adsabs.harvard.edu/abs/2012AJ....144...16K} {144, 16}

\bibitem[\protect\citeauthoryear{{Kurapati} et~al.,}{{Kurapati}
  et~al.}{2024}]{Kurapati+2023}
{Kurapati} S.,  et~al., 2024, \mn@doi [\mnras] {10.1093/mnras/stad3823}, \href
  {https://ui.adsabs.harvard.edu/abs/2024MNRAS.528..542K} {528, 542}

\bibitem[\protect\citeauthoryear{{Kurtz}}{{Kurtz}}{2005}]{Kurtz2005}
{Kurtz} S.,  2005, in {Cesaroni} R.,  {Felli} M.,  {Churchwell} E.,
  {Walmsley} M.,  eds, ~ Vol. 227, Massive Star Birth: A Crossroads of
  Astrophysics. pp 111--119, \mn@doi{10.1017/S1743921305004424}

\bibitem[\protect\citeauthoryear{{Lang}, {Morris}  \& {Echevarria}}{{Lang}
  et~al.}{1999}]{LangME1999}
{Lang} C.~C.,  {Morris} M.,   {Echevarria} L.,  1999, \mn@doi [\apj]
  {10.1086/308012}, \href
  {https://ui.adsabs.harvard.edu/abs/1999ApJ...526..727L} {526, 727}

\bibitem[\protect\citeauthoryear{{Lang}, {Wang}, {Lu}  \& {Clubb}}{{Lang}
  et~al.}{2010}]{Lang+2010}
{Lang} C.~C.,  {Wang} Q.~D.,  {Lu} F.,   {Clubb} K.~I.,  2010, \mn@doi [\apj]
  {10.1088/0004-637X/709/2/1125}, \href
  {https://ui.adsabs.harvard.edu/abs/2010ApJ...709.1125L} {709, 1125}

\bibitem[\protect\citeauthoryear{{Langer}, {Hamann}, {Lennon}, {Najarro},
  {Pauldrach}  \& {Puls}}{{Langer} et~al.}{1994}]{Langer+1994}
{Langer} N.,  {Hamann} W.~R.,  {Lennon} M.,  {Najarro} F.,  {Pauldrach}
  A.~W.~A.,   {Puls} J.,  1994, \aap, \href
  {https://ui.adsabs.harvard.edu/abs/1994A&A...290..819L} {290, 819}

\bibitem[\protect\citeauthoryear{{Langston}, {Minter}, {D'Addario},
  {Eberhardt}, {Koski}  \& {Zuber}}{{Langston} et~al.}{2000}]{Langston+2000}
{Langston} G.,  {Minter} A.,  {D'Addario} L.,  {Eberhardt} K.,  {Koski} K.,
  {Zuber} J.,  2000, \mn@doi [\aj] {10.1086/301382}, \href
  {https://ui.adsabs.harvard.edu/abs/2000AJ....119.2801L} {119, 2801}

\bibitem[\protect\citeauthoryear{{Li}, {Wheeler}, {Bash}  \& {Jefferys}}{{Li}
  et~al.}{1991}]{Li+1991}
{Li} Z.,  {Wheeler} J.~C.,  {Bash} F.~N.,   {Jefferys} W.~H.,  1991, \mn@doi
  [\apj] {10.1086/170409}, \href
  {https://ui.adsabs.harvard.edu/abs/1991ApJ...378...93L} {378, 93}

\bibitem[\protect\citeauthoryear{{Li}, {Urquhart}, {Leurini}, {Csengeri},
  {Wyrowski}, {Menten}  \& {Schuller}}{{Li} et~al.}{2016}]{Li+2016}
{Li} G.-X.,  {Urquhart} J.~S.,  {Leurini} S.,  {Csengeri} T.,  {Wyrowski} F.,
  {Menten} K.~M.,   {Schuller} F.,  2016, \mn@doi [\aap]
  {10.1051/0004-6361/201527468}, \href
  {https://ui.adsabs.harvard.edu/abs/2016A&A...591A...5L} {591, A5}

\bibitem[\protect\citeauthoryear{{Low} et~al.,}{{Low} et~al.}{1984}]{Low+1984}
{Low} F.~J.,  et~al., 1984, \mn@doi [\apjl] {10.1086/184213}, \href
  {https://ui.adsabs.harvard.edu/abs/1984ApJ...278L..19L} {278, L19}

\bibitem[\protect\citeauthoryear{{Makhathini}}{{Makhathini}}{2018}]{Makhathini2018}
{Makhathini} S.,  2018, PhD thesis, Rhodes University, South Africa

\bibitem[\protect\citeauthoryear{{Manchester}, {Hobbs}, {Teoh}  \&
  {Hobbs}}{{Manchester} et~al.}{2005}]{Manchester+2005}
{Manchester} R.~N.,  {Hobbs} G.~B.,  {Teoh} A.,   {Hobbs} M.,  2005, \mn@doi
  [\aj] {10.1086/428488}, \href
  {https://ui.adsabs.harvard.edu/abs/2005AJ....129.1993M} {129, 1993}

\bibitem[\protect\citeauthoryear{{Maswanganye}}{{Maswanganye}}{2017}]{Maswanganye2017}
{Maswanganye} J.~P.,  2017, PhD thesis, North-West University

\bibitem[\protect\citeauthoryear{{Mauch}, {Murphy}, {Buttery}, {Curran},
  {Hunstead}, {Piestrzynski}, {Robertson}  \& {Sadler}}{{Mauch}
  et~al.}{2003}]{Mauch+2003}
{Mauch} T.,  {Murphy} T.,  {Buttery} H.~J.,  {Curran} J.,  {Hunstead} R.~W.,
  {Piestrzynski} B.,  {Robertson} J.~G.,   {Sadler} E.~M.,  2003, \mn@doi
  [\mnras] {10.1046/j.1365-8711.2003.06605.x}, \href
  {https://ui.adsabs.harvard.edu/abs/2003MNRAS.342.1117M} {342, 1117}

\bibitem[\protect\citeauthoryear{{Mauch} et~al.,}{{Mauch}
  et~al.}{2020}]{Mauch+2020}
{Mauch} T.,  et~al., 2020, \mn@doi [\apj] {10.3847/1538-4357/ab5d2d}, \href
  {https://ui.adsabs.harvard.edu/abs/2020ApJ...888...61M} {888, 61}

\bibitem[\protect\citeauthoryear{{Mayer} \& {Becker}}{{Mayer} \&
  {Becker}}{2021}]{MayerB2021}
{Mayer} M. G.~F.,  {Becker} W.,  2021, \mn@doi [\aap]
  {10.1051/0004-6361/202141119}, \href
  {https://ui.adsabs.harvard.edu/abs/2021A&A...651A..40M} {651, A40}

\bibitem[\protect\citeauthoryear{{McConnell} et~al.,}{{McConnell}
  et~al.}{2020}]{McConnell+2020}
{McConnell} D.,  et~al., 2020, \mn@doi [\pasa] {10.1017/pasa.2020.41}, \href
  {https://ui.adsabs.harvard.edu/abs/2020PASA...37...48M} {37, e048}

\bibitem[\protect\citeauthoryear{{Melrose} \& {Dulk}}{{Melrose} \&
  {Dulk}}{1982}]{Melrose1982}
{Melrose} D.~B.,  {Dulk} G.~A.,  1982, \mn@doi [\apj] {10.1086/160219}, \href
  {https://ui.adsabs.harvard.edu/abs/1982ApJ...259..844M} {259, 844}

\bibitem[\protect\citeauthoryear{{Miller} et~al.,}{{Miller}
  et~al.}{2010}]{Miller+2010}
{Miller} A.~A.,  et~al., 2010, \mn@doi [\mnras]
  {10.1111/j.1365-2966.2010.16280.x}, \href
  {https://ui.adsabs.harvard.edu/abs/2010MNRAS.404..305M} {404, 305}

\bibitem[\protect\citeauthoryear{{Mizuno} et~al.,}{{Mizuno}
  et~al.}{2010}]{Mizuno+2010}
{Mizuno} D.~R.,  et~al., 2010, \mn@doi [\aj] {10.1088/0004-6256/139/4/1542},
  \href {https://ui.adsabs.harvard.edu/abs/2010AJ....139.1542M} {139, 1542}

\bibitem[\protect\citeauthoryear{{Molinari} et~al.,}{{Molinari}
  et~al.}{2010}]{Molinari+2010}
{Molinari} S.,  et~al., 2010, \mn@doi [\pasp] {10.1086/651314}, \href
  {https://ui.adsabs.harvard.edu/abs/2010PASP..122..314M} {122, 314}

\bibitem[\protect\citeauthoryear{{Morris} \& {Serabyn}}{{Morris} \&
  {Serabyn}}{1996}]{MorrisS1996}
{Morris} M.,  {Serabyn} E.,  1996, \mn@doi [\araa]
  {10.1146/annurev.astro.34.1.645}, \href
  {https://ui.adsabs.harvard.edu/abs/1996ARA&A..34..645M} {34, 645}

\bibitem[\protect\citeauthoryear{{Nowak}, {Flagey}, {Noriega-Crespo}, {Billot},
  {Carey}, {Paladini}  \& {Van Dyk}}{{Nowak} et~al.}{2014}]{Nowak+2014}
{Nowak} M.,  {Flagey} N.,  {Noriega-Crespo} A.,  {Billot} N.,  {Carey} S.~J.,
  {Paladini} R.,   {Van Dyk} S.~D.,  2014, \mn@doi [\apj]
  {10.1088/0004-637X/796/2/116}, \href
  {https://ui.adsabs.harvard.edu/abs/2014ApJ...796..116N} {796, 116}

\bibitem[\protect\citeauthoryear{{Padmanabh} et~al.,}{{Padmanabh}
  et~al.}{2023}]{Padmanabh+2023}
{Padmanabh} P.~V.,  et~al., 2023, \mn@doi [\mnras] {10.1093/mnras/stad1900},
  \href {https://ui.adsabs.harvard.edu/abs/2023MNRAS.524.1291P} {524, 1291}

\bibitem[\protect\citeauthoryear{{Parker}, {Boji{\v{c}}i{\'c}}  \&
  {Frew}}{{Parker} et~al.}{2016}]{ParkerBF2016}
{Parker} Q.~A.,  {Boji{\v{c}}i{\'c}} I.~S.,   {Frew} D.~J.,  2016, in Journal
  of Physics Conference Series. p. 032008 (\mn@eprint {arXiv} {1603.07042}),
  \mn@doi{10.1088/1742-6596/728/3/032008}

\bibitem[\protect\citeauthoryear{{Pastorello} et~al.,}{{Pastorello}
  et~al.}{2018}]{Pastorello+2018}
{Pastorello} A.,  et~al., 2018, \mn@doi [\mnras] {10.1093/mnras/stx2668}, \href
  {https://ui.adsabs.harvard.edu/abs/2018MNRAS.474..197P} {474, 197}

\bibitem[\protect\citeauthoryear{Plavin, Cotton  \& Mauch}{Plavin
  et~al.}{2020}]{plavin+2020}
Plavin A.,  Cotton W.~D.,   Mauch T.,  2020, Obit Development Memo Series, 62,
  1

\bibitem[\protect\citeauthoryear{{Price}, {Egan}, {Carey}, {Mizuno}  \&
  {Kuchar}}{{Price} et~al.}{2001}]{Price+2001}
{Price} S.~D.,  {Egan} M.~P.,  {Carey} S.~J.,  {Mizuno} D.~R.,   {Kuchar}
  T.~A.,  2001, \mn@doi [\aj] {10.1086/320404}, \href
  {https://ui.adsabs.harvard.edu/abs/2001AJ....121.2819P} {121, 2819}

\bibitem[\protect\citeauthoryear{{Pritchard} et~al.,}{{Pritchard}
  et~al.}{2021}]{Pritchard2021}
{Pritchard} J.,  et~al., 2021, \mn@doi [\mnras] {10.1093/mnras/stab299}, \href
  {https://ui.adsabs.harvard.edu/abs/2021MNRAS.502.5438P} {502, 5438}

\bibitem[\protect\citeauthoryear{{Ranasinghe} \& {Leahy}}{{Ranasinghe} \&
  {Leahy}}{2022}]{RanasingheL2022}
{Ranasinghe} S.,  {Leahy} D.,  2022, \mn@doi [\apj] {10.3847/1538-4357/ac940a},
  \href {https://ui.adsabs.harvard.edu/abs/2022ApJ...940...63R} {940, 63}

\bibitem[\protect\citeauthoryear{{Ranasinghe}, {Leahy}  \& {Stil}}{{Ranasinghe}
  et~al.}{2021}]{Ranasinghe+2021}
{Ranasinghe} S.,  {Leahy} D.,   {Stil} J.,  2021, \mn@doi [Universe]
  {10.3390/universe7090338}, \href
  {https://ui.adsabs.harvard.edu/abs/2021Univ....7..338R} {7, 338}

\bibitem[\protect\citeauthoryear{{Reich}, {Reich}  \& {Fuerst}}{{Reich}
  et~al.}{1990}]{ReichRF1990}
{Reich} W.,  {Reich} P.,   {Fuerst} E.,  1990, \aaps, \href
  {https://ui.adsabs.harvard.edu/abs/1990A&AS...83..539R} {83, 539}

\bibitem[\protect\citeauthoryear{{Reynolds}}{{Reynolds}}{1994}]{Reynolds1994}
{Reynolds} J.~E.,  1994, {ATNF Memo}, {AT/39.3/040}

\bibitem[\protect\citeauthoryear{{Reynolds} et~al.,}{{Reynolds}
  et~al.}{2013}]{Reynolds+2013}
{Reynolds} M.~T.,  et~al., 2013, \mn@doi [\apj] {10.1088/0004-637X/766/2/112},
  \href {https://ui.adsabs.harvard.edu/abs/2013ApJ...766..112R} {766, 112}

\bibitem[\protect\citeauthoryear{{Richardson} \& {Mehner}}{{Richardson} \&
  {Mehner}}{2018}]{Richardson2018}
{Richardson} N.~D.,  {Mehner} A.,  2018, \mn@doi [Research Notes of the
  American Astronomical Society] {10.3847/2515-5172/aad1f3}, \href
  {https://ui.adsabs.harvard.edu/abs/2018RNAAS...2..121R} {2, 121}

\bibitem[\protect\citeauthoryear{{Roberts}, {Romani}, {Johnston}  \&
  {Green}}{{Roberts} et~al.}{1999}]{Roberts+1999}
{Roberts} M. S.~E.,  {Romani} R.~W.,  {Johnston} S.,   {Green} A.~J.,  1999,
  \mn@doi [\apj] {10.1086/307058}, \href
  {https://ui.adsabs.harvard.edu/abs/1999ApJ...515..712R} {515, 712}

\bibitem[\protect\citeauthoryear{{Romani} et~al.,}{{Romani}
  et~al.}{2023}]{Romani+2023}
{Romani} R.~W.,  et~al., 2023, \mn@doi [\apj] {10.3847/1538-4357/acfa02}, \href
  {https://ui.adsabs.harvard.edu/abs/2023ApJ...957...23R} {957, 23}

\bibitem[\protect\citeauthoryear{{Rosner} \& {Bodo}}{{Rosner} \&
  {Bodo}}{1996}]{RosnerB1996}
{Rosner} R.,  {Bodo} G.,  1996, \mn@doi [\apjl] {10.1086/310286}, \href
  {https://ui.adsabs.harvard.edu/abs/1996ApJ...470L..49R} {470, L49}

\bibitem[\protect\citeauthoryear{{Sabin} et~al.,}{{Sabin}
  et~al.}{2014}]{Sabin+2014}
{Sabin} L.,  et~al., 2014, \mn@doi [\mnras] {10.1093/mnras/stu1404}, \href
  {https://ui.adsabs.harvard.edu/abs/2014MNRAS.443.3388S} {443, 3388}

\bibitem[\protect\citeauthoryear{{Salvato} et~al.,}{{Salvato}
  et~al.}{2018}]{salvato2018}
{Salvato} M.,  et~al., 2018, \mn@doi [\mnras] {10.1093/mnras/stx2651}, \href
  {https://ui.adsabs.harvard.edu/abs/2018MNRAS.473.4937S} {473, 4937}

\bibitem[\protect\citeauthoryear{{S{\'a}nchez-Ayaso}, {Combi}, {Albacete
  Colombo}, {L{\'o}pez-Santiago}, {Mart{\'\i}}  \&
  {Mu{\~n}oz-Arjonilla}}{{S{\'a}nchez-Ayaso} et~al.}{2012}]{Sanchez-Ayoso+2012}
{S{\'a}nchez-Ayaso} E.,  {Combi} J.~A.,  {Albacete Colombo} J.~F.,
  {L{\'o}pez-Santiago} J.,  {Mart{\'\i}} J.,   {Mu{\~n}oz-Arjonilla} A.~J.,
  2012, \mn@doi [\apss] {10.1007/s10509-011-0886-4}, \href
  {https://ui.adsabs.harvard.edu/abs/2012Ap&SS.337..573S} {337, 573}

\bibitem[\protect\citeauthoryear{{Schisano} et~al.,}{{Schisano}
  et~al.}{2020}]{Schisano+2020}
{Schisano} E.,  et~al., 2020, \mn@doi [\mnras] {10.1093/mnras/stz3466}, \href
  {https://ui.adsabs.harvard.edu/abs/2020MNRAS.492.5420S} {492, 5420}

\bibitem[\protect\citeauthoryear{{Schuller} et~al.,}{{Schuller}
  et~al.}{2009}]{Schuller+2009}
{Schuller} F.,  et~al., 2009, \mn@doi [\aap] {10.1051/0004-6361/200811568},
  \href {https://ui.adsabs.harvard.edu/abs/2009A&A...504..415S} {504, 415}

\bibitem[\protect\citeauthoryear{{Silva}, {Flagey}, {Noriega-Crespo}, {Carey}
  \& {Ingallinera}}{{Silva} et~al.}{2017}]{Silva+2017}
{Silva} K.~M.,  {Flagey} N.,  {Noriega-Crespo} A.,  {Carey} S.,   {Ingallinera}
  A.,  2017, \mn@doi [\aj] {10.3847/1538-3881/153/3/115}, \href
  {https://ui.adsabs.harvard.edu/abs/2017AJ....153..115S} {153, 115}

\bibitem[\protect\citeauthoryear{{Slane}}{{Slane}}{2017}]{Slane2017}
{Slane} P.,  2017, in {Alsabti} A.~W.,  {Murdin} P.,  eds, , Handbook of
  Supernovae.
Berlin: Springer, p.~2159, \mn@doi{10.1007/978-3-319-21846-5_95}

\bibitem[\protect\citeauthoryear{{Soler} et~al.,}{{Soler}
  et~al.}{2020}]{Soler+2020}
{Soler} J.~D.,  et~al., 2020, \mn@doi [\aap] {10.1051/0004-6361/202038882},
  \href {https://ui.adsabs.harvard.edu/abs/2020A&A...642A.163S} {642, A163}

\bibitem[\protect\citeauthoryear{{Springob} et~al.,}{{Springob}
  et~al.}{2016}]{Springob+2016}
{Springob} C.~M.,  et~al., 2016, \mn@doi [\mnras] {10.1093/mnras/stv2648},
  \href {https://ui.adsabs.harvard.edu/abs/2016MNRAS.456.1886S} {456, 1886}

\bibitem[\protect\citeauthoryear{{Staveley-Smith}, {Kraan-Korteweg},
  {Schr{\"o}der}, {Henning}, {Koribalski}, {Stewart}  \&
  {Heald}}{{Staveley-Smith} et~al.}{2016}]{Staveley-Smith+2016}
{Staveley-Smith} L.,  {Kraan-Korteweg} R.~C.,  {Schr{\"o}der} A.~C.,  {Henning}
  P.~A.,  {Koribalski} B.~S.,  {Stewart} I.~M.,   {Heald} G.,  2016, \mn@doi
  [\aj] {10.3847/0004-6256/151/3/52}, \href
  {http://cdsads.u-strasbg.fr/abs/2016AJ....151...52S} {151, 52}

\bibitem[\protect\citeauthoryear{{Sternberg}, {Hoffmann}  \&
  {Pauldrach}}{{Sternberg} et~al.}{2003}]{Sternberg+2003}
{Sternberg} A.,  {Hoffmann} T.~L.,   {Pauldrach} A.~W.~A.,  2003, \mn@doi
  [\apj] {10.1086/379506}, \href
  {https://ui.adsabs.harvard.edu/abs/2003ApJ...599.1333S} {599, 1333}

\bibitem[\protect\citeauthoryear{{Steyn} et~al.,}{{Steyn}
  et~al.}{2024}]{Steyn+2023}
{Steyn} N.,  et~al., 2024, \mn@doi [\mnras] {10.1093/mnrasl/slad196}, \href
  {https://ui.adsabs.harvard.edu/abs/2024MNRAS.529L..88S} {529, L88}

\bibitem[\protect\citeauthoryear{{Taddia} et~al.,}{{Taddia}
  et~al.}{2020}]{Taddia+2020}
{Taddia} F.,  et~al., 2020, \mn@doi [\aap] {10.1051/0004-6361/201936654}, \href
  {https://ui.adsabs.harvard.edu/abs/2020A&A...638A..92T} {638, A92}

\bibitem[\protect\citeauthoryear{{Tammann}, {Loeffler}  \&
  {Schroeder}}{{Tammann} et~al.}{1994}]{TammannLS1994}
{Tammann} G.~A.,  {Loeffler} W.,   {Schroeder} A.,  1994, \mn@doi [\apjs]
  {10.1086/192002}, \href
  {https://ui.adsabs.harvard.edu/abs/1994ApJS...92..487T} {92, 487}

\bibitem[\protect\citeauthoryear{{Thompson} et~al.,}{{Thompson}
  et~al.}{2015}]{Thompson2015}
{Thompson} M.,  et~al., 2015, in Advancing Astrophysics with the Square
  Kilometre Array (AASKA14). p.~126 (\mn@eprint {arXiv} {1412.5554}),
  \mn@doi{10.22323/1.215.0126}

\bibitem[\protect\citeauthoryear{{Trigilio} et~al.,}{{Trigilio}
  et~al.}{2018}]{Trigilio+2018}
{Trigilio} C.,  et~al., 2018, \mn@doi [\mnras] {10.1093/mnras/sty2280}, \href
  {https://ui.adsabs.harvard.edu/abs/2018MNRAS.481..217T} {481, 217}

\bibitem[\protect\citeauthoryear{{Trushkin}}{{Trushkin}}{1999}]{Trushkin+1999}
{Trushkin} S.~A.,  1999, Odessa Astronomical Publications, \href
  {https://ui.adsabs.harvard.edu/abs/1999OAP....12..144T} {12, 144}

\bibitem[\protect\citeauthoryear{{Tully}}{{Tully}}{2008}]{Tully2008}
{Tully} R.~B.,  2008, in {Davies} J.~I.,  {Disney} M.~J.,  eds, ~ Vol. 244,
  Dark Galaxies and Lost Baryons. pp 146--151 (\mn@eprint {arXiv} {0708.0864}),
  \mn@doi{10.1017/S1743921307013932}

\bibitem[\protect\citeauthoryear{{Tully}, {Pomar{\`e}de}, {Graziani},
  {Courtois}, {Hoffman}  \& {Shaya}}{{Tully} et~al.}{2019}]{Tully+2019}
{Tully} R.~B.,  {Pomar{\`e}de} D.,  {Graziani} R.,  {Courtois} H.~M.,
  {Hoffman} Y.,   {Shaya} E.~J.,  2019, \mn@doi [\apj]
  {10.3847/1538-4357/ab2597}, \href
  {https://ui.adsabs.harvard.edu/abs/2019ApJ...880...24T} {880, 24}

\bibitem[\protect\citeauthoryear{{Umana}, {Buemi}, {Trigilio}, {Leto}, {Hora}
  \& {Fazio}}{{Umana} et~al.}{2011a}]{Umana+2011a}
{Umana} G.,  {Buemi} C.~S.,  {Trigilio} C.,  {Leto} P.,  {Hora} J.~L.,
  {Fazio} G.,  2011a, Bulletin de la Societe Royale des Sciences de Liege,
  \href {https://ui.adsabs.harvard.edu/abs/2011BSRSL..80..335U} {80, 335}

\bibitem[\protect\citeauthoryear{{Umana}, {Buemi}, {Trigilio}, {Leto},
  {Agliozzo}, {Ingallinera}, {Noriega-Crespo}  \& {Hora}}{{Umana}
  et~al.}{2011b}]{Umana+2011b}
{Umana} G.,  {Buemi} C.~S.,  {Trigilio} C.,  {Leto} P.,  {Agliozzo} C.,
  {Ingallinera} A.,  {Noriega-Crespo} A.,   {Hora} J.~L.,  2011b, \mn@doi
  [\apjl] {10.1088/2041-8205/739/1/L11}, \href
  {https://ui.adsabs.harvard.edu/abs/2011ApJ...739L..11U} {739, L11}

\bibitem[\protect\citeauthoryear{{Umana} et~al.,}{{Umana}
  et~al.}{2012}]{Umana+2012}
{Umana} G.,  et~al., 2012, \mn@doi [\mnras] {10.1111/j.1365-2966.2012.22018.x},
  \href {https://ui.adsabs.harvard.edu/abs/2012MNRAS.427.2975U} {427, 2975}

\bibitem[\protect\citeauthoryear{{Umana} et~al.,}{{Umana}
  et~al.}{2015a}]{Umana+2015b}
{Umana} G.,  et~al., 2015a, in Advancing Astrophysics with the Square Kilometre
  Array (AASKA14). p.~118 (\mn@eprint {arXiv} {1412.5833}),
  \mn@doi{10.22323/1.215.0118}

\bibitem[\protect\citeauthoryear{{Umana} et~al.,}{{Umana}
  et~al.}{2015b}]{Umana+2015}
{Umana} G.,  et~al., 2015b, \mn@doi [\mnras] {10.1093/mnras/stv1976}, \href
  {https://ui.adsabs.harvard.edu/abs/2015MNRAS.454..902U} {454, 902}

\bibitem[\protect\citeauthoryear{{Umana} et~al.,}{{Umana}
  et~al.}{2021}]{Umana+2021}
{Umana} G.,  et~al., 2021, \mn@doi [\mnras] {10.1093/mnras/stab1279}, \href
  {https://ui.adsabs.harvard.edu/abs/2021MNRAS.506.2232U} {506, 2232}

\bibitem[\protect\citeauthoryear{{Urquhart} et~al.,}{{Urquhart}
  et~al.}{2013}]{Urquhart+2013}
{Urquhart} J.~S.,  et~al., 2013, \mn@doi [\mnras] {10.1093/mnras/stt1310},
  \href {https://ui.adsabs.harvard.edu/abs/2013MNRAS.435..400U} {435, 400}

\bibitem[\protect\citeauthoryear{{Urquhart} et~al.,}{{Urquhart}
  et~al.}{2018}]{Urquhart+2018}
{Urquhart} J.~S.,  et~al., 2018, \mn@doi [\mnras] {10.1093/mnras/stx2258},
  \href {https://ui.adsabs.harvard.edu/abs/2018MNRAS.473.1059U} {473, 1059}

\bibitem[\protect\citeauthoryear{{Ustamujic} et~al.,}{{Ustamujic}
  et~al.}{2021}]{Ustamujic+2021}
{Ustamujic} S.,  et~al., 2021, \mn@doi [\aap] {10.1051/0004-6361/202141569},
  \href {https://ui.adsabs.harvard.edu/abs/2021A&A...654A.167U} {654, A167}

\bibitem[\protect\citeauthoryear{{Wachter}, {van Dyk}, {Hoard}  \&
  {Morris}}{{Wachter} et~al.}{2010}]{Wachter+2010}
{Wachter} S.,  {van Dyk} S.,  {Hoard} D.~W.,   {Morris} P.,  2010, in
  {Leitherer} C.,  {Bennett} P.~D.,  {Morris} P.~W.,   {Van Loon} J.~T.,  eds,
  Astronomical Society of the Pacific Conference Series Vol. 425, Hot and Cool:
  Bridging Gaps in Massive Star Evolution. p.~291

\bibitem[\protect\citeauthoryear{{Wang}, {Zhang}, {Jiang}, {Zhao}, {Chen},
  {Chen}, {Gao}  \& {Liu}}{{Wang} et~al.}{2020}]{Wang+2020}
{Wang} S.,  {Zhang} C.,  {Jiang} B.,  {Zhao} H.,  {Chen} B.,  {Chen} X.,  {Gao}
  J.,   {Liu} J.,  2020, \mn@doi [\aap] {10.1051/0004-6361/201936868}, \href
  {https://ui.adsabs.harvard.edu/abs/2020A&A...639A..72W} {639, A72}

\bibitem[\protect\citeauthoryear{{Weis} \& {Bomans}}{{Weis} \&
  {Bomans}}{2020}]{WeisB2020}
{Weis} K.,  {Bomans} D.~J.,  2020, \mn@doi [Galaxies]
  {10.3390/galaxies8010020}, \href
  {https://ui.adsabs.harvard.edu/abs/2020Galax...8...20W} {8, 20}

\bibitem[\protect\citeauthoryear{{Wells}, {Urquhart}, {Moore}, {Browning},
  {Ragan}, {Rigby}, {Eden}  \& {Thompson}}{{Wells} et~al.}{2022}]{Wells+2022}
{Wells} M.~R.~A.,  {Urquhart} J.~S.,  {Moore} T.~J.~T.,  {Browning} K.~E.,
  {Ragan} S.~E.,  {Rigby} A.~J.,  {Eden} D.~J.,   {Thompson} M.~A.,  2022,
  \mn@doi [\mnras] {10.1093/mnras/stac2420}, \href
  {https://ui.adsabs.harvard.edu/abs/2022MNRAS.516.4245W} {516, 4245}

\bibitem[\protect\citeauthoryear{{Wenger} et~al.,}{{Wenger}
  et~al.}{2000}]{Wenger:2000}
{Wenger} M.,  et~al., 2000, \mn@doi [Astronomy and Astrophysics Supplement
  Series] {10.1051/aas:2000332}, \href
  {https://ui.adsabs.harvard.edu/abs/2000A&AS..143....9W} {143, 9}

\bibitem[\protect\citeauthoryear{{Wenger} et~al.,}{{Wenger}
  et~al.}{2021}]{Wenger+2021}
{Wenger} T.~V.,  et~al., 2021, \mn@doi [\apjs] {10.3847/1538-4365/abf4d4},
  \href {https://ui.adsabs.harvard.edu/abs/2021ApJS..254...36W} {254, 36}

\bibitem[\protect\citeauthoryear{{Westmeier} et~al.,}{{Westmeier}
  et~al.}{2021}]{Westmeier+2021}
{Westmeier} T.,  et~al., 2021, \mn@doi [\mnras] {10.1093/mnras/stab1881}, \href
  {https://ui.adsabs.harvard.edu/abs/2021MNRAS.506.3962W} {506, 3962}

\bibitem[\protect\citeauthoryear{{Whiteoak}}{{Whiteoak}}{1993}]{Whiteoak1993}
{Whiteoak} J. B.~Z.,  1993, \mn@doi [\apj] {10.1086/173194}, \href
  {https://ui.adsabs.harvard.edu/abs/1993ApJ...415..701W} {415, 701}

\bibitem[\protect\citeauthoryear{{Whiteoak} \& {Green}}{{Whiteoak} \&
  {Green}}{1996}]{WhiteoakG1996}
{Whiteoak} J.~B.~Z.,  {Green} A.~J.,  1996, \aaps, \href
  {https://ui.adsabs.harvard.edu/abs/1996A&AS..118..329W} {118, 329}

\bibitem[\protect\citeauthoryear{{Woudt}, {Kraan-Korteweg}, {Cayatte},
  {Balkowski}  \& {Felenbok}}{{Woudt} et~al.}{2004}]{Woudt+2004}
{Woudt} P.~A.,  {Kraan-Korteweg} R.~C.,  {Cayatte} V.,  {Balkowski} C.,
  {Felenbok} P.,  2004, \mn@doi [\aap] {10.1051/0004-6361:20034600}, \href
  {https://ui.adsabs.harvard.edu/abs/2004A&A...415....9W} {415, 9}

\bibitem[\protect\citeauthoryear{{Wright} et~al.,}{{Wright}
  et~al.}{2010}]{Wright+2010}
{Wright} E.~L.,  et~al., 2010, \mn@doi [\aj] {10.1088/0004-6256/140/6/1868},
  \href {http://adsabs.harvard.edu/abs/2010AJ....140.1868W} {140, 1868}

\bibitem[\protect\citeauthoryear{{Yang}, {Thompson}, {Tian}, {Bihr}, {Beuther}
  \& {Hindson}}{{Yang} et~al.}{2019}]{Yang+2019}
{Yang} A.~Y.,  {Thompson} M.~A.,  {Tian} W.~W.,  {Bihr} S.,  {Beuther} H.,
  {Hindson} L.,  2019, \mn@doi [\mnras] {10.1093/mnras/sty2811}, \href
  {https://ui.adsabs.harvard.edu/abs/2019MNRAS.482.2681Y} {482, 2681}

\bibitem[\protect\citeauthoryear{{Yu}, {Zijlstra}  \& {Jiang}}{{Yu}
  et~al.}{2021}]{Yu:2021}
{Yu} B.,  {Zijlstra} A.,   {Jiang} B.,  2021, \mn@doi [Universe]
  {10.3390/universe7050119}, \href
  {https://ui.adsabs.harvard.edu/abs/2021Univ....7..119Y} {7, 119}

\bibitem[\protect\citeauthoryear{{Yusef-Zadeh}}{{Yusef-Zadeh}}{1990}]{Yusef-Zadeh1990}
{Yusef-Zadeh} F.,  1990, \mn@doi [\apjl] {10.1086/185817}, \href
  {https://ui.adsabs.harvard.edu/abs/1990ApJ...361L..19Y} {361, L19}

\bibitem[\protect\citeauthoryear{{Yusef-Zadeh}, {Morris}  \&
  {Chance}}{{Yusef-Zadeh} et~al.}{1984}]{Yusef-ZadehMC1984}
{Yusef-Zadeh} F.,  {Morris} M.,   {Chance} D.,  1984, \mn@doi [\nat]
  {10.1038/310557a0}, \href
  {https://ui.adsabs.harvard.edu/abs/1984Natur.310..557Y} {310, 557}

\bibitem[\protect\citeauthoryear{{Yusef-Zadeh}, {Hewitt}  \&
  {Cotton}}{{Yusef-Zadeh} et~al.}{2004}]{Yusef-ZadehHC2004}
{Yusef-Zadeh} F.,  {Hewitt} J.~W.,   {Cotton} W.,  2004, \mn@doi [\apjs]
  {10.1086/425257}, \href
  {https://ui.adsabs.harvard.edu/abs/2004ApJS..155..421Y} {155, 421}

\bibitem[\protect\citeauthoryear{{Yusef-Zadeh}, {Arendt}, {Wardle}, {Heywood}
  \& {Cotton}}{{Yusef-Zadeh} et~al.}{2022a}]{Yusef-Zadeh+2022D}
{Yusef-Zadeh} F.,  {Arendt} R.~G.,  {Wardle} M.,  {Heywood} I.,   {Cotton} W.,
  2022a, \mn@doi [\mnras] {10.1093/mnras/stac2415}, \href
  {https://ui.adsabs.harvard.edu/abs/2022MNRAS.517..294Y} {517, 294}

\bibitem[\protect\citeauthoryear{{Yusef-Zadeh}, {Arendt}, {Wardle}, {Heywood},
  {Cotton}  \& {Camilo}}{{Yusef-Zadeh} et~al.}{2022b}]{Yusef-Zadeh+2022A}
{Yusef-Zadeh} F.,  {Arendt} R.~G.,  {Wardle} M.,  {Heywood} I.,  {Cotton} W.,
  {Camilo} F.,  2022b, \mn@doi [\apjl] {10.3847/2041-8213/ac4802}, \href
  {https://ui.adsabs.harvard.edu/abs/2022ApJ...925L..18Y} {925, L18}

\bibitem[\protect\citeauthoryear{{Zijlstra} \& {Pottasch}}{{Zijlstra} \&
  {Pottasch}}{1991}]{ZijlstraP1991}
{Zijlstra} A.~A.,  {Pottasch} S.~R.,  1991, \aap, \href
  {https://ui.adsabs.harvard.edu/abs/1991A&A...243..478Z} {243, 478}

\makeatother
\end{thebibliography}




\appendix

\section{The polarization of supernova remnant W44}
\label{sec:W44}

Although for the majority of the \SMGPS\ data no polarization calibration
was done, all four correlation products (XX, YY, XY and YX) were obtained,
and generally sufficient calibrators were observed for full polarization
calibration, so a full polarization calibration can be realized using the
raw visibility data from the SARAO archive.  As an example, we show here
some results from a single pointing which was processed in this manner.

The \MKGPS\ pointing N12R01C01 in Session 82 of the survey
is centered near SNR W44 (G34.7$-$0.4). This pointing was 
recalibrated including polarization.
The data were re-imaged using
multiresolution CLEAN in Stokes $I, Q, U$ and $V$, with 
1\% fractional bandwidth.  This should give good sensitivity to
Faraday depths $\pm 1700$ rad m$^{-2}$ 
(Obit memo 82\footnote{
\url{https://www.cv.nrao.edu/~bcotton/ObitDoc/RMGain.pdf}}).
    
\begin{figure*}
\centering
\includegraphics[width=0.49\linewidth, trim=0.9in 1.7in 0.9in 2.2in,clip]{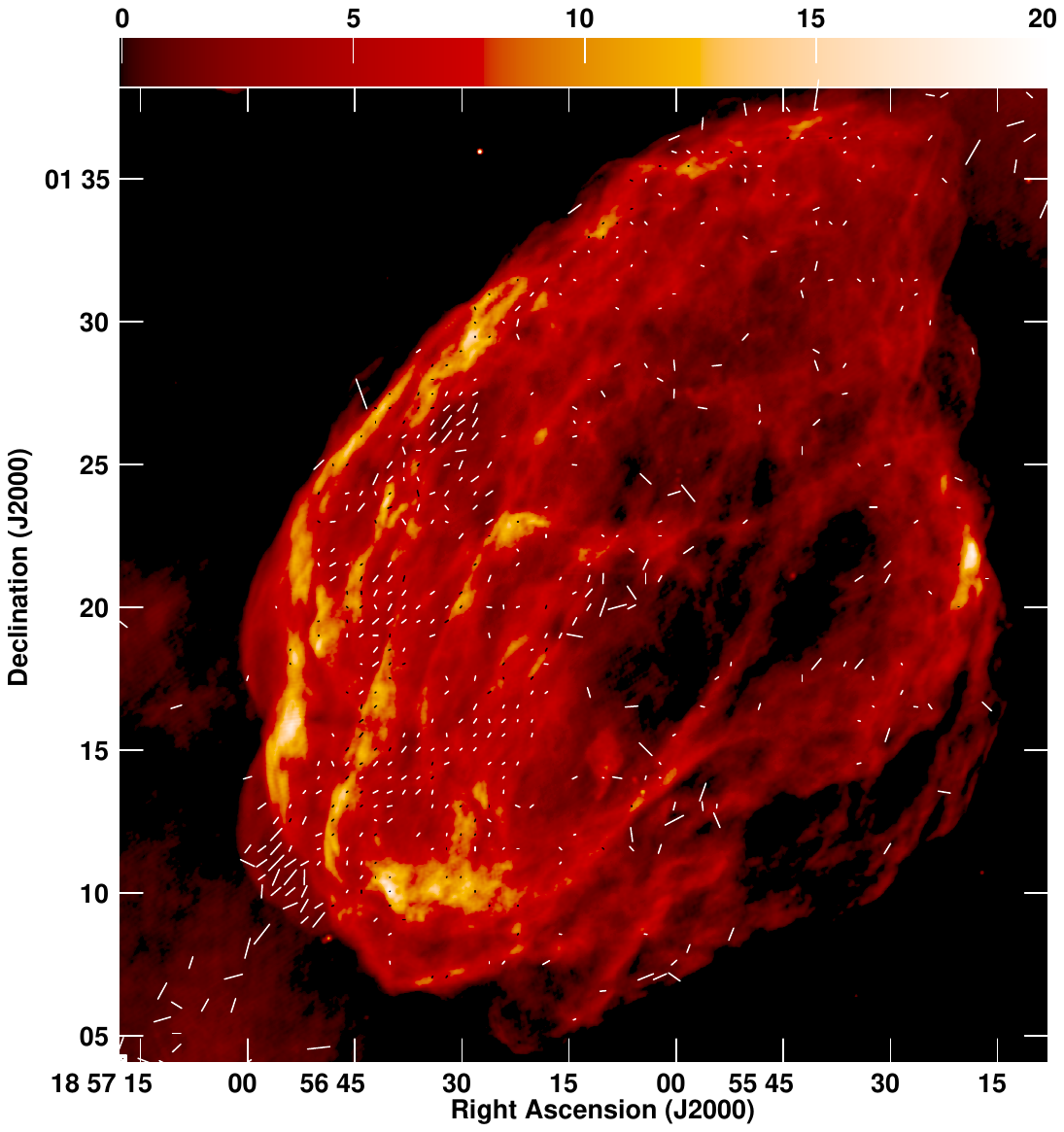}
\includegraphics[width=0.49\linewidth, trim=0.9in 1.7in 0.9in 2.2in,clip]{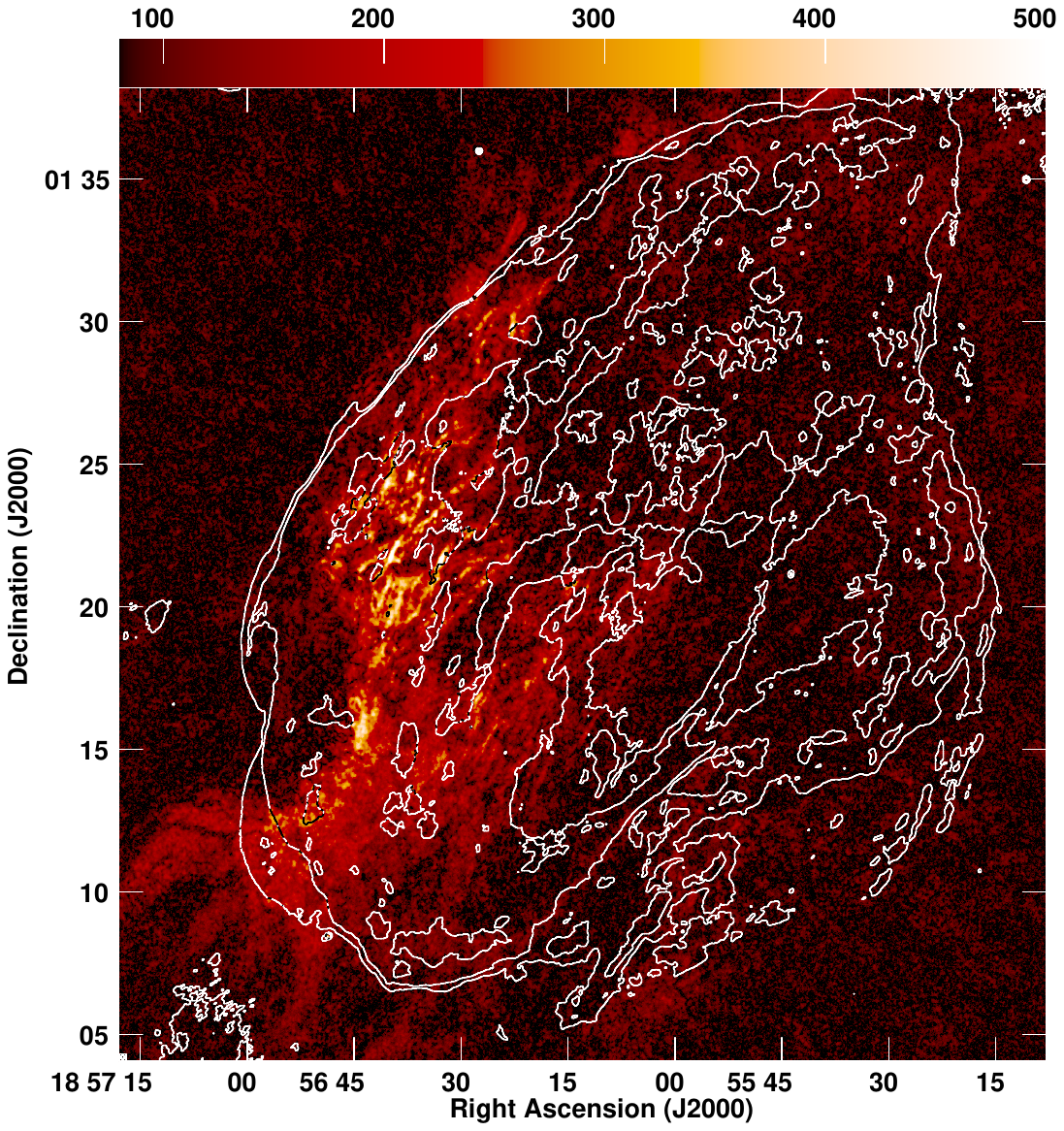}
\includegraphics[width=0.49\linewidth, trim=0.9in 1.7in 0.9in 2.2in,clip]{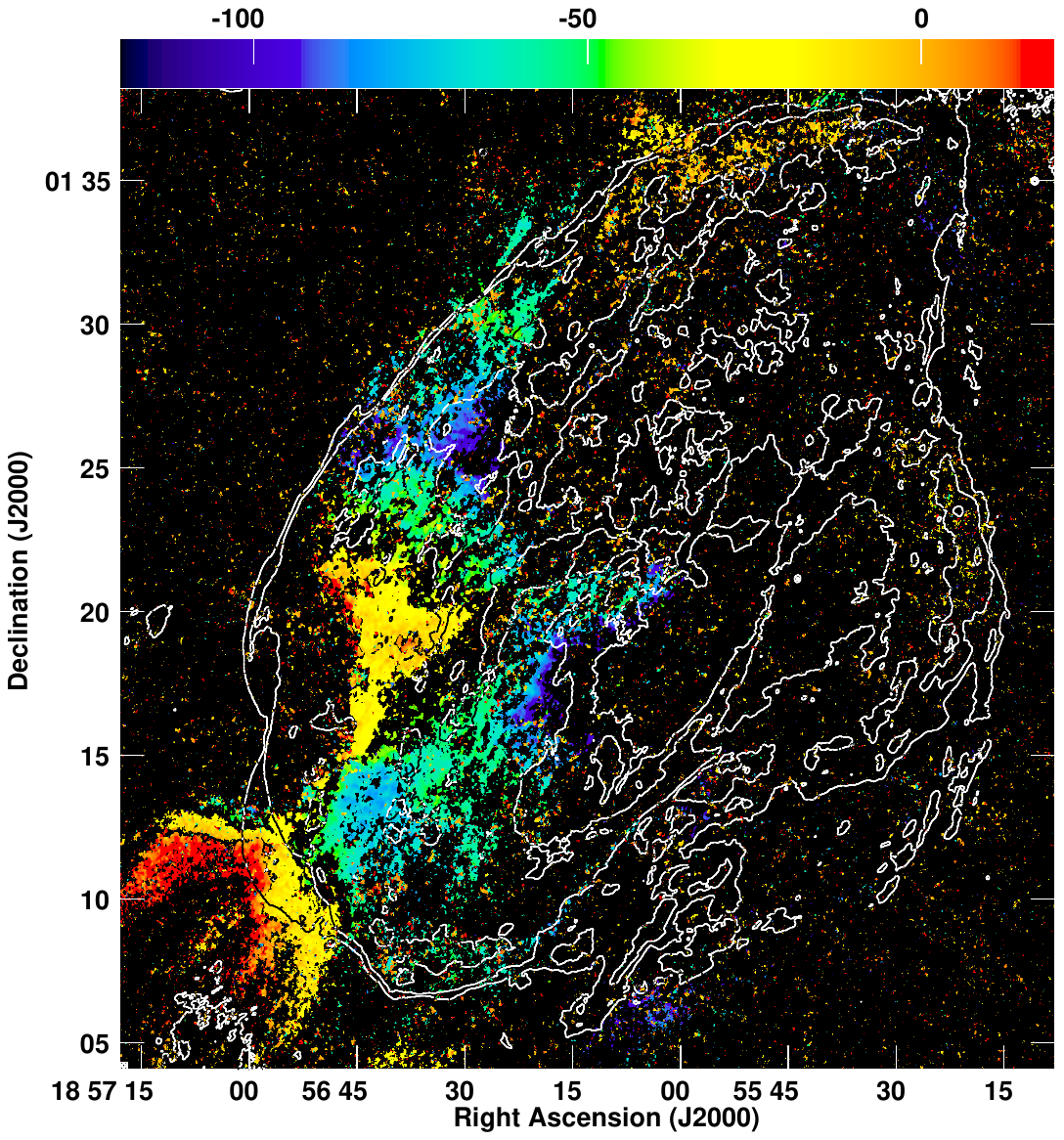}
\caption{\emph{Left:} Radio image of the SNR W44, with the colorscale showing Stokes $I$,
and labelled in m\Jb, 
and the white or black line segments indicating polarized emission.  The 
direction of the line segments indicates the direction of projected magnetic field,
while the length indicates the amount of polarized emission.
\emph{Right:} The polarized intensity in SNR W44\@. The colorscale
shows the polarized brightness, labelled in $\mu$\Jb. The contours show the Stokes
$I$ brightness at 1, 4, and 16 m\Jb.
\emph{Bottom:} The rotation measure (RM) of the SNR W44\@. The colorscale
shows the RM, labelled in rad~m$^{-2}$. The contours show the Stokes
$I$ brightness at 1, 4, and 16 m\Jb.}
\label{fig:W44pol}
\end{figure*}

For the Stokes $I$ imaging, we used resolutions of 7\farcs8, 
65\arcsec, and  128\arcsec, and CLEANed
1,414,479 components (loop gain 0.03), resulting in a total
CLEANed flux density of 81.6 Jy, with CLEANing stopped when
the residuals were  $\lesssim$2\,m\Jb. 
The Stokes $I$ image is shown in Fig.\ \ref{fig:W44pol}.
The $Q$ and $U$ cubes were subjected to a
peak Faraday depth least squares fitting search (Obit function
RMFit.Cube) in the range $\pm$2000 rad m$^{-2}$, to determine a single best-fit rotation measure (RM). A portion of the cube was
examined using a Faraday Synthesis, to determine a Faraday spectrum, in the range $\pm$800 rad m$^{-2}$.
The direction of the projected magnetic field, the polarized intensity image,
and the RM are also shown in Fig.~\ref{fig:W44pol}.

The south eastern portion of the SNR has a region of relatively
bright Stokes $I$, but no detectable polarized emission in spite of
a search over a wide range of Faraday depth.  This is plausibly the 
result of
beam depolarization in which the range of Faraday depths inside the
resolution element is large enough to depolarize the emission.

\section{Author Affiliations}
\label{sec:affiliations}

$^{1}$South African Radio Astronomy Observatory, 2 Fir Street, Observatory 7925, South Africa\\
$^{2}$SKA Observatory, 2 Fir Street, Observatory 7925, South Africa\\
$^{3}$National Radio Astronomy Observatory, 520 Edgemont Road, Charlottesville, VA 22903, USA\\
$^{4}$School of Physics and Astronomy, University of Leeds, Leeds LS2 9JT, UK\\
$^{5}$INAF -- Osservatorio Astrofisico di Catania, Via S. Sofia 78, I-95123 Catania, Italy\\
$^{6}$SARAO/Hartebeesthoek Radio Astronomy Observatory, PO Box 443, Krugersdorp 1740, South Africa\\
$^{7}$Department of Physics and Astronomy, York University, Toronto, M3J 1P3, Ontario, Canada\\
$^{8}$Department of Astronomy, University of Cape Town, Private Bag X3, Rondebosch 7701, South Africa\\
$^{9}$Department of Physics and Astronomy, West Virginia University, Morgantown, WV 26506, USA\\
$^{10}$Adjunct Astronomer at the Green Bank Observatory, P.O. Box 2, Green Bank, WV 24944, USA\\
$^{11}$Center for Gravitational Waves and Cosmology, West Virginia University, Chestnut Ridge Research Building, Morgantown, WV 26505, USA\\
$^{12}$South African Astronomical Observatory, PO Box 9, Observatory 7935, Cape Town, South Africa\\
$^{13}$Department of Physics, University of the Free State, PO Box 339, Bloemfontein 9300, South Africa\\
$^{14}$Research Center for Intelligent Computing Platforms, Zhejiang Laboratory, Hangzhou 311100, PR China\\
$^{15}$Department of Mathematical Sciences, University of South Africa, Cnr Christian de Wet Rd and Pioneer Avenue, Florida Park, 1709, Roodepoort, South Africa\\
$^{16}$Centre for Space Research, Physics Department, North-West University, Potchefstroom 2520, South Africa\\
$^{17}$Department of Physics and Astronomy, Faculty of Physical Sciences, University of Nigeria, Carver Building, 1 University Road, Nsukka 410001, Nigeria\\
$^{18}$The Inter-University Institute for Data Intensive Astronomy (IDIA), and University of Cape Town, Private Bag X3, Rondebosch 7701, South Africa\\
$^{19}$UK Astronomy Technology Centre, Royal Observatory Edinburgh, Blackford Hill, Edinburgh EH9 3HJ, UK\\
$^{20}$Department of Astronomy and Space Science, Technical University of Kenya, Nairobi, Kenya\\
$^{21}$Astro Space Center of Lebedev Physical Institute, Profsouznaya str. 84/32, Moscow 117997, Russia\\
$^{22}$Black Hole Initiative at Harvard University, 20 Garden Street, Cambridge, MA 02138, USA\\
$^{23}$INAF -- Osservatorio Astronomico di Cagliari, Via della Scienza 5, 09047 Selargius, CA, Italy\\
$^{24}$Department of Astronomy, University of Washington, Seattle, WA 98195, USA\\
$^{25}$Department of Physics, Astronomy, and Mathematics, University of Hertfordshire, Hatfield AL10 9AB, UK\\
$^{26}$Jodrell Bank Centre for Astrophysics, Department of Physics and Astronomy, The University of Manchester, Manchester M13 9PL, UK\\
$^{27}$International Centre for Radio Astronomy Research (ICRAR), The University of Western Australia, 35 Stirling Highway, Australia\\
$^{28}$Department of Physics, Indian Institute of Science Education and Research Bhopal, Bhopal Bypass Road, Bhauri, Bhopal 462 066, Madhya Pradesh, India\\
$^{29}$Department of Physics, Aberystwyth University, Ceredigion, Cymru, SY23 3BZ, UK\\
$^{30}$SKA Observatory, Jodrell Bank, Lower Withington, Macclesfield, Cheshire, SK11 9FT, UK\\
$^{31}$Department of Physical Sciences, Independent University, Bangladesh, Bashundhara RA, Dhaka 1229, Bangladesh\\
$^{32}$Department of Electrical and Electronic Engineering, Stellenbosch University, Stellenbosch 7600, South Africa\\
$^{33}$EMSS Antennas, 18 Techno Avenue, Technopark, Stellenbosch 7600, South Africa\\
$^{34}$High Energy Physics, Cosmology and Astrophysics Theory (HEPCAT) Group, Department of Mathematics and Applied Mathematics, University of Cape Town, Rondebosch 7701, South Africa\\
$^{35}$Oxford Astrophysics, Denys Wilkinson Building, Keble Road, Oxford OX1 3RH, UK\\
$^{36}$Department of Physics and Electronics, Rhodes University, PO Box 94, Grahamstown 6140, South Africa\\
$^{37}$DeepAlert (Pty) Ltd., 12 Blaauwklippen Rd, Kirstenhof 7945, Cape Town, South Africa\\
$^{38}$Presidential Infrastructure Coordinating Commission, 77 Meintjies Street, Sunnyside, Pretoria 0001, South Africa\\
$^{39}$African Institute for Mathematical Sciences, 6 Melrose Road, Muizenberg 7945, South Africa\\
$^{40}$Tsolo Storage Systems (Pty) Ltd., 12 Links Drive, Pinelands 7405, South Africa\\
$^{41}$Department of Physics and Astronomy, University of the Western Cape, Bellville, Cape Town 7535, South Africa\\
$^{42}$INAF -- Istituto di Radioastronomia, Via Gobetti 101, 40129 Bologna, Italy\\


\bsp	
\label{lastpage}
\end{document}